\documentclass[aps,superscriptaddress,showpacs,letterpaper,twocolumn,reprint,nofootinbib,longbibliography]{revtex4-1}
\usepackage{graphicx} 
\usepackage{amsmath} 
\usepackage{amssymb}   
\usepackage{amsfonts}
\usepackage{amsthm}
\usepackage{color}  
\usepackage{subfig}

\graphicspath{{./Figures/}}

\newcommand{\ket}[1]{| #1 \rangle}

\usepackage{qcircuit}

\begin{document}

\title{Demonstration of multi-qubit entanglement and  algorithms on a \\programmable neutral atom quantum computer}
\author{T. M. Graham}
\author{Y. Song}
\author{J. Scott}
\author{C. Poole}
\author{L. Phuttitarn}
\author{K. Jooya}
\author{P. Eichler}
\author{X. Jiang}
\author{A. Marra}
\altaffiliation{Present address: Department of Physics, University of Central Florida}
\author{B. Grinkemeyer}
\altaffiliation{Present address: Department of Physics, Harvard University}
\author{M. Kwon}
\altaffiliation{Present address: Department of Physics, Columbia University}
\affiliation{
Department of Physics,  University of Wisconsin-Madison, 
Madison, WI, 53706, USA
}
\author{M. Ebert}
\affiliation{ColdQuanta, Inc., 111 N Fairchild St, Madison, WI 53703, USA }
\author{J. Cherek}
\affiliation{ColdQuanta, Inc., 3030 Sterling Circle, Boulder, CO 80301, USA }
\author{M. T. Lichtman}
\author{M. Gillette}
\affiliation{ColdQuanta, Inc., 111 N Fairchild St, Madison, WI 53703, USA }
\author {J. Gilbert}
\affiliation{ColdQuanta, Inc., 3030 Sterling Circle, Boulder, CO 80301, USA }
\author{D. Bowman}
\author{T. Ballance}
\affiliation{ColdQuanta UK, Oxford Centre for Innovation, Oxford OX1 1BY, UK}
\author{C. Campbell}
\author{E. D. Dahl}
\affiliation{ColdQuanta, Inc., 3030 Sterling Circle, Boulder, CO 80301, USA }
\author{O. Crawford}
\author{N. S. Blunt}
\author{B. Rogers}
\affiliation{Riverlane, Cambridge CB2 3BZ,  UK}
\author{T. Noel}
\affiliation{ColdQuanta, Inc., 3030 Sterling Circle, Boulder, CO 80301, USA }
\author{M. Saffman}
\affiliation{
Department of Physics,  University of Wisconsin-Madison, 
Madison, WI, 53706, USA
}
\affiliation{ColdQuanta, Inc., 111 N Fairchild St, Madison, WI 53703, USA }

 \date{\today}

\begin{abstract} {\bf Gate model quantum computers promise to solve currently intractable computational problems if they can be operated at scale with long coherence times and high fidelity logic. Neutral atom hyperfine qubits provide inherent scalability due to their identical characteristics, long coherence times, and ability to be trapped in dense multi-dimensional arrays\cite{Saffman2010}. Combined with the strong entangling interactions provided by Rydberg states\cite{Jaksch2000,Gaetan2009,Urban2009}, all the necessary characteristics for  quantum computation are available.  
Here we demonstrate several quantum algorithms on a programmable gate model neutral atom quantum computer in an architecture based on individual addressing of single atoms with tightly focused optical beams scanned across a two-dimensional array of qubits. Preparation of entangled Greenberger-Horne-Zeilinger (GHZ) states\cite{Greenberger1989} with up to 6 qubits, quantum phase estimation for a chemistry problem\cite{Aspuru-Guzik2005}, and the Quantum Approximate Optimization Algorithm (QAOA)\cite{Farhi2014} for the MaxCut graph problem are demonstrated. These results highlight the emergent capability of neutral atom qubit arrays for universal, programmable quantum computation, as well as  preparation of non-classical states of use for quantum enhanced sensing. }
\end{abstract}

\maketitle


Remarkable progress was made in recent years in the development of quantum computers which use quantum states and operations to  encode and process information. Such quantum computers promise to solve certain classes of computing problems exponentially faster than modern transistor-based computers. However, quantum bits (qubits) are fragile and degrade if not isolated from environmental noise, yet must interact with other qubits to perform calculations. Many physical systems have been used to address these challenges. Digital quantum circuits have been demonstrated with trapped ion\cite{Martinez2016,Figgatt2017}, superconducting\cite{DiCarlo2009,Harrigan2021}, quantum dot\cite{Watson2018}, and optical\cite{XQZhou2013} processors. Neutral atom arrays have been used for analog quantum simulation with up to hundreds of interacting spins\cite{Scholl2021,Ebadi2021}. Although powerful, the reliability of analog simulation techniques without error correction for complex problems with large qubit numbers remains an open question\cite{Hauke2012}.  Digital gate model quantum circuits are provably compatible with error correction which enables large scale computation\cite{Aharonov2008,Knill1998}. We demonstrate here, for the first time, quantum algorithms encoded in gate model digital  circuits on a programmable neutral atom processor. 

 \begin{figure*}[!t] 
 \includegraphics[width=15.cm]{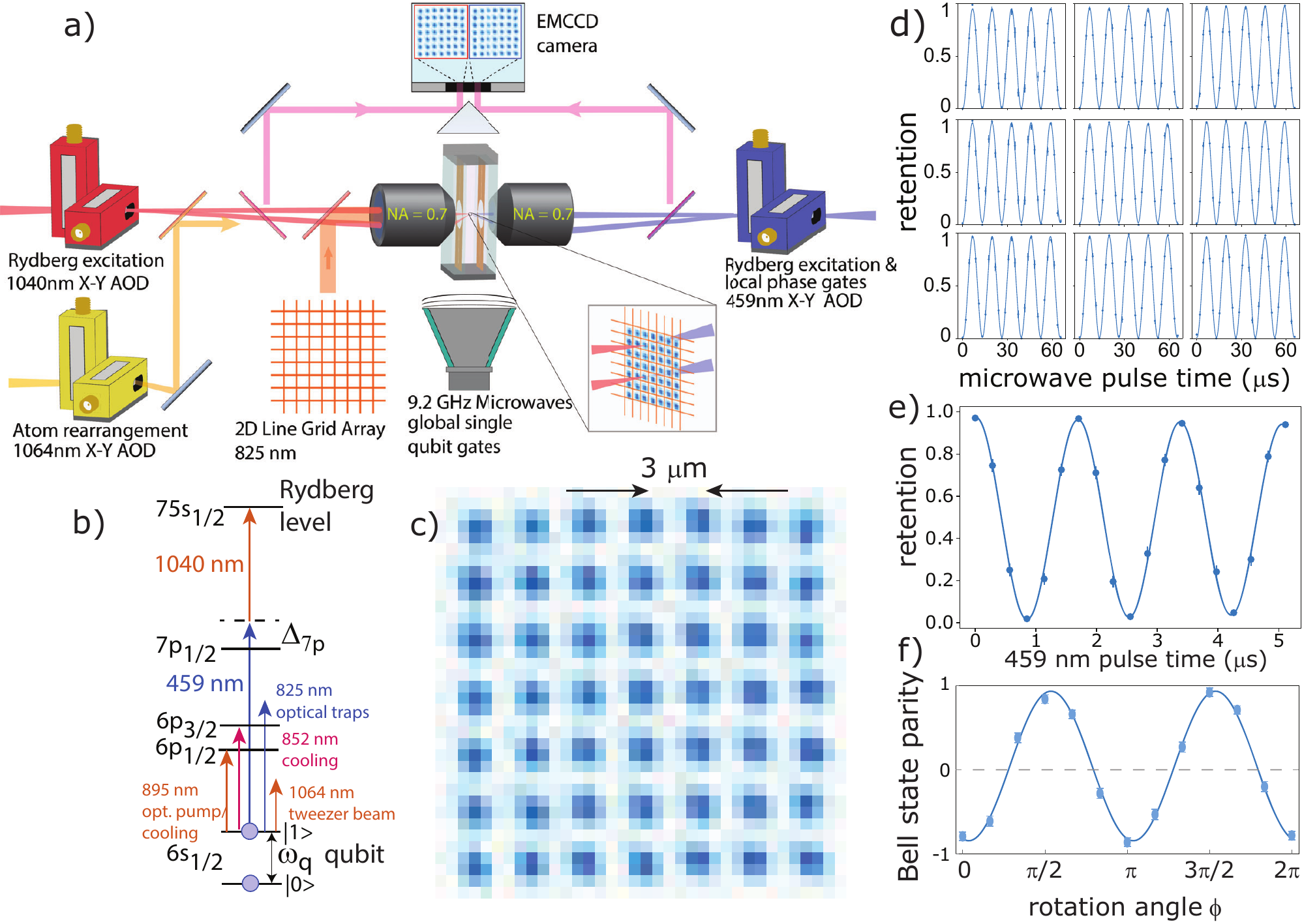}
     \caption{\label{fig.Full_Assembly} \textbf{Experimental Quantum Computing Platform} \textbf{a)} Experimental layout for trapping and addressing atomic qubits. Atoms are trapped in a blue-detuned line grid array (see Methods and \cite{SM2021circuits} for details), which is imaged onto the atom trapping region with a NA$=0.7$ lens. Atom occupation is determined by collecting atomic fluorescence using NA$=0.7$ lenses at opposite faces of the cell and imaging the light onto two separate regions of an EMCCD camera.  A 1064 nm tweezer beam  is used to rearrange atoms into desired sites for circuit operation. Circuits are decomposed into a universal gate set consisting of global ${\sf R_\phi}(\theta)$ rotations about an axis in the $x-y$ plane driven by microwaves, local ${\sf R_Z}(\theta)$ rotations driven by the 459 nm beam, and $\sf C_Z$  entangling gates using simultaneous Rydberg excitation of atom pairs by the 459 and 1040 nm beams (see Methods). 
\textbf{b)} Atomic level diagram and wavelengths used for cooling, trapping, and qubit control.   
     \textbf{c)} Averaged atomic fluorescence image of the 49 site array with spacing $3~\mu\rm m$. Each camera pixel is $0.6\times0.6 ~\mu\rm m$ at the atoms. 
     \textbf{d)} Global microwave Rabi rotations on a block of 9 qubits at 76.5 kHz. The microwave phase, amplitude, and frequency are controlled by an arbitrary waveform generator. 
     \textbf{e)} A Ramsey experiment with microwave $\pi/2$ pulses and the focused 459 beam providing a ${\sf R_Z}(\theta)$ rotation on a single site. Stark shifts of $\sim 600~\rm  kHz$ are used so that the 15 ns rise/fall time of the on/off acousto-optic modulators (not pictured) can be neglected when calculating the pulse time for ${\sf R_Z}(\theta)$ gates.  
     \textbf{f)} Parity oscillation of a 2-qubit Bell state created using a $\sf C_Z$ gate. A de facto measure of an entangling gate's performance is its ability to generate a Bell state. the measured and uncorrected Bell state fidelity was $92.7(1.3)\%$ for an optimized qubit pair  ($\sim 95.5\%$ corrected for state preparation and measurement (SPAM) errors), with the average for all connected qubit pairs used in circuits $90\%$ ($\sim92.5\%$ SPAM corrected).  
     }
\end{figure*}

Qubits encoded on neutral atoms trapped in an optical lattice provide a scalable architecture for digital quantum computing\cite{Saffman2010}. One- and two-qubit gate operations have previously been demonstrated in large arrays \cite{Xia2015,YWang2016,Graham2019} using qubits that have excellent coherence properties\cite{YWang2016} and can be reliably measured\cite{Madjarov2020}. In the last few years, techniques have been introduced which have enabled atomic rearrangement for deterministic array loading\cite{Barredo2016,Endres2016,Kim2016}. Our approach, as shown in Fig. \ref{fig.Full_Assembly}, combines these recent advances to provide multi-qubit circuit capability in an architecture based on rapid scanning of tightly focused optical control beams.  Atoms are laser cooled and then trapped in a blue-detuned optical lattice. Atom occupancy and quantum state measurements are based on imaging near resonant scattered light onto an  electron multiplying CCD camera. A red-detuned optical tweezer rearranges the detected atoms to deterministically load a subset of atom traps which are used for computation. After state preparation, we perform quantum computations using a universal gate set consisting of global microwave rotations, local $\sf R_Z$ phase gates, and two-qubit $\sf C_Z$ gates (see methods). With this platform, we created 2-6 qubit GHZ states, demonstrated the quantum phase estimation algorithm, and implemented QAOA for the Maximum Cut (MaxCut) problem.      

\begin{figure*}[!ht]
 \includegraphics[width=0.9\textwidth]{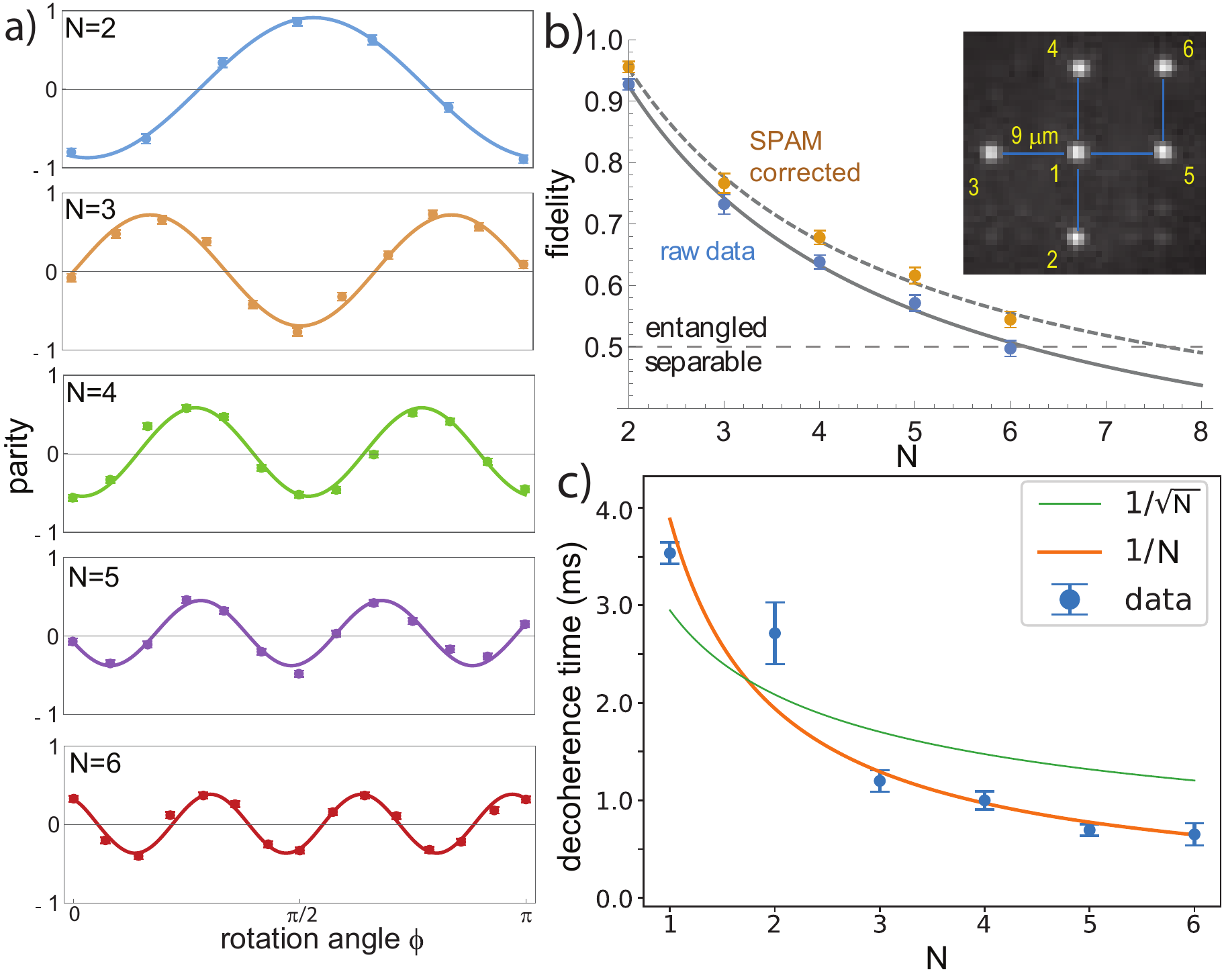}
    \caption{\textbf{Preparation of GHZ states} \textbf{a)} Parity oscillations for GHZ states with 2-6 qubits. The oscillation frequency shows the characteristic linear dependence on the number of qubits. \textbf{b)} The fidelity of the created GHZ state versus qubit number. GHZ states were prepared by applying a Hadamard gate to the central qubit 1 followed by a sequence of {\sf CNOT} gates on pairs 1-2, 1-3, 1-4, 1-5, 5-6, as indicated by the blue lines.  Compiled circuits expressed in  our native gate set are shown in \cite{SM2021circuits}.  
    GHZ states with 2-5 qubits have an uncorrected fidelity indicating they are entangled. Assuming a $1.5\%$ per qubit SPAM error(see \cite{SM2021circuits}), and applying a fidelity correction of $1/0.985^N$ all corrected GHZ state fidelities are over $50\%$ including $N=6$. The fidelity decay follows an approximate displaced $1/N$ scaling. The solid and dashed lines are curve fits to $a+b/(N-c)$
    with $a=0.192, b=2.21, c=-1.014$ for the raw data and
    $a=0.269, b=1.96, c=-0.872$ for the SPAM corrected data. 
    \textbf{c)} Decoherence time of GHZ states measured by Ramsey interference. The lifetime fits well to a $1/N$ scaling.  The $N=1$ data point is the $T_2^*$ time for a single qubit averaged over all six sites.  }
    \label{fig:GHZ_Data}
  \end{figure*}
  
\section*{GHZ state preparation}

Entanglement is perhaps the quintessential feature of quantum information science.  The non-local correlations present in an entangled quantum state can be stronger than is classically possible. These correlations are leveraged as a resource in quantum computing algorithms, quantum metrology, and many quantum communication protocols.  Entangled states can be composed of any number of particles, and there are many classes of entangled states with various properties. Greenberger–Horne–Zeilinger (GHZ) states, also known as cat-states, compose one such class and are of the form $\ket{GHZ}_{N} = \frac{1}{\sqrt{2}} \left(\ket{00...0}_N + e^{\imath \phi} \ket{11...1}_N \right)$, where $N$ is the number of particles occupying the state and $\phi$ is a phase shift between the two terms. GHZ states provide the strongest non-local correlations possible for an $N$-particle entangled state \cite{Gisin1998}. However, GHZ states are very fragile as loss of a single particle completely destroys the entanglement. Also, because all particles contribute to the phase evolution, the dephasing time decreases with the particle number.  Such states are  challenging to create, requiring either many particles to interact with each other or a series of  two-particle interactions performed in sequence.  These  properties have made GHZ production a standard benchmark for quantifying the performance of a quantum computer. GHZ states with 18 particles have been produced using  superconducting qubits\cite{CSong2019}, and 24 particles using trapped ion 
qubits\cite{Pogorelov2021}. GHZ states have also been produced using up to 20 neutral atom qubits\cite{Omran2019}; however, these GHZ states were encoded on a ground-Rydberg state transition and were correspondingly short lived (coherence lifetimes are less than $2 ~\mu \rm s$ for $N \geq 4$) due to decay and the high sensitivity of Rydberg states to environmental perturbations.  We have created and measured the first $N>2$ GHZ states that are encoded on the long-lived hyperfine ground state qubits of neutral atoms.

Using quantum circuits consisting of global microwaves, local $\sf R_Z$ gates, and $\sf C_Z$ gates\cite{SM2021circuits} we have created  GHZ states with up to $N=6$ qubits. To quantify how accurately these states were created, we measured their quantum state fidelity. The fidelity of a GHZ state can be determined from the population, $P_{\ket{0}_N}$ and $P_{\ket{1}_N}$ for states $\ket{0}_{N}=\ket{00...0}_N$ and $\ket{1}_{N}=\ket{11...1}_N$, respectively, and the coherence between these states. We determined the population from a direct measurement in the qubit basis and the coherence from a parity oscillation measurement\cite{Sackett2000}. To measure the parity, we used a microwave pulse to implement the global unitary, $\Pi^{N}_{j=1}e^{\imath \frac{\pi}{4} \sigma^{j}_{\phi}}$ where $\sigma^{j}_{\phi}=\cos(\phi) {\sf X}_j +  \sin(\phi) {\sf Y}_j$ and ${\sf X}_j, {\sf Y}_j$ are  Pauli operators on  qubit $j$.  After this rotation, the atoms are measured in the logical basis and the parity is computed from  
$P = P_{\textrm{even}}-P_{\textrm{odd}}$, where $ P_{\textrm{even(odd)}}$ is the probability  of observing an even (odd) state.  By measuring the parity for various values of $\phi$, we obtain parity oscillation curves for GHZ states up to $N=6$ as shown in Fig.  \ref{fig:GHZ_Data}.  The fidelity of a GHZ state is $F_{GHZ_N}=(P_{\ket{0}_{N}} + P_{\ket{1}_{N}} + C_N)/2$, where $C_N$ is the amplitude of the $N$ qubit parity oscillation. We observe the expected factor of $N$ scaling in parity oscillation frequency\cite{Wineland1992}.  This enhanced collective oscillation rate has applications in quantum metrology\cite{Giovannetti2004}, but also leads to a faster dephasing. 

 \begin{figure*}[!ht]
\includegraphics[width=15.5cm]{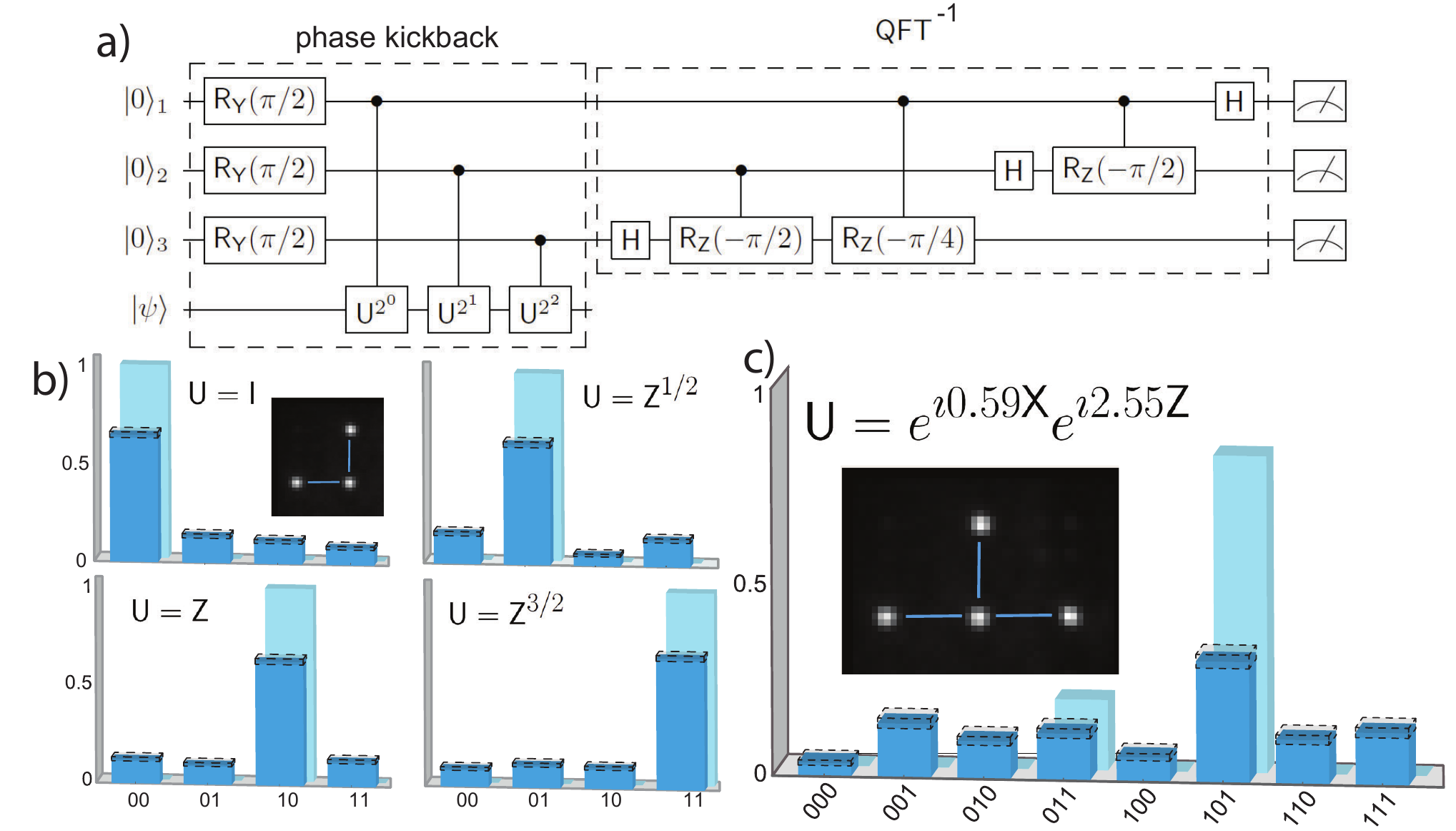}
      \caption{\textbf{Quantum phase estimation using 3 and 4 qubits} \textbf{a)} Phase estimation circuit  using 4 qubits. An eigenstate (or near eigenstate) of the operator $U$ is encoded on the state qubit, $\ket{\psi}$. Controlled unitary operators are then performed between  qubits 1-3 in the measurement register and the state qubit. The two qubit controlled unitaries  each require two {\sf CNOT} gates.  A quantum inverse Fourier transform is then performed on the measurement qubits and their output is measured.  The measured bit string encodes the phase shift on $\ket{\psi}$ when acted on by $U$.  \textbf{b)} Quantum phase estimation results using 3 qubits. For this demonstration each of the four unitary operators measured had $\ket{1}$ as an eigenstate. The phase shift imparted by $U$ was able to be perfectly represented by a single two bit value. The theoretical value (light blue bars) was $100\%$ probability in the target bit string.  In all cases the measured (dark blue) target bit strings had  $>60\%$ probability.  \textbf{c)} Quantum phase estimation with four qubits to estimate the molecular energy of a H$_2$ molecule. In this calculation, we use the Hartree-Fock state $\ket{\psi}=\ket{1}$ (see main text for details). Note that even theoretically, none of the bit strings has a $100\%$ probability. This is due to two factors.  First, the target eigenvalue cannot be exactly represented by a three bit number. Second, the Hartree-Fock state does not have a $100\%$ overlap with the ground eigenstate. Partial overlap with both eigenstates of $\tilde {\sf U}$ leads to nonzero probability for the 101 and 011 bitstrings. Note that the former effect has a very small influence on the output probabilities, so most of the probability is contained in 101 and 011 bitstrings.
      }
    \label{fig.phaseestimation}
  \end{figure*}

The scaling of the coherence time with the size of the GHZ state  depends on the properties of the relevant dephasing sources. The coherence of optically trapped neutral atom hyperfine qubits is primarily limited by three mechanisms\cite{Saffman2005a}: magnetic field noise, fluctuations of the trap light intensity, and atomic motion. Fluctuations of the trap intensity and the magnetic fields,  cause differential frequency shifts on the qubit levels\cite{Carr2016}. These correlated and non-Markovian perturbations  lead to a $1/N$ scaling of the GHZ coherence time\cite{Monz2011}. This scaling is observed in Fig. \ref{fig:GHZ_Data} despite the use of blue detuned traps where the atoms are localized at a local minimum of the optical intensity, and $m=0$ clock states which have only a weak quadratic Zeeman sensitivity.
 All GHZ states up to $N=6$ retain coherence for more than $600 ~\mu\rm s$, about 500 times longer than previously reported neutral atom GHZ states\cite{Omran2019}.
 This increased coherence is due to the fact that the GHZ states prepared  here are encoded on a ground hyperfine qubit basis rather than the ground-Rydberg basis used in previous experiments. 
 
The third mechanism, atomic motion, is also non-Markovian, but is not collective since the phase of the atomic motion in different traps is not correlated. This should lead to a slower $1/\sqrt N$ scaling in addition to the $1/N$ contributions mentioned above.  This motion can be reduced and the coherence extended through deeper cooling. Alternatively, dynamical decoupling sequences can be applied to suppress all of the non-Markovian sources of dephasing. For single qubits, we have observed coherence times as long as 1 s using XY8 pulse 
sequences\cite{Gullion1990} and more than a factor of five improvement of the coherence time of GHZ$_N$ states. The achievable GHZ coherence time, and the resulting scaling exponent using optimized decoupling sequences, is left for future studies.

\section*{Phase estimation algorithm}

Quantum phase estimation was one of the original algorithms responsible for the rapid growth of interest in quantum computing\cite{Abrams1999}.  This algorithm is used to estimate the complex phase of an operator acting on an eigenstate and has broad applications as a subroutine in other quantum algorithms, including factoring and quantum chemistry.  Quantum phase estimation is one of a class of related algorithms which achieves a quantum advantage via the exponential speedup of the quantum Fourier transform over the classical Fourier transform algorithm. In this algorithm, there is a state register and a measurement register. The state register consists of a set of qubits that are in an eigenstate $\ket{\psi}$ of a unitary operator, $\sf U$, such that ${\sf U} \ket{\psi} = e^{\imath \phi} \ket{\psi}$.  To perform phase estimation, information about the action of $\sf U$ on the state register is encoded on the measurement register through a series of controlled unitary operations shown in Fig. \ref{fig.phaseestimation}. In this procedure, the state of qubit  $j$ in the measurement register controls if a unitary ${\sf U}^{2^{j-1}}$ is applied to the state register.  After these controlled unitary operations, an inverse quantum Fourier transform is performed on the qubits in the measurement register which  are then measured in the computational basis.  The phase, $\phi$, can be approximately determined from the measured bit-string. Each bit-string value corresponds to a particular phase value on the interval between $[0,2\pi)$. If $\phi$ is between these values, then multiple bit-strings will be measured at the end of the circuit. Similarly, if the state $\ket{\psi}$ is not an exact eigenstate of $\sf U$, then phase signatures of the eigenvalues of each eigenstate composing $\ket{\psi}$ will be present in the output measurements.  As more qubits are used in the measurement register, $\phi$ can be determined with greater accuracy, since there are more unique bit-strings to represent phases on the $[0, 2 \pi)$ interval.  

As a first test, shown in Fig. \ref{fig.phaseestimation}, we performed  phase estimation with 3 qubits (1 qubit in the state register and 2 in the measurement register) with ${\sf U}={\sf I}$, ${\sf Z}^{1/2}$, ${\sf Z}$, ${\sf Z}^{3/2}$ which act on state $\ket{1}$ with phase shifts $\phi=0,\pi/2,\pi,3\pi/2$. These phase shifts can be exactly represented with two bits. The measured probabilities of the desired output states were $>64\%$ in all cases. The deviation from the ideal 100\% output probability is due to accumulation of gate errors (see \cite{SM2021circuits} for further details). 

As a second example, we performed phase estimation for a prototypical quantum chemistry calculation, the molecular energy of a Hydrogen molecule. 
An eigenstate of a time-independent Hamiltonian acquires a phase shift that is proportional to its energy, 
${\sf U} \ket{\psi}=e^{\imath {\sf H} t} \ket{\psi} = e^{\imath \phi} \ket{\psi}$.  Quantum phase estimation is then used to measure the phase for a particular chosen time ($t_0$), and the state energy can be determined from the measured phase, $E= \phi / t_0$. 
The time required for a complete classical calculation of molecular energies scales exponentially with the number of electronic orbitals. However, quantum phase estimation allows polynomial time energy estimates
\cite{Aspuru-Guzik2005}.  We consider a Hamiltonian which represents a hydrogen molecule in the STO-3G basis making use of the Bravyi-Kitaev transformation\cite{Bravyi2002} and tapering qubits corresponding to the total number of electrons, the $z$ component of the spin, and a reflection symmetry\cite{Bravyi2017}. With these approximations, the molecular energy estimation reduces to a single qubit problem. The Hamiltonian has the form ${\sf H} = a_0 + a_1 {\sf Z} + a_2 {\sf X}$. If we assume a bond length of $0.7414$ angstroms, then $a_0 = -0.328717$, $a_1 = 0.787967$, and $a_2 = 0.181289,$ all in units of Hartrees. This Hamiltonian has a ground state of $0.112828\ket{0}+0.993615\ket{1}$ with an energy of $-1.13727 ~\rm Ha$. The $a_0$ energy offset is applied classically and can be neglected from the quantum calculation.
The eigenvalues then lie between $-E_{\rm max}$ and $+E_{\rm max}$ with $E_{\rm max}=|a_1|+|a_2|=0.9693~\rm Ha$ and we choose $t_0=\pi/E_{\rm max}$ such that the phases corresponding to the eigenvalues of ${\sf H}t$ lie between $-\pi$ and $\pi$.
We approximated the operator, $\sf U$, using first-order Trotterization as $\tilde{\sf U} = e^{\imath a_2 t {\sf X}} e^{\imath a_1 t {\sf Z}}$. In the Bravyi-Kitaev basis, the Hartree-Fock state is the product state of spin up or down qubits which gives the lowest energy expectation. For this Hamiltonian, the Hartree-Fock state is $\ket{1}$, which we used as the initial state for the state estimation.  This state has a probability overlap of 0.82 and 0.18 with the eigenstates of $\tilde{\sf U}$ corresponding to energies of $-1.0495$ and $0.3920~\rm Ha$ (note that $a_0$ was added back to the energy obtained from the eigenvalues). 

For the computation, we use four qubits (one qubit in the state register and three qubits in the measurement register). In an ideal circuit with infinite resolution the measured phase values would be 0.6282 and  0.3718 which are close to 0.625 (0.101 in binary)  and 0.375 (0.011 in binary). Thus in a noise free circuit we expect to observe bit strings 101 and 011 82\% and 18\% of the time, respectively. 
After compiling the circuit into our native gate set, we ran the circuit 700 times.  The most frequently observed bit string was 101 corresponding to an energy estimate of $-1.06~\rm Ha$ (again $a_0$ was added to obtain the final result).  Using more sophisticated methods, the molecular energy of hydrogen was found to be $-1.174476~\rm Ha$ \cite{Kolos1986}. The difference between the more accurate value and the experimental result arises from the limited number of qubits used for phase estimation, using the minimal STO-3G basis rather than a larger basis set, and the approximations used in the circuit implementation.  Further improvements in precision can be obtained by using more qubits to represent the phase and using a Hamiltonian which more accurately represents the molecular energy of Hydrogen.

  \begin{figure*}[!ht]
\includegraphics[width=0.9\textwidth]{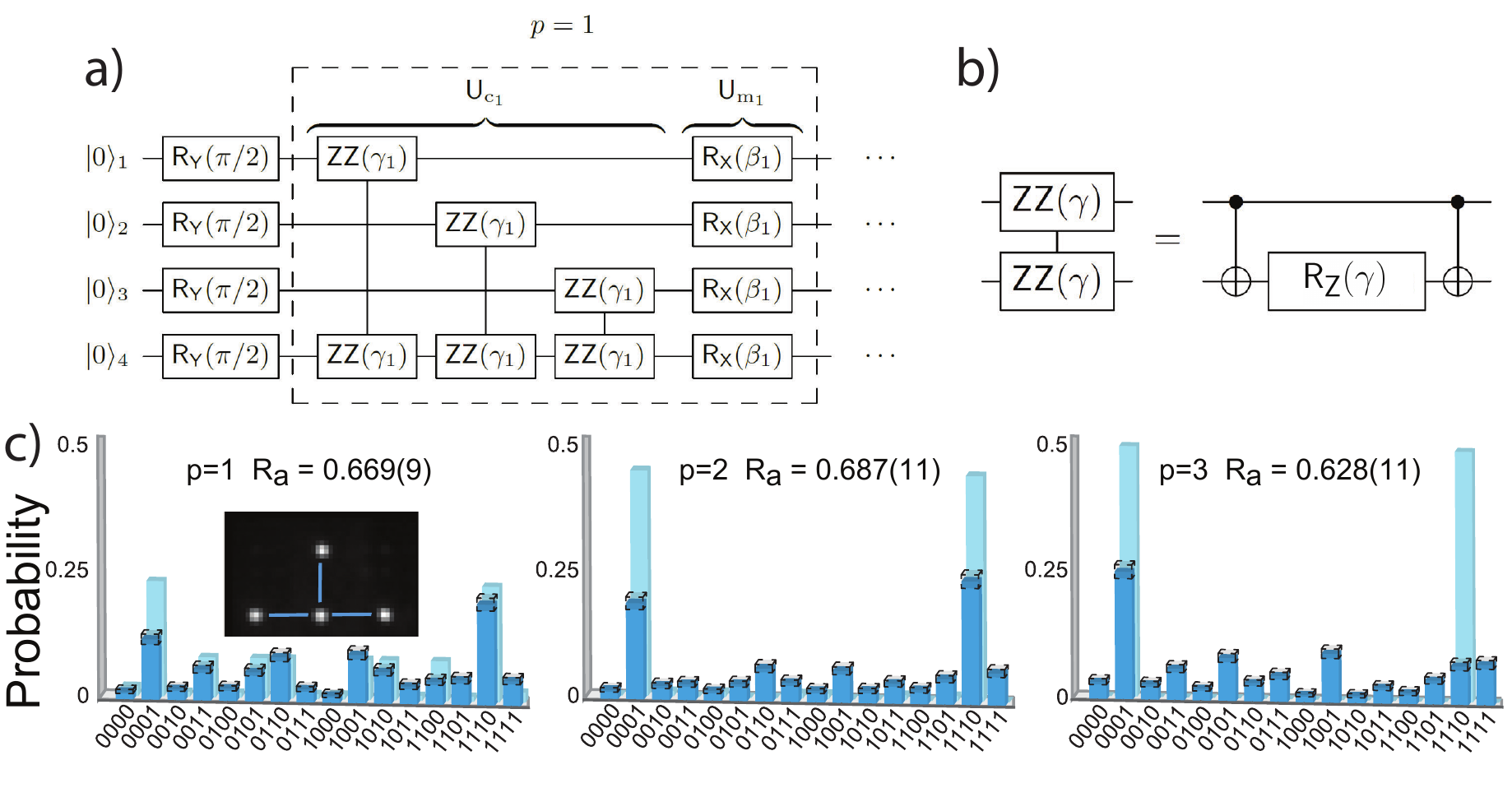}
  \caption{  \label{fig.QAOA}\textbf{QAOA algorithm for solving MaxCut} \textbf{a} Circuit diagram for 4-qubit QAOA MaxCut showing a single $p=1$ cycle.  \textbf{b} Decomposition of the ${\sf ZZ}(\gamma)=e^{-\imath \frac{\gamma}{2}{\sf Z}\otimes {\sf Z}}$ interaction into two {\sf CNOT} gates and a ${\sf R_Z}$ rotation. \textbf{c} Circuit results for $p=1,2,3$ with optimized $\gamma$ and $\beta$ values\cite{SM2021circuits}. The inset in the $p=1$  bar chart shows the four qubits used  with blue lines indicating $\sf C_Z$ gate connections between qubit pairs. Dark (light) blue bars show the experimental (ideal theoretical) output probabilities.  The experimental approximation ratios are indicated in the bar charts. The ideal theoretical approximation ratios for $p=1,2,3$ are $0.772, 0.934, 1.0$. As the circuit gets longer, the approximation ratio increases, but so does the circuit depth (and accumulation of gate errors). These factors led to an increase in approximation ratio for $p=2$ followed by a drop at  $p=3$.}
    \end{figure*}
    
\section*{QAOA algorithm}

There has been a large effort to design quantum algorithms which leverage both quantum and classical computing power to solve problems with fewer operations than would be required by a classical computer alone. Hybrid quantum-classical algorithms, which seek to achieve useful computational results without requiring a full error-corrected quantum computer\cite{Preskill2018}, combine a quantum core that efficiently  generates high-dimensional quantum states,  a task that requires exponential resources on a classical machine, with a classical outer loop that selects parameters of the quantum circuit to optimize the value of the quantum state for solving the problem at hand. Primary examples of such hybrid algorithms are the  Variational Quantum Eigensolver\cite{Peruzzo2014}	 and the related Quantum Approximate Optimization Algorithm (QAOA) \cite{Farhi2014}.

QAOA is particularly well suited for solving combinatorial optimization problems that admit a Hamiltonian formulation.   An  ansatz state is parameterized in terms of unitary evolution operations representing a cost function (${\sf U}_{\rm c}=e^{\imath \gamma {\sf H}_{\rm c}}$) and state mixing (${\sf U}_{\rm m}=e^{\imath \beta {\sf H}_{\rm m}}$) where $\gamma$ and $\beta$ parameters are set by the classical optimizer and where ${\sf H}_{\rm c}$ and ${\sf H}_{\rm m}$ represent cost and mixing Hamiltonians respectively. The QAOA circuit consists of $p$ repeated layers of cost and mixing Hamiltonians acting on an $N$ qubit initial state $\ket{s}=(\ket{0}+\ket{1})^{\otimes N}$. This evolution results in the final state $\ket{{\boldsymbol \gamma},{\boldsymbol \beta}} = {\sf U}_{\rm m}(\beta_{p}){\sf U}_{\rm c}(\gamma_{p}) \cdots {\sf U}_{\rm m}(\beta_{1}){\sf U}_{\rm c}(\gamma_{1}) \ket{s}$. The repeated application of mixing and cost Hamiltonians can be regarded  as a Trotterized version of adiabatic evolution of the initial state to the ground state of ${\sf H}_{\rm c}$. In the $p \rightarrow \infty$ limit, these two processes are equivalent, while the availability of $2p$ variational parameters provides more degrees of freedom for optimizing the rate of convergence compared to a simple adiabatic ramp. After preparing the ansatz state, the expectation value of the cost Hamiltonian is measured  and fed into a classical optimizer.  The cost Hamiltonian is designed to be diagonal in the computational basis, so after the classical optimizer finds optimal settings for all $\gamma_i$ and $\beta_i$, a computational basis measurement  yields a bit-string that corresponds to the optimized combinatorial problem solution if $p$ is sufficiently large.

The MaxCut problem is an example of an NP-hard problem to which QAOA can be readily applied. MaxCut seeks to partition the vertices of a graph into two sets such that the maximum number of edges are cut. A partition $z$ of a graph with $m$ edges and $n$ vertices can be quantified with a cost function, $C(z)=\sum^{m}_{\alpha=1} C_{\alpha}(z)$ where $C_{\alpha}(z)=1$ if the $\alpha$ edge is cut (i.e. the two vertices of the edge are in different sets) and $C_{\alpha}(z)=0$ for non-cut edges. The maximum cut is found when a partition maximizes $C(z)$. This cost function can be readily translated into a Hamiltonian operating on a set of qubits representing a graph, ${\sf H}_{\rm c} = \frac{1}{2}\sum^{m}_{\alpha=1} (1-{\sf Z}_{\alpha_1} {\sf Z}_{\alpha_2})$,  where $\alpha_1$ and $\alpha_2$ are qubit indices representing the vertices of the edge $\alpha$\cite{Farhi2014}. The two basis states of the qubit then map onto the two sets into which the two vertices are grouped.

We have  implemented QAOA for the MaxCut problem  on three and four node graphs. The first graph measured was a three vertex line graph with the center vertex connected to the two outer vertices. There are two degenerate MaxCut solutions with the center and outer vertices in different sets, which gives two cuts.  Using optimized values of $\gamma$ and $\beta$\cite{SM2021circuits}, we  measured output bit strings for  $p=1$ and $p=2$.  The results can be scored as an approximation ratio $R_a = \frac{1}{S_{\textrm{max}}} \sum_i{p_i S_i}$ where $p_i$ is the probability of a particular bit-string, $S_i$ is the number of edge cuts for the  bit string, and $S_{\textrm{max}}$ is the maximum cut number.  For the line graph, the $p=1$ circuit achieves an approximation ratio of $0.65(1)$ (theoretical 0.825) and the $p=2$ circuit achieves an approximation ratio of $0.71(1)$ (theoretical 1.0). We have also implemented MaxCut  for graphs with four-qubits, as shown in Fig. \ref{fig.QAOA}. We see a clear gain in approximation ratio when increasing from $p=1$ to $p=2$; however, the approximation ratio drops for $p=3$ though the theoretical approximation ratio improves. This approximation ratio drop is due to limitations in the two-qubit gate error, which degrades the approximation ratio more than the theoretical improvement from incrementing $p$. Further improvements in $\sf C_Z$ gate fidelity will enable larger graphs and higher approximation ratios\cite{Harrigan2021}.

\section*{Outlook}

The experimental results described above demonstrate that an array of neutral atoms trapped in an optical lattice form a programmable circuit-model quantum computer.  We demonstrated the ability of this prototype computer to create entangled GHZ states with up to six qubits and demonstrated quantum algorithms on four qubits with a circuit depth of up to 18 $\sf C_Z$ gates \cite{SM2021circuits}. This capability opens the door on a vast collection of applications. The creation of long-lived GHZ states has utility in entanglement enhanced 
sensing\cite{Giovannetti2004}.   The ability to perform  quantum phase estimation enables a suite of algorithms in addition to the quantum chemistry applications discussed above. Such applications include integer factoring\cite{Shor1994} and estimating solutions to linear equations\cite{Harrow2009}. Indeed quantum phase estimation underlies all the known avenues to exponential quantum speed-up\cite{OBrien2019}. 
 Hybrid quantum/classical algorithms, including QAOA demonstrated here,  have found wide use in a number of applications\cite{Endo2021}.
 
 Although the experiments presented above are far from providing a  quantum advantage over classical computation, they represent an important milestone for the development of neutral atom qubit based processors. While the current two qubit gate fidelity is limited when compared to more mature computing platforms, this neutral atom platform provides a unique combination of properties which facilitate scalability. In particular, the ability of this platform to increase the qubit number simply by adding more laser power and changing the number of RF tones driving the trap AODs is in stark contrast to other technologies which require fabrication of completely new chips or traps to increase qubit number.  The factors limiting two qubit gate fidelity and algorithmic performance today are well understood, as is the engineering roadmap to reach higher performance. Of primary importance are improved laser cooling to reach the atomic motional ground state, spatial shaping of the optical control beams for reduced gate errors, optimization of optical trap parameters for improved localization and coherence\cite{Robicheaux2021,Carr2016}, and higher laser power for reduced scattering from the intermediate $7p_{1/2}$ state.  Combining these advances in a single, scalable qubit array will lead to a neutral atom platform  for high performance digital quantum computation.  

While finalizing this manuscript we became aware of related work demonstrating encoding of logical qubits with a complementary neutral atom architecture\cite{Bluvstein2021b}.

\bibliography{atomic,saffman_refs,rydberg,qc_refs,optics}

\begin{thebibliography}{56}%
\makeatletter
\providecommand \@ifxundefined [1]{%
 \@ifx{#1\undefined}
}%
\providecommand \@ifnum [1]{%
 \ifnum #1\expandafter \@firstoftwo
 \else \expandafter \@secondoftwo
 \fi
}%
\providecommand \@ifx [1]{%
 \ifx #1\expandafter \@firstoftwo
 \else \expandafter \@secondoftwo
 \fi
}%
\providecommand \natexlab [1]{#1}%
\providecommand \enquote  [1]{``#1''}%
\providecommand \bibnamefont  [1]{#1}%
\providecommand \bibfnamefont [1]{#1}%
\providecommand \citenamefont [1]{#1}%
\providecommand \href@noop [0]{\@secondoftwo}%
\providecommand \href [0]{\begingroup \@sanitize@url \@href}%
\providecommand \@href[1]{\@@startlink{#1}\@@href}%
\providecommand \@@href[1]{\endgroup#1\@@endlink}%
\providecommand \@sanitize@url [0]{\catcode `\\12\catcode `\$12\catcode
  `\&12\catcode `\#12\catcode `\^12\catcode `\_12\catcode `\%12\relax}%
\providecommand \@@startlink[1]{}%
\providecommand \@@endlink[0]{}%
\providecommand \url  [0]{\begingroup\@sanitize@url \@url }%
\providecommand \@url [1]{\endgroup\@href {#1}{\urlprefix }}%
\providecommand \urlprefix  [0]{URL }%
\providecommand \Eprint [0]{\href }%
\providecommand \doibase [0]{http://dx.doi.org/}%
\providecommand \selectlanguage [0]{\@gobble}%
\providecommand \bibinfo  [0]{\@secondoftwo}%
\providecommand \bibfield  [0]{\@secondoftwo}%
\providecommand \translation [1]{[#1]}%
\providecommand \BibitemOpen [0]{}%
\providecommand \bibitemStop [0]{}%
\providecommand \bibitemNoStop [0]{.\EOS\space}%
\providecommand \EOS [0]{\spacefactor3000\relax}%
\providecommand \BibitemShut  [1]{\csname bibitem#1\endcsname}%
\let\auto@bib@innerbib\@empty
\bibitem [{\citenamefont {Saffman}\ \emph {et~al.}(2010)\citenamefont
  {Saffman}, \citenamefont {Walker},\ and\ \citenamefont
  {M\o{}lmer}}]{Saffman2010}%
  \BibitemOpen
  \bibfield  {author} {\bibinfo {author} {\bibfnamefont {M.}~\bibnamefont
  {Saffman}}, \bibinfo {author} {\bibfnamefont {T.~G.}\ \bibnamefont {Walker}},
  \ and\ \bibinfo {author} {\bibfnamefont {K.}~\bibnamefont {M\o{}lmer}},\
  }\bibfield  {title} {\enquote {\bibinfo {title} {Quantum information with
  {R}ydberg atoms},}\ }\href@noop {} {\bibfield  {journal} {\bibinfo  {journal}
  {Rev. Mod. Phys.}\ }\textbf {\bibinfo {volume} {82}},\ \bibinfo {pages}
  {2313} (\bibinfo {year} {2010})}\BibitemShut {NoStop}%
\bibitem [{\citenamefont {Jaksch}\ \emph {et~al.}(2000)\citenamefont {Jaksch},
  \citenamefont {Cirac}, \citenamefont {Zoller}, \citenamefont {Rolston},
  \citenamefont {C\^ot\'e},\ and\ \citenamefont {Lukin}}]{Jaksch2000}%
  \BibitemOpen
  \bibfield  {author} {\bibinfo {author} {\bibfnamefont {D.}~\bibnamefont
  {Jaksch}}, \bibinfo {author} {\bibfnamefont {J.~I.}\ \bibnamefont {Cirac}},
  \bibinfo {author} {\bibfnamefont {P.}~\bibnamefont {Zoller}}, \bibinfo
  {author} {\bibfnamefont {S.~L.}\ \bibnamefont {Rolston}}, \bibinfo {author}
  {\bibfnamefont {R.}~\bibnamefont {C\^ot\'e}}, \ and\ \bibinfo {author}
  {\bibfnamefont {M.~D.}\ \bibnamefont {Lukin}},\ }\bibfield  {title} {\enquote
  {\bibinfo {title} {Fast quantum gates for neutral atoms},}\ }\href@noop {}
  {\bibfield  {journal} {\bibinfo  {journal} {Phys. Rev. Lett.}\ }\textbf
  {\bibinfo {volume} {85}},\ \bibinfo {pages} {2208--2211} (\bibinfo {year}
  {2000})}\BibitemShut {NoStop}%
\bibitem [{\citenamefont {Ga\"etan}\ \emph {et~al.}(2009)\citenamefont
  {Ga\"etan}, \citenamefont {Miroshnychenko}, \citenamefont {Wilk},
  \citenamefont {Chotia}, \citenamefont {Viteau}, \citenamefont {Comparat},
  \citenamefont {Pillet}, \citenamefont {Browaeys},\ and\ \citenamefont
  {Grangier}}]{Gaetan2009}%
  \BibitemOpen
  \bibfield  {author} {\bibinfo {author} {\bibfnamefont {A.}~\bibnamefont
  {Ga\"etan}}, \bibinfo {author} {\bibfnamefont {Y.}~\bibnamefont
  {Miroshnychenko}}, \bibinfo {author} {\bibfnamefont {T.}~\bibnamefont
  {Wilk}}, \bibinfo {author} {\bibfnamefont {A.}~\bibnamefont {Chotia}},
  \bibinfo {author} {\bibfnamefont {M.}~\bibnamefont {Viteau}}, \bibinfo
  {author} {\bibfnamefont {D.}~\bibnamefont {Comparat}}, \bibinfo {author}
  {\bibfnamefont {P.}~\bibnamefont {Pillet}}, \bibinfo {author} {\bibfnamefont
  {A.}~\bibnamefont {Browaeys}}, \ and\ \bibinfo {author} {\bibfnamefont
  {P.}~\bibnamefont {Grangier}},\ }\bibfield  {title} {\enquote {\bibinfo
  {title} {Observation of collective excitation of two individual atoms in the
  {R}ydberg blockade regime},}\ }\href@noop {} {\bibfield  {journal} {\bibinfo
  {journal} {Nature Phys.}\ }\textbf {\bibinfo {volume} {5}},\ \bibinfo {pages}
  {115} (\bibinfo {year} {2009})}\BibitemShut {NoStop}%
\bibitem [{\citenamefont {Urban}\ \emph {et~al.}(2009)\citenamefont {Urban},
  \citenamefont {Johnson}, \citenamefont {Henage}, \citenamefont {Isenhower},
  \citenamefont {Yavuz}, \citenamefont {Walker},\ and\ \citenamefont
  {Saffman}}]{Urban2009}%
  \BibitemOpen
  \bibfield  {author} {\bibinfo {author} {\bibfnamefont {E.}~\bibnamefont
  {Urban}}, \bibinfo {author} {\bibfnamefont {T.~A.}\ \bibnamefont {Johnson}},
  \bibinfo {author} {\bibfnamefont {T.}~\bibnamefont {Henage}}, \bibinfo
  {author} {\bibfnamefont {L.}~\bibnamefont {Isenhower}}, \bibinfo {author}
  {\bibfnamefont {D.~D.}\ \bibnamefont {Yavuz}}, \bibinfo {author}
  {\bibfnamefont {T.~G.}\ \bibnamefont {Walker}}, \ and\ \bibinfo {author}
  {\bibfnamefont {M.}~\bibnamefont {Saffman}},\ }\bibfield  {title} {\enquote
  {\bibinfo {title} {Observation of {R}ydberg blockade between two atoms},}\
  }\href@noop {} {\bibfield  {journal} {\bibinfo  {journal} {Nature Phys.}\
  }\textbf {\bibinfo {volume} {5}},\ \bibinfo {pages} {110} (\bibinfo {year}
  {2009})}\BibitemShut {NoStop}%
\bibitem [{\citenamefont {Greenberger}\ \emph {et~al.}(1989)\citenamefont
  {Greenberger}, \citenamefont {Horne},\ and\ \citenamefont
  {Zeilinger}}]{Greenberger1989}%
  \BibitemOpen
  \bibfield  {author} {\bibinfo {author} {\bibfnamefont {D.~M.}\ \bibnamefont
  {Greenberger}}, \bibinfo {author} {\bibfnamefont {M.~A.}\ \bibnamefont
  {Horne}}, \ and\ \bibinfo {author} {\bibfnamefont {A.}~\bibnamefont
  {Zeilinger}},\ }\bibfield  {title} {\enquote {\bibinfo {title} {Going beyond
  bell's theorem},}\ }in\ \href@noop {} {\emph {\bibinfo {booktitle} {Bell's
  Theorem, Quantum Theory and Conceptions of the Universe}}},\ \bibinfo
  {editor} {edited by\ \bibinfo {editor} {\bibfnamefont {M.}~\bibnamefont
  {Kafatos}}}\ (\bibinfo  {publisher} {Springer},\ \bibinfo {year} {1989})\
  p.~\bibinfo {pages} {69}\BibitemShut {NoStop}%
\bibitem [{\citenamefont {Aspuru-Guzik}\ \emph {et~al.}(2005)\citenamefont
  {Aspuru-Guzik}, \citenamefont {Dutoi}, \citenamefont {Love},\ and\
  \citenamefont {Head-Gordon}}]{Aspuru-Guzik2005}%
  \BibitemOpen
  \bibfield  {author} {\bibinfo {author} {\bibfnamefont {A.}~\bibnamefont
  {Aspuru-Guzik}}, \bibinfo {author} {\bibfnamefont {A.~D.}\ \bibnamefont
  {Dutoi}}, \bibinfo {author} {\bibfnamefont {P.~J.}\ \bibnamefont {Love}}, \
  and\ \bibinfo {author} {\bibfnamefont {M.}~\bibnamefont {Head-Gordon}},\
  }\bibfield  {title} {\enquote {\bibinfo {title} {Simulated quantum
  computation of molecular energies},}\ }\href@noop {} {\bibfield  {journal}
  {\bibinfo  {journal} {Science}\ }\textbf {\bibinfo {volume} {309}},\ \bibinfo
  {pages} {1704} (\bibinfo {year} {2005})}\BibitemShut {NoStop}%
\bibitem [{\citenamefont {Farhi}\ \emph {et~al.}(2014)\citenamefont {Farhi},
  \citenamefont {Goldstone},\ and\ \citenamefont {Gutmann}}]{Farhi2014}%
  \BibitemOpen
  \bibfield  {author} {\bibinfo {author} {\bibfnamefont {E.}~\bibnamefont
  {Farhi}}, \bibinfo {author} {\bibfnamefont {J.}~\bibnamefont {Goldstone}}, \
  and\ \bibinfo {author} {\bibfnamefont {S.}~\bibnamefont {Gutmann}},\
  }\bibfield  {title} {\enquote {\bibinfo {title} {A quantum approximate
  optimization algorithm},}\ }\href@noop {} {\bibfield  {journal} {\bibinfo
  {journal} {arXiv:1411.4028}\ } (\bibinfo {year} {2014})}\BibitemShut
  {NoStop}%
\bibitem [{\citenamefont {Martinez}\ \emph {et~al.}(2016)\citenamefont
  {Martinez}, \citenamefont {Muschik}, \citenamefont {Schindler}, \citenamefont
  {Nigg}, \citenamefont {Erhard}, \citenamefont {Heyl}, \citenamefont {Hauke},
  \citenamefont {Dalmonte}, \citenamefont {Monz}, \citenamefont {Zoller},\ and\
  \citenamefont {Blatt}}]{Martinez2016}%
  \BibitemOpen
  \bibfield  {author} {\bibinfo {author} {\bibfnamefont {E.~A.}\ \bibnamefont
  {Martinez}}, \bibinfo {author} {\bibfnamefont {C.~A.}\ \bibnamefont
  {Muschik}}, \bibinfo {author} {\bibfnamefont {P.}~\bibnamefont {Schindler}},
  \bibinfo {author} {\bibfnamefont {D.}~\bibnamefont {Nigg}}, \bibinfo {author}
  {\bibfnamefont {A.}~\bibnamefont {Erhard}}, \bibinfo {author} {\bibfnamefont
  {M.}~\bibnamefont {Heyl}}, \bibinfo {author} {\bibfnamefont {P.}~\bibnamefont
  {Hauke}}, \bibinfo {author} {\bibfnamefont {M.}~\bibnamefont {Dalmonte}},
  \bibinfo {author} {\bibfnamefont {T.}~\bibnamefont {Monz}}, \bibinfo {author}
  {\bibfnamefont {P.}~\bibnamefont {Zoller}}, \ and\ \bibinfo {author}
  {\bibfnamefont {R.}~\bibnamefont {Blatt}},\ }\bibfield  {title} {\enquote
  {\bibinfo {title} {Real-time dynamics of lattice gauge theories with a
  few-qubit quantum computer},}\ }\href@noop {} {\bibfield  {journal} {\bibinfo
   {journal} {Nature}\ }\textbf {\bibinfo {volume} {534}},\ \bibinfo {pages}
  {516} (\bibinfo {year} {2016})}\BibitemShut {NoStop}%
\bibitem [{\citenamefont {Figgatt}\ \emph {et~al.}(2017)\citenamefont
  {Figgatt}, \citenamefont {Maslov}, \citenamefont {Landsman}, \citenamefont
  {Linke}, \citenamefont {Debnath},\ and\ \citenamefont
  {Monroe}}]{Figgatt2017}%
  \BibitemOpen
  \bibfield  {author} {\bibinfo {author} {\bibfnamefont {C.}~\bibnamefont
  {Figgatt}}, \bibinfo {author} {\bibfnamefont {D.}~\bibnamefont {Maslov}},
  \bibinfo {author} {\bibfnamefont {K.A.}\ \bibnamefont {Landsman}}, \bibinfo
  {author} {\bibfnamefont {N.M.}\ \bibnamefont {Linke}}, \bibinfo {author}
  {\bibfnamefont {S.}~\bibnamefont {Debnath}}, \ and\ \bibinfo {author}
  {\bibfnamefont {C.}~\bibnamefont {Monroe}},\ }\bibfield  {title} {\enquote
  {\bibinfo {title} {Complete 3-qubit {G}rover search on a programmable quantum
  computer},}\ }\href@noop {} {\bibfield  {journal} {\bibinfo  {journal} {Nat.
  Commun.}\ }\textbf {\bibinfo {volume} {8}},\ \bibinfo {pages} {1918}
  (\bibinfo {year} {2017})}\BibitemShut {NoStop}%
\bibitem [{\citenamefont {DiCarlo}\ \emph {et~al.}(2009)\citenamefont
  {DiCarlo}, \citenamefont {Chow}, \citenamefont {Gambetta}, \citenamefont
  {Bishop}, \citenamefont {Johnson}, \citenamefont {Schuster}, \citenamefont
  {Majer}, \citenamefont {Blais}, \citenamefont {Frunzio}, \citenamefont
  {Girvin},\ and\ \citenamefont {Schoelkopf}}]{DiCarlo2009}%
  \BibitemOpen
  \bibfield  {author} {\bibinfo {author} {\bibfnamefont {L.}~\bibnamefont
  {DiCarlo}}, \bibinfo {author} {\bibfnamefont {J.~M.}\ \bibnamefont {Chow}},
  \bibinfo {author} {\bibfnamefont {J.~M.}\ \bibnamefont {Gambetta}}, \bibinfo
  {author} {\bibfnamefont {Lev~S.}\ \bibnamefont {Bishop}}, \bibinfo {author}
  {\bibfnamefont {B.~R.}\ \bibnamefont {Johnson}}, \bibinfo {author}
  {\bibfnamefont {D.~I.}\ \bibnamefont {Schuster}}, \bibinfo {author}
  {\bibfnamefont {J.}~\bibnamefont {Majer}}, \bibinfo {author} {\bibfnamefont
  {A.}~\bibnamefont {Blais}}, \bibinfo {author} {\bibfnamefont
  {L.}~\bibnamefont {Frunzio}}, \bibinfo {author} {\bibfnamefont {S.~M.}\
  \bibnamefont {Girvin}}, \ and\ \bibinfo {author} {\bibfnamefont {R.~J.}\
  \bibnamefont {Schoelkopf}},\ }\bibfield  {title} {\enquote {\bibinfo {title}
  {Demonstration of two-qubit algorithms with a superconducting quantum
  processor},}\ }\href@noop {} {\bibfield  {journal} {\bibinfo  {journal}
  {Nature (London)}\ }\textbf {\bibinfo {volume} {460}},\ \bibinfo {pages}
  {240} (\bibinfo {year} {2009})}\BibitemShut {NoStop}%
\bibitem [{\citenamefont {Harrigan}\ \emph {et~al.}(2021)\citenamefont
  {Harrigan}, \citenamefont {Sung}, \citenamefont {Neeley}, \citenamefont
  {Satzinger}, \citenamefont {Arute}, \citenamefont {Arya}, \citenamefont
  {Atalaya}, \citenamefont {Bardin}, \citenamefont {Barends}, \citenamefont
  {Boixo}, \citenamefont {Broughton}, \citenamefont {Buckley}, \citenamefont
  {Buell}, \citenamefont {Burkett}, \citenamefont {Bushnell}, \citenamefont
  {Chen}, \citenamefont {Chen}, \citenamefont {Chiaro}, \citenamefont
  {Collins}, \citenamefont {Courtney}, \citenamefont {Demura}, \citenamefont
  {Dunsworth}, \citenamefont {Eppens}, \citenamefont {Fowler}, \citenamefont
  {Foxen}, \citenamefont {Gidney}, \citenamefont {Giustina}, \citenamefont
  {Graff}, \citenamefont {Habegger}, \citenamefont {Ho}, \citenamefont {Hong},
  \citenamefont {Huang}, \citenamefont {Ioffe}, \citenamefont {Isakov},
  \citenamefont {Jeffrey}, \citenamefont {Jiang}, \citenamefont {Jones},
  \citenamefont {Kafri}, \citenamefont {Kechedzhi}, \citenamefont {Kelly},
  \citenamefont {Kim}, \citenamefont {Klimov}, \citenamefont {Korotkov},
  \citenamefont {Kostritsa}, \citenamefont {Landhuis}, \citenamefont {Laptev},
  \citenamefont {Lindmark}, \citenamefont {Leib}, \citenamefont {Martin},
  \citenamefont {Martinis}, \citenamefont {McClean}, \citenamefont {McEwen},
  \citenamefont {Megrant}, \citenamefont {Mi}, \citenamefont {Mohseni},
  \citenamefont {Mruczkiewicz}, \citenamefont {Mutus}, \citenamefont {Naaman},
  \citenamefont {Neill}, \citenamefont {Neukart}, \citenamefont {Niu},
  \citenamefont {O'Brien}, \citenamefont {O'Gorman}, \citenamefont {Ostby},
  \citenamefont {Petukhov}, \citenamefont {Putterman}, \citenamefont
  {Quintana}, \citenamefont {Roushan}, \citenamefont {Rubin}, \citenamefont
  {Sank}, \citenamefont {Skolik}, \citenamefont {Smelyanskiy}, \citenamefont
  {Strain}, \citenamefont {Streif}, \citenamefont {Szalay}, \citenamefont
  {Vainsencher}, \citenamefont {White}, \citenamefont {Yao}, \citenamefont
  {Yeh}, \citenamefont {Zalcman}, \citenamefont {Zhou}, \citenamefont {Neven},
  \citenamefont {Bacon}, \citenamefont {Lucero}, \citenamefont {Farhi},\ and\
  \citenamefont {Babbush}}]{Harrigan2021}%
  \BibitemOpen
  \bibfield  {author} {\bibinfo {author} {\bibfnamefont {M.P.}\ \bibnamefont
  {Harrigan}}, \bibinfo {author} {\bibfnamefont {K.~J.}\ \bibnamefont {Sung}},
  \bibinfo {author} {\bibfnamefont {M.}~\bibnamefont {Neeley}}, \bibinfo
  {author} {\bibfnamefont {K.~J.}\ \bibnamefont {Satzinger}}, \bibinfo {author}
  {\bibfnamefont {F.}~\bibnamefont {Arute}}, \bibinfo {author} {\bibfnamefont
  {K.}~\bibnamefont {Arya}}, \bibinfo {author} {\bibfnamefont {J.}~\bibnamefont
  {Atalaya}}, \bibinfo {author} {\bibfnamefont {J.~C.}\ \bibnamefont {Bardin}},
  \bibinfo {author} {\bibfnamefont {R.}~\bibnamefont {Barends}}, \bibinfo
  {author} {\bibfnamefont {S.}~\bibnamefont {Boixo}}, \bibinfo {author}
  {\bibfnamefont {M.}~\bibnamefont {Broughton}}, \bibinfo {author}
  {\bibfnamefont {B.~B.}\ \bibnamefont {Buckley}}, \bibinfo {author}
  {\bibfnamefont {D.~A.}\ \bibnamefont {Buell}}, \bibinfo {author}
  {\bibfnamefont {B.}~\bibnamefont {Burkett}}, \bibinfo {author} {\bibfnamefont
  {N.}~\bibnamefont {Bushnell}}, \bibinfo {author} {\bibfnamefont
  {Y.}~\bibnamefont {Chen}}, \bibinfo {author} {\bibfnamefont {Z.}~\bibnamefont
  {Chen}}, \bibinfo {author} {\bibfnamefont {B.}~\bibnamefont {Chiaro}},
  \bibinfo {author} {\bibfnamefont {R.}~\bibnamefont {Collins}}, \bibinfo
  {author} {\bibfnamefont {W.}~\bibnamefont {Courtney}}, \bibinfo {author}
  {\bibfnamefont {S.}~\bibnamefont {Demura}}, \bibinfo {author} {\bibfnamefont
  {A.}~\bibnamefont {Dunsworth}}, \bibinfo {author} {\bibfnamefont
  {D.}~\bibnamefont {Eppens}}, \bibinfo {author} {\bibfnamefont
  {A.}~\bibnamefont {Fowler}}, \bibinfo {author} {\bibfnamefont
  {B.}~\bibnamefont {Foxen}}, \bibinfo {author} {\bibfnamefont
  {C.}~\bibnamefont {Gidney}}, \bibinfo {author} {\bibfnamefont
  {M.}~\bibnamefont {Giustina}}, \bibinfo {author} {\bibfnamefont
  {R.}~\bibnamefont {Graff}}, \bibinfo {author} {\bibfnamefont
  {S.}~\bibnamefont {Habegger}}, \bibinfo {author} {\bibfnamefont
  {A.}~\bibnamefont {Ho}}, \bibinfo {author} {\bibfnamefont {S.}~\bibnamefont
  {Hong}}, \bibinfo {author} {\bibfnamefont {T.}~\bibnamefont {Huang}},
  \bibinfo {author} {\bibfnamefont {L.~B.}\ \bibnamefont {Ioffe}}, \bibinfo
  {author} {\bibfnamefont {S.~V.}\ \bibnamefont {Isakov}}, \bibinfo {author}
  {\bibfnamefont {E.}~\bibnamefont {Jeffrey}}, \bibinfo {author} {\bibfnamefont
  {Z.}~\bibnamefont {Jiang}}, \bibinfo {author} {\bibfnamefont
  {C.}~\bibnamefont {Jones}}, \bibinfo {author} {\bibfnamefont
  {D.}~\bibnamefont {Kafri}}, \bibinfo {author} {\bibfnamefont
  {K.}~\bibnamefont {Kechedzhi}}, \bibinfo {author} {\bibfnamefont
  {J.}~\bibnamefont {Kelly}}, \bibinfo {author} {\bibfnamefont
  {S.}~\bibnamefont {Kim}}, \bibinfo {author} {\bibfnamefont {P.~V.}\
  \bibnamefont {Klimov}}, \bibinfo {author} {\bibfnamefont {A.~N.}\
  \bibnamefont {Korotkov}}, \bibinfo {author} {\bibfnamefont {F.}~\bibnamefont
  {Kostritsa}}, \bibinfo {author} {\bibfnamefont {D.}~\bibnamefont {Landhuis}},
  \bibinfo {author} {\bibfnamefont {P.}~\bibnamefont {Laptev}}, \bibinfo
  {author} {\bibfnamefont {M.}~\bibnamefont {Lindmark}}, \bibinfo {author}
  {\bibfnamefont {M.}~\bibnamefont {Leib}}, \bibinfo {author} {\bibfnamefont
  {O.}~\bibnamefont {Martin}}, \bibinfo {author} {\bibfnamefont {J.~M.}\
  \bibnamefont {Martinis}}, \bibinfo {author} {\bibfnamefont {J.~R.}\
  \bibnamefont {McClean}}, \bibinfo {author} {\bibfnamefont {M.}~\bibnamefont
  {McEwen}}, \bibinfo {author} {\bibfnamefont {A.}~\bibnamefont {Megrant}},
  \bibinfo {author} {\bibfnamefont {X.}~\bibnamefont {Mi}}, \bibinfo {author}
  {\bibfnamefont {M.}~\bibnamefont {Mohseni}}, \bibinfo {author} {\bibfnamefont
  {W.}~\bibnamefont {Mruczkiewicz}}, \bibinfo {author} {\bibfnamefont
  {J.}~\bibnamefont {Mutus}}, \bibinfo {author} {\bibfnamefont
  {O.}~\bibnamefont {Naaman}}, \bibinfo {author} {\bibfnamefont
  {C.}~\bibnamefont {Neill}}, \bibinfo {author} {\bibfnamefont
  {F.}~\bibnamefont {Neukart}}, \bibinfo {author} {\bibfnamefont {M.~Y.}\
  \bibnamefont {Niu}}, \bibinfo {author} {\bibfnamefont {T.~E.}\ \bibnamefont
  {O'Brien}}, \bibinfo {author} {\bibfnamefont {B.}~\bibnamefont {O'Gorman}},
  \bibinfo {author} {\bibfnamefont {E.}~\bibnamefont {Ostby}}, \bibinfo
  {author} {\bibfnamefont {A.}~\bibnamefont {Petukhov}}, \bibinfo {author}
  {\bibfnamefont {H.}~\bibnamefont {Putterman}}, \bibinfo {author}
  {\bibfnamefont {C.}~\bibnamefont {Quintana}}, \bibinfo {author}
  {\bibfnamefont {P.}~\bibnamefont {Roushan}}, \bibinfo {author} {\bibfnamefont
  {N.~C.}\ \bibnamefont {Rubin}}, \bibinfo {author} {\bibfnamefont
  {D.}~\bibnamefont {Sank}}, \bibinfo {author} {\bibfnamefont {A.}~\bibnamefont
  {Skolik}}, \bibinfo {author} {\bibfnamefont {V.}~\bibnamefont {Smelyanskiy}},
  \bibinfo {author} {\bibfnamefont {D.}~\bibnamefont {Strain}}, \bibinfo
  {author} {\bibfnamefont {M.}~\bibnamefont {Streif}}, \bibinfo {author}
  {\bibfnamefont {M.}~\bibnamefont {Szalay}}, \bibinfo {author} {\bibfnamefont
  {A.}~\bibnamefont {Vainsencher}}, \bibinfo {author} {\bibfnamefont
  {T.}~\bibnamefont {White}}, \bibinfo {author} {\bibfnamefont {Z.~J.}\
  \bibnamefont {Yao}}, \bibinfo {author} {\bibfnamefont {P.}~\bibnamefont
  {Yeh}}, \bibinfo {author} {\bibfnamefont {A.}~\bibnamefont {Zalcman}},
  \bibinfo {author} {\bibfnamefont {L.}~\bibnamefont {Zhou}}, \bibinfo {author}
  {\bibfnamefont {H.}~\bibnamefont {Neven}}, \bibinfo {author} {\bibfnamefont
  {D.}~\bibnamefont {Bacon}}, \bibinfo {author} {\bibfnamefont
  {E.}~\bibnamefont {Lucero}}, \bibinfo {author} {\bibfnamefont
  {E.}~\bibnamefont {Farhi}}, \ and\ \bibinfo {author} {\bibfnamefont
  {R.}~\bibnamefont {Babbush}},\ }\bibfield  {title} {\enquote {\bibinfo
  {title} {Quantum approximate optimization of non-planar graph problems on a
  planar superconducting processor},}\ }\href@noop {} {\bibfield  {journal}
  {\bibinfo  {journal} {Nat. Phys.}\ }\textbf {\bibinfo {volume} {17}},\
  \bibinfo {pages} {332} (\bibinfo {year} {2021})}\BibitemShut {NoStop}%
\bibitem [{\citenamefont {Watson}\ \emph {et~al.}(2018)\citenamefont {Watson},
  \citenamefont {Philips}, \citenamefont {Kawakami}, \citenamefont {Ward},
  \citenamefont {Scarlino}, \citenamefont {Veldhorst}, \citenamefont {Savage},
  \citenamefont {Lagally}, \citenamefont {Friesen}, \citenamefont
  {Coppersmith}, \citenamefont {Eriksson},\ and\ \citenamefont
  {Vandersypen}}]{Watson2018}%
  \BibitemOpen
  \bibfield  {author} {\bibinfo {author} {\bibfnamefont {T.~F.}\ \bibnamefont
  {Watson}}, \bibinfo {author} {\bibfnamefont {S.~G.~J.}\ \bibnamefont
  {Philips}}, \bibinfo {author} {\bibfnamefont {E.}~\bibnamefont {Kawakami}},
  \bibinfo {author} {\bibfnamefont {D.~R.}\ \bibnamefont {Ward}}, \bibinfo
  {author} {\bibfnamefont {P.}~\bibnamefont {Scarlino}}, \bibinfo {author}
  {\bibfnamefont {M.}~\bibnamefont {Veldhorst}}, \bibinfo {author}
  {\bibfnamefont {D.~E.}\ \bibnamefont {Savage}}, \bibinfo {author}
  {\bibfnamefont {M.~G.}\ \bibnamefont {Lagally}}, \bibinfo {author}
  {\bibfnamefont {Mark}\ \bibnamefont {Friesen}}, \bibinfo {author}
  {\bibfnamefont {S.~N.}\ \bibnamefont {Coppersmith}}, \bibinfo {author}
  {\bibfnamefont {M.~A.}\ \bibnamefont {Eriksson}}, \ and\ \bibinfo {author}
  {\bibfnamefont {L.~M.~K.}\ \bibnamefont {Vandersypen}},\ }\bibfield  {title}
  {\enquote {\bibinfo {title} {A programmable two-qubit quantum processor in
  silicon},}\ }\href@noop {} {\bibfield  {journal} {\bibinfo  {journal}
  {Nature}\ }\textbf {\bibinfo {volume} {555}},\ \bibinfo {pages} {633}
  (\bibinfo {year} {2018})}\BibitemShut {NoStop}%
\bibitem [{\citenamefont {Zhou}\ \emph {et~al.}(2013)\citenamefont {Zhou},
  \citenamefont {Kalasuwan}, \citenamefont {Ralph},\ and\ \citenamefont
  {O'Brien}}]{XQZhou2013}%
  \BibitemOpen
  \bibfield  {author} {\bibinfo {author} {\bibfnamefont {X.-Q.}\ \bibnamefont
  {Zhou}}, \bibinfo {author} {\bibfnamefont {P.}~\bibnamefont {Kalasuwan}},
  \bibinfo {author} {\bibfnamefont {T.~C.}\ \bibnamefont {Ralph}}, \ and\
  \bibinfo {author} {\bibfnamefont {J.~L.}\ \bibnamefont {O'Brien}},\
  }\bibfield  {title} {\enquote {\bibinfo {title} {Calculating unknown
  eigenvalues with a quantum algorithm},}\ }\href@noop {} {\bibfield  {journal}
  {\bibinfo  {journal} {Nat. Phot.}\ }\textbf {\bibinfo {volume} {7}},\
  \bibinfo {pages} {223} (\bibinfo {year} {2013})}\BibitemShut {NoStop}%
\bibitem [{\citenamefont {Scholl}\ \emph {et~al.}(2021)\citenamefont {Scholl},
  \citenamefont {Schuler}, \citenamefont {Williams}, \citenamefont
  {Eberharter}, \citenamefont {Barredo}, \citenamefont {Schymik}, \citenamefont
  {Lienhard}, \citenamefont {Henry}, \citenamefont {Lang}, \citenamefont
  {Lahaye}, \citenamefont {L\"auchli},\ and\ \citenamefont
  {Browaeys}}]{Scholl2021}%
  \BibitemOpen
  \bibfield  {author} {\bibinfo {author} {\bibfnamefont {P.}~\bibnamefont
  {Scholl}}, \bibinfo {author} {\bibfnamefont {M.}~\bibnamefont {Schuler}},
  \bibinfo {author} {\bibfnamefont {H.~J.}\ \bibnamefont {Williams}}, \bibinfo
  {author} {\bibfnamefont {A.~A.}\ \bibnamefont {Eberharter}}, \bibinfo
  {author} {\bibfnamefont {D.}~\bibnamefont {Barredo}}, \bibinfo {author}
  {\bibfnamefont {K.-N.}\ \bibnamefont {Schymik}}, \bibinfo {author}
  {\bibfnamefont {V.}~\bibnamefont {Lienhard}}, \bibinfo {author}
  {\bibfnamefont {L.-P.}\ \bibnamefont {Henry}}, \bibinfo {author}
  {\bibfnamefont {T.~C.}\ \bibnamefont {Lang}}, \bibinfo {author}
  {\bibfnamefont {T.}~\bibnamefont {Lahaye}}, \bibinfo {author} {\bibfnamefont
  {A.~M.}\ \bibnamefont {L\"auchli}}, \ and\ \bibinfo {author} {\bibfnamefont
  {A.}~\bibnamefont {Browaeys}},\ }\bibfield  {title} {\enquote {\bibinfo
  {title} {Quantum simulation of 2{D} antiferromagnets with hundreds of
  {R}ydberg atoms},}\ }\href@noop {} {\bibfield  {journal} {\bibinfo  {journal}
  {Nature}\ }\textbf {\bibinfo {volume} {595}},\ \bibinfo {pages} {233}
  (\bibinfo {year} {2021})}\BibitemShut {NoStop}%
\bibitem [{\citenamefont {Ebadi}\ \emph {et~al.}(2021)\citenamefont {Ebadi},
  \citenamefont {Wang}, \citenamefont {Levine}, \citenamefont {Keesling},
  \citenamefont {Semeghini}, \citenamefont {Omran}, \citenamefont {Bluvstein},
  \citenamefont {Samajdar}, \citenamefont {Pichler}, \citenamefont {Ho},
  \citenamefont {Choi}, \citenamefont {Sachdev}, \citenamefont {Greiner},
  \citenamefont {Vuleti\'c},\ and\ \citenamefont {Lukin}}]{Ebadi2021}%
  \BibitemOpen
  \bibfield  {author} {\bibinfo {author} {\bibfnamefont {S.}~\bibnamefont
  {Ebadi}}, \bibinfo {author} {\bibfnamefont {T.~T.}\ \bibnamefont {Wang}},
  \bibinfo {author} {\bibfnamefont {H.}~\bibnamefont {Levine}}, \bibinfo
  {author} {\bibfnamefont {A.}~\bibnamefont {Keesling}}, \bibinfo {author}
  {\bibfnamefont {G.}~\bibnamefont {Semeghini}}, \bibinfo {author}
  {\bibfnamefont {A.}~\bibnamefont {Omran}}, \bibinfo {author} {\bibfnamefont
  {D.}~\bibnamefont {Bluvstein}}, \bibinfo {author} {\bibfnamefont
  {R.}~\bibnamefont {Samajdar}}, \bibinfo {author} {\bibfnamefont
  {H.}~\bibnamefont {Pichler}}, \bibinfo {author} {\bibfnamefont {W.~W.}\
  \bibnamefont {Ho}}, \bibinfo {author} {\bibfnamefont {S.}~\bibnamefont
  {Choi}}, \bibinfo {author} {\bibfnamefont {S.}~\bibnamefont {Sachdev}},
  \bibinfo {author} {\bibfnamefont {M.}~\bibnamefont {Greiner}}, \bibinfo
  {author} {\bibfnamefont {V.}~\bibnamefont {Vuleti\'c}}, \ and\ \bibinfo
  {author} {\bibfnamefont {M.~D.}\ \bibnamefont {Lukin}},\ }\bibfield  {title}
  {\enquote {\bibinfo {title} {Quantum phases of matter on a 256-atom
  programmable quantum simulator},}\ }\href@noop {} {\bibfield  {journal}
  {\bibinfo  {journal} {Nature}\ }\textbf {\bibinfo {volume} {595}},\ \bibinfo
  {pages} {227} (\bibinfo {year} {2021})}\BibitemShut {NoStop}%
\bibitem [{\citenamefont {Hauke}\ \emph {et~al.}(2012)\citenamefont {Hauke},
  \citenamefont {Cucchietti}, \citenamefont {Tagliacozzo}, \citenamefont
  {Deutsch},\ and\ \citenamefont {Lewenstein}}]{Hauke2012}%
  \BibitemOpen
  \bibfield  {author} {\bibinfo {author} {\bibfnamefont {P.}~\bibnamefont
  {Hauke}}, \bibinfo {author} {\bibfnamefont {F.~M.}\ \bibnamefont
  {Cucchietti}}, \bibinfo {author} {\bibfnamefont {L.}~\bibnamefont
  {Tagliacozzo}}, \bibinfo {author} {\bibfnamefont {I.}~\bibnamefont
  {Deutsch}}, \ and\ \bibinfo {author} {\bibfnamefont {M.}~\bibnamefont
  {Lewenstein}},\ }\bibfield  {title} {\enquote {\bibinfo {title} {Can one
  trust quantum simulators?}}\ }\href@noop {} {\bibfield  {journal} {\bibinfo
  {journal} {Rep. Progr. Phys.}\ }\textbf {\bibinfo {volume} {75}},\ \bibinfo
  {pages} {082401} (\bibinfo {year} {2012})}\BibitemShut {NoStop}%
\bibitem [{\citenamefont {Aharonov}\ and\ \citenamefont
  {Ben-Or}(2008)}]{Aharonov2008}%
  \BibitemOpen
  \bibfield  {author} {\bibinfo {author} {\bibfnamefont {D.}~\bibnamefont
  {Aharonov}}\ and\ \bibinfo {author} {\bibfnamefont {M.}~\bibnamefont
  {Ben-Or}},\ }\bibfield  {title} {\enquote {\bibinfo {title} {Fault-tolerant
  quantum computation with constant error rate},}\ }\href@noop {} {\bibfield
  {journal} {\bibinfo  {journal} {SIAM Jour. Comput.}\ }\textbf {\bibinfo
  {volume} {38}},\ \bibinfo {pages} {1207--1282} (\bibinfo {year}
  {2008})}\BibitemShut {NoStop}%
\bibitem [{\citenamefont {Knill}\ \emph {et~al.}(1998)\citenamefont {Knill},
  \citenamefont {Laflamme},\ and\ \citenamefont {Zurek}}]{Knill1998}%
  \BibitemOpen
  \bibfield  {author} {\bibinfo {author} {\bibfnamefont {E.}~\bibnamefont
  {Knill}}, \bibinfo {author} {\bibfnamefont {R.}~\bibnamefont {Laflamme}}, \
  and\ \bibinfo {author} {\bibfnamefont {W.~H.}\ \bibnamefont {Zurek}},\
  }\bibfield  {title} {\enquote {\bibinfo {title} {Resilient quantum
  computation},}\ }\href@noop {} {\bibfield  {journal} {\bibinfo  {journal}
  {Science}\ }\textbf {\bibinfo {volume} {279}},\ \bibinfo {pages} {342}
  (\bibinfo {year} {1998})}\BibitemShut {NoStop}%
\bibitem [{SM2()}]{SM2021circuits}%
  \BibitemOpen
  \href@noop {} {}\bibinfo {note} {Supplementary material at ... which includes
  references \cite{YFHsiao2018,
  Gillen-Christandl2016,Kuhr2005,Maller2015,Levine2019,Saffman2020,SZhang2011}.}\BibitemShut
  {Stop}%
\bibitem [{\citenamefont {Xia}\ \emph {et~al.}(2015)\citenamefont {Xia},
  \citenamefont {Lichtman}, \citenamefont {Maller}, \citenamefont {Carr},
  \citenamefont {Piotrowicz}, \citenamefont {Isenhower},\ and\ \citenamefont
  {Saffman}}]{Xia2015}%
  \BibitemOpen
  \bibfield  {author} {\bibinfo {author} {\bibfnamefont {T.}~\bibnamefont
  {Xia}}, \bibinfo {author} {\bibfnamefont {M.}~\bibnamefont {Lichtman}},
  \bibinfo {author} {\bibfnamefont {K.}~\bibnamefont {Maller}}, \bibinfo
  {author} {\bibfnamefont {A.~W.}\ \bibnamefont {Carr}}, \bibinfo {author}
  {\bibfnamefont {M.~J.}\ \bibnamefont {Piotrowicz}}, \bibinfo {author}
  {\bibfnamefont {L.}~\bibnamefont {Isenhower}}, \ and\ \bibinfo {author}
  {\bibfnamefont {M.}~\bibnamefont {Saffman}},\ }\bibfield  {title} {\enquote
  {\bibinfo {title} {Randomized benchmarking of single-qubit gates in a {2D}
  array of neutral-atom qubits},}\ }\href@noop {} {\bibfield  {journal}
  {\bibinfo  {journal} {Phys. Rev. Lett.}\ }\textbf {\bibinfo {volume} {114}},\
  \bibinfo {pages} {100503} (\bibinfo {year} {2015})}\BibitemShut {NoStop}%
\bibitem [{\citenamefont {Wang}\ \emph {et~al.}(2016)\citenamefont {Wang},
  \citenamefont {Kumar}, \citenamefont {Wu},\ and\ \citenamefont
  {Weiss}}]{YWang2016}%
  \BibitemOpen
  \bibfield  {author} {\bibinfo {author} {\bibfnamefont {Y.}~\bibnamefont
  {Wang}}, \bibinfo {author} {\bibfnamefont {A.}~\bibnamefont {Kumar}},
  \bibinfo {author} {\bibfnamefont {T.-Y.}\ \bibnamefont {Wu}}, \ and\ \bibinfo
  {author} {\bibfnamefont {D.~S.}\ \bibnamefont {Weiss}},\ }\bibfield  {title}
  {\enquote {\bibinfo {title} {Single-qubit gates based on targeted phase
  shifts in a 3{D} neutral atom array},}\ }\href@noop {} {\bibfield  {journal}
  {\bibinfo  {journal} {Science}\ }\textbf {\bibinfo {volume} {352}},\ \bibinfo
  {pages} {1562} (\bibinfo {year} {2016})}\BibitemShut {NoStop}%
\bibitem [{\citenamefont {Graham}\ \emph {et~al.}(2019)\citenamefont {Graham},
  \citenamefont {Kwon}, \citenamefont {Grinkemeyer}, \citenamefont {Marra},
  \citenamefont {Jiang}, \citenamefont {Lichtman}, \citenamefont {Sun},
  \citenamefont {Ebert},\ and\ \citenamefont {Saffman}}]{Graham2019}%
  \BibitemOpen
  \bibfield  {author} {\bibinfo {author} {\bibfnamefont {T.}~\bibnamefont
  {Graham}}, \bibinfo {author} {\bibfnamefont {M.}~\bibnamefont {Kwon}},
  \bibinfo {author} {\bibfnamefont {B.}~\bibnamefont {Grinkemeyer}}, \bibinfo
  {author} {\bibfnamefont {A.}~\bibnamefont {Marra}}, \bibinfo {author}
  {\bibfnamefont {X.}~\bibnamefont {Jiang}}, \bibinfo {author} {\bibfnamefont
  {M.}~\bibnamefont {Lichtman}}, \bibinfo {author} {\bibfnamefont
  {Y.}~\bibnamefont {Sun}}, \bibinfo {author} {\bibfnamefont {M.}~\bibnamefont
  {Ebert}}, \ and\ \bibinfo {author} {\bibfnamefont {M.}~\bibnamefont
  {Saffman}},\ }\bibfield  {title} {\enquote {\bibinfo {title} {Rydberg
  mediated entanglement in a two-dimensional neutral atom qubit array},}\
  }\href@noop {} {\bibfield  {journal} {\bibinfo  {journal} {Phys. Rev. Lett.}\
  }\textbf {\bibinfo {volume} {123}},\ \bibinfo {pages} {230501} (\bibinfo
  {year} {2019})}\BibitemShut {NoStop}%
\bibitem [{\citenamefont {Madjarov}\ \emph {et~al.}(2020)\citenamefont
  {Madjarov}, \citenamefont {Covey}, \citenamefont {Shaw}, \citenamefont
  {Choi}, \citenamefont {Kale}, \citenamefont {Cooper}, \citenamefont
  {Pichler}, \citenamefont {Schkolnik}, \citenamefont {Williams},\ and\
  \citenamefont {Endres}}]{Madjarov2020}%
  \BibitemOpen
  \bibfield  {author} {\bibinfo {author} {\bibfnamefont {I.~S.}\ \bibnamefont
  {Madjarov}}, \bibinfo {author} {\bibfnamefont {J.~P.}\ \bibnamefont {Covey}},
  \bibinfo {author} {\bibfnamefont {A.~L.}\ \bibnamefont {Shaw}}, \bibinfo
  {author} {\bibfnamefont {J.}~\bibnamefont {Choi}}, \bibinfo {author}
  {\bibfnamefont {A.}~\bibnamefont {Kale}}, \bibinfo {author} {\bibfnamefont
  {A.}~\bibnamefont {Cooper}}, \bibinfo {author} {\bibfnamefont
  {H.}~\bibnamefont {Pichler}}, \bibinfo {author} {\bibfnamefont
  {V.}~\bibnamefont {Schkolnik}}, \bibinfo {author} {\bibfnamefont {J.~R.}\
  \bibnamefont {Williams}}, \ and\ \bibinfo {author} {\bibfnamefont
  {M.}~\bibnamefont {Endres}},\ }\bibfield  {title} {\enquote {\bibinfo {title}
  {High-fidelity entanglement and detection of alkaline-earth {R}ydberg
  atoms},}\ }\href@noop {} {\bibfield  {journal} {\bibinfo  {journal} {Nat.
  Phys.}\ }\textbf {\bibinfo {volume} {16}},\ \bibinfo {pages} {857} (\bibinfo
  {year} {2020})}\BibitemShut {NoStop}%
\bibitem [{\citenamefont {Barredo}\ \emph {et~al.}(2016)\citenamefont
  {Barredo}, \citenamefont {de~Les\'el\'euc}, \citenamefont {Lienhard},
  \citenamefont {Lahaye},\ and\ \citenamefont {Browaeys}}]{Barredo2016}%
  \BibitemOpen
  \bibfield  {author} {\bibinfo {author} {\bibfnamefont {D.}~\bibnamefont
  {Barredo}}, \bibinfo {author} {\bibfnamefont {S.}~\bibnamefont
  {de~Les\'el\'euc}}, \bibinfo {author} {\bibfnamefont {V.}~\bibnamefont
  {Lienhard}}, \bibinfo {author} {\bibfnamefont {T.}~\bibnamefont {Lahaye}}, \
  and\ \bibinfo {author} {\bibfnamefont {A.}~\bibnamefont {Browaeys}},\
  }\bibfield  {title} {\enquote {\bibinfo {title} {An atom-by-atom assembler of
  defect-free arbitrary two-dimensional atomic arrays},}\ }\href@noop {}
  {\bibfield  {journal} {\bibinfo  {journal} {Science}\ }\textbf {\bibinfo
  {volume} {354}},\ \bibinfo {pages} {1021} (\bibinfo {year}
  {2016})}\BibitemShut {NoStop}%
\bibitem [{\citenamefont {Endres}\ \emph {et~al.}(2016)\citenamefont {Endres},
  \citenamefont {Bernien}, \citenamefont {Keesling}, \citenamefont {Levine},
  \citenamefont {Anschuetz}, \citenamefont {Krajenbrink}, \citenamefont
  {Senko}, \citenamefont {Vuletic}, \citenamefont {Greiner},\ and\
  \citenamefont {Lukin}}]{Endres2016}%
  \BibitemOpen
  \bibfield  {author} {\bibinfo {author} {\bibfnamefont {M.}~\bibnamefont
  {Endres}}, \bibinfo {author} {\bibfnamefont {H.}~\bibnamefont {Bernien}},
  \bibinfo {author} {\bibfnamefont {A.}~\bibnamefont {Keesling}}, \bibinfo
  {author} {\bibfnamefont {H.}~\bibnamefont {Levine}}, \bibinfo {author}
  {\bibfnamefont {E.~R.}\ \bibnamefont {Anschuetz}}, \bibinfo {author}
  {\bibfnamefont {A.}~\bibnamefont {Krajenbrink}}, \bibinfo {author}
  {\bibfnamefont {C.}~\bibnamefont {Senko}}, \bibinfo {author} {\bibfnamefont
  {V.}~\bibnamefont {Vuletic}}, \bibinfo {author} {\bibfnamefont
  {M.}~\bibnamefont {Greiner}}, \ and\ \bibinfo {author} {\bibfnamefont
  {M.~D.}\ \bibnamefont {Lukin}},\ }\bibfield  {title} {\enquote {\bibinfo
  {title} {Atom-by-atom assembly of defect-free one-dimensional cold atom
  arrays},}\ }\href@noop {} {\bibfield  {journal} {\bibinfo  {journal}
  {Science}\ }\textbf {\bibinfo {volume} {354}},\ \bibinfo {pages} {1024}
  (\bibinfo {year} {2016})}\BibitemShut {NoStop}%
\bibitem [{\citenamefont {Kim}\ \emph {et~al.}(2016)\citenamefont {Kim},
  \citenamefont {Lee}, \citenamefont {g.~Lee}, \citenamefont {Jo},
  \citenamefont {Song},\ and\ \citenamefont {Ahn}}]{Kim2016}%
  \BibitemOpen
  \bibfield  {author} {\bibinfo {author} {\bibfnamefont {H.}~\bibnamefont
  {Kim}}, \bibinfo {author} {\bibfnamefont {W.}~\bibnamefont {Lee}}, \bibinfo
  {author} {\bibfnamefont {H.}~\bibnamefont {g.~Lee}}, \bibinfo {author}
  {\bibfnamefont {H.}~\bibnamefont {Jo}}, \bibinfo {author} {\bibfnamefont
  {Y.}~\bibnamefont {Song}}, \ and\ \bibinfo {author} {\bibfnamefont
  {J.}~\bibnamefont {Ahn}},\ }\bibfield  {title} {\enquote {\bibinfo {title}
  {In situ single-atom array synthesis using dynamic holographic optical
  tweezers},}\ }\href@noop {} {\bibfield  {journal} {\bibinfo  {journal} {Nat.
  Commun.}\ }\textbf {\bibinfo {volume} {7}},\ \bibinfo {pages} {13317}
  (\bibinfo {year} {2016})}\BibitemShut {NoStop}%
\bibitem [{\citenamefont {Gisin}\ and\ \citenamefont
  {Bechmann-Pasquinucci}(1998)}]{Gisin1998}%
  \BibitemOpen
  \bibfield  {author} {\bibinfo {author} {\bibfnamefont {N.}~\bibnamefont
  {Gisin}}\ and\ \bibinfo {author} {\bibfnamefont {H.}~\bibnamefont
  {Bechmann-Pasquinucci}},\ }\bibfield  {title} {\enquote {\bibinfo {title}
  {Bell inequality, {B}ell states and maximally entangled states for $n$
  qubits},}\ }\href {\doibase https://doi.org/10.1016/S0375-9601(98)00516-7}
  {\bibfield  {journal} {\bibinfo  {journal} {Phys. Lett. A}\ }\textbf
  {\bibinfo {volume} {246}},\ \bibinfo {pages} {1} (\bibinfo {year}
  {1998})}\BibitemShut {NoStop}%
\bibitem [{\citenamefont {Song}\ \emph {et~al.}(2019)\citenamefont {Song},
  \citenamefont {Xu}, \citenamefont {Li}, \citenamefont {Zhang}, \citenamefont
  {Zhang}, \citenamefont {Liu}, \citenamefont {Guo}, \citenamefont {Wang},
  \citenamefont {Ren}, \citenamefont {Hao}, \citenamefont {Feng}, \citenamefont
  {Fan}, \citenamefont {Zheng}, \citenamefont {Wang}, \citenamefont {Wang},\
  and\ \citenamefont {Zhu}}]{CSong2019}%
  \BibitemOpen
  \bibfield  {author} {\bibinfo {author} {\bibfnamefont {C.}~\bibnamefont
  {Song}}, \bibinfo {author} {\bibfnamefont {K.}~\bibnamefont {Xu}}, \bibinfo
  {author} {\bibfnamefont {H.}~\bibnamefont {Li}}, \bibinfo {author}
  {\bibfnamefont {Y.-R.}\ \bibnamefont {Zhang}}, \bibinfo {author}
  {\bibfnamefont {X.}~\bibnamefont {Zhang}}, \bibinfo {author} {\bibfnamefont
  {W.}~\bibnamefont {Liu}}, \bibinfo {author} {\bibfnamefont {Q.}~\bibnamefont
  {Guo}}, \bibinfo {author} {\bibfnamefont {Z.}~\bibnamefont {Wang}}, \bibinfo
  {author} {\bibfnamefont {W.}~\bibnamefont {Ren}}, \bibinfo {author}
  {\bibfnamefont {J.}~\bibnamefont {Hao}}, \bibinfo {author} {\bibfnamefont
  {H.}~\bibnamefont {Feng}}, \bibinfo {author} {\bibfnamefont {H.}~\bibnamefont
  {Fan}}, \bibinfo {author} {\bibfnamefont {D.}~\bibnamefont {Zheng}}, \bibinfo
  {author} {\bibfnamefont {D.-W.}\ \bibnamefont {Wang}}, \bibinfo {author}
  {\bibfnamefont {H.}~\bibnamefont {Wang}}, \ and\ \bibinfo {author}
  {\bibfnamefont {S.-Y.}\ \bibnamefont {Zhu}},\ }\bibfield  {title} {\enquote
  {\bibinfo {title} {Generation of multicomponent atomic {S}chr\"odinger cat
  states of up to 20 qubits},}\ }\href {\doibase 10.1126/science.aay0600}
  {\bibfield  {journal} {\bibinfo  {journal} {Science}\ }\textbf {\bibinfo
  {volume} {365}},\ \bibinfo {pages} {574–577} (\bibinfo {year}
  {2019})}\BibitemShut {NoStop}%
\bibitem [{\citenamefont {Pogorelov}\ \emph {et~al.}(2021)\citenamefont
  {Pogorelov}, \citenamefont {Feldker}, \citenamefont {Marciniak},
  \citenamefont {Postler}, \citenamefont {Jacob}, \citenamefont
  {Krieglsteiner}, \citenamefont {Podlesnic}, \citenamefont {Meth},
  \citenamefont {Negnevitsky}, \citenamefont {Stadler}, \citenamefont
  {H\"ofer}, \citenamefont {W\"achter}, \citenamefont {Lakhmanskiy},
  \citenamefont {Blatt}, \citenamefont {Schindler},\ and\ \citenamefont
  {Monz}}]{Pogorelov2021}%
  \BibitemOpen
  \bibfield  {author} {\bibinfo {author} {\bibfnamefont {I.}~\bibnamefont
  {Pogorelov}}, \bibinfo {author} {\bibfnamefont {T.}~\bibnamefont {Feldker}},
  \bibinfo {author} {\bibfnamefont {Ch.~D.}\ \bibnamefont {Marciniak}},
  \bibinfo {author} {\bibfnamefont {L.}~\bibnamefont {Postler}}, \bibinfo
  {author} {\bibfnamefont {G.}~\bibnamefont {Jacob}}, \bibinfo {author}
  {\bibfnamefont {O.}~\bibnamefont {Krieglsteiner}}, \bibinfo {author}
  {\bibfnamefont {V.}~\bibnamefont {Podlesnic}}, \bibinfo {author}
  {\bibfnamefont {M.}~\bibnamefont {Meth}}, \bibinfo {author} {\bibfnamefont
  {V.}~\bibnamefont {Negnevitsky}}, \bibinfo {author} {\bibfnamefont
  {M.}~\bibnamefont {Stadler}}, \bibinfo {author} {\bibfnamefont
  {B.}~\bibnamefont {H\"ofer}}, \bibinfo {author} {\bibfnamefont
  {C.}~\bibnamefont {W\"achter}}, \bibinfo {author} {\bibfnamefont
  {K.}~\bibnamefont {Lakhmanskiy}}, \bibinfo {author} {\bibfnamefont
  {R.}~\bibnamefont {Blatt}}, \bibinfo {author} {\bibfnamefont
  {P.}~\bibnamefont {Schindler}}, \ and\ \bibinfo {author} {\bibfnamefont
  {T.}~\bibnamefont {Monz}},\ }\bibfield  {title} {\enquote {\bibinfo {title}
  {Compact ion-trap quantum computing demonstrator},}\ }\href {\doibase
  10.1103/PRXQuantum.2.020343} {\bibfield  {journal} {\bibinfo  {journal} {PRX
  Quantum}\ }\textbf {\bibinfo {volume} {2}},\ \bibinfo {pages} {020343}
  (\bibinfo {year} {2021})}\BibitemShut {NoStop}%
\bibitem [{\citenamefont {Omran}\ \emph {et~al.}(2019)\citenamefont {Omran},
  \citenamefont {Levine}, \citenamefont {Keesling}, \citenamefont {Semeghini},
  \citenamefont {Wang}, \citenamefont {Ebadi}, \citenamefont {Bernien},
  \citenamefont {Zibrov}, \citenamefont {Pichler}, \citenamefont {Choi},
  \citenamefont {Cui}, \citenamefont {Rossignolo}, \citenamefont {Rembold},
  \citenamefont {Montangero}, \citenamefont {Calarco}, \citenamefont {Endres},
  \citenamefont {Greiner}, \citenamefont {Vuleti\'c},\ and\ \citenamefont
  {Lukin}}]{Omran2019}%
  \BibitemOpen
  \bibfield  {author} {\bibinfo {author} {\bibfnamefont {A.}~\bibnamefont
  {Omran}}, \bibinfo {author} {\bibfnamefont {H.}~\bibnamefont {Levine}},
  \bibinfo {author} {\bibfnamefont {A.}~\bibnamefont {Keesling}}, \bibinfo
  {author} {\bibfnamefont {G.}~\bibnamefont {Semeghini}}, \bibinfo {author}
  {\bibfnamefont {T.~T.}\ \bibnamefont {Wang}}, \bibinfo {author}
  {\bibfnamefont {S.}~\bibnamefont {Ebadi}}, \bibinfo {author} {\bibfnamefont
  {H.}~\bibnamefont {Bernien}}, \bibinfo {author} {\bibfnamefont {A.~S.}\
  \bibnamefont {Zibrov}}, \bibinfo {author} {\bibfnamefont {H.}~\bibnamefont
  {Pichler}}, \bibinfo {author} {\bibfnamefont {S.}~\bibnamefont {Choi}},
  \bibinfo {author} {\bibfnamefont {J.}~\bibnamefont {Cui}}, \bibinfo {author}
  {\bibfnamefont {M.}~\bibnamefont {Rossignolo}}, \bibinfo {author}
  {\bibfnamefont {P.}~\bibnamefont {Rembold}}, \bibinfo {author} {\bibfnamefont
  {S.}~\bibnamefont {Montangero}}, \bibinfo {author} {\bibfnamefont
  {T.}~\bibnamefont {Calarco}}, \bibinfo {author} {\bibfnamefont
  {M.}~\bibnamefont {Endres}}, \bibinfo {author} {\bibfnamefont
  {M.}~\bibnamefont {Greiner}}, \bibinfo {author} {\bibfnamefont
  {V.}~\bibnamefont {Vuleti\'c}}, \ and\ \bibinfo {author} {\bibfnamefont
  {M.~D.}\ \bibnamefont {Lukin}},\ }\bibfield  {title} {\enquote {\bibinfo
  {title} {Generation and manipulation of {S}chr\"odinger cat states in
  {R}ydberg atom arrays},}\ }\href@noop {} {\bibfield  {journal} {\bibinfo
  {journal} {Science}\ }\textbf {\bibinfo {volume} {365}},\ \bibinfo {pages}
  {570} (\bibinfo {year} {2019})}\BibitemShut {NoStop}%
\bibitem [{\citenamefont {Sackett}\ \emph {et~al.}(2000)\citenamefont
  {Sackett}, \citenamefont {Kielpinski}, \citenamefont {King}, \citenamefont
  {Langer}, \citenamefont {Meyer}, \citenamefont {Myatt}, \citenamefont {Rowe},
  \citenamefont {Turchette}, \citenamefont {Itano}, \citenamefont {Wineland},\
  and\ \citenamefont {Monroe}}]{Sackett2000}%
  \BibitemOpen
  \bibfield  {author} {\bibinfo {author} {\bibfnamefont {C.~A.}\ \bibnamefont
  {Sackett}}, \bibinfo {author} {\bibfnamefont {D.}~\bibnamefont {Kielpinski}},
  \bibinfo {author} {\bibfnamefont {B.~E.}\ \bibnamefont {King}}, \bibinfo
  {author} {\bibfnamefont {C.}~\bibnamefont {Langer}}, \bibinfo {author}
  {\bibfnamefont {V.}~\bibnamefont {Meyer}}, \bibinfo {author} {\bibfnamefont
  {C.~J.}\ \bibnamefont {Myatt}}, \bibinfo {author} {\bibfnamefont
  {M.}~\bibnamefont {Rowe}}, \bibinfo {author} {\bibfnamefont {Q.~A.}\
  \bibnamefont {Turchette}}, \bibinfo {author} {\bibfnamefont {W.~M.}\
  \bibnamefont {Itano}}, \bibinfo {author} {\bibfnamefont {D.~J.}\ \bibnamefont
  {Wineland}}, \ and\ \bibinfo {author} {\bibfnamefont {C.}~\bibnamefont
  {Monroe}},\ }\bibfield  {title} {\enquote {\bibinfo {title} {Experimental
  entanglement of four particles},}\ }\href@noop {} {\bibfield  {journal}
  {\bibinfo  {journal} {Nature (London)}\ }\textbf {\bibinfo {volume} {404}},\
  \bibinfo {pages} {256} (\bibinfo {year} {2000})}\BibitemShut {NoStop}%
\bibitem [{\citenamefont {Wineland}\ \emph {et~al.}(1992)\citenamefont
  {Wineland}, \citenamefont {Bollinger}, \citenamefont {Itano}, \citenamefont
  {Moore},\ and\ \citenamefont {Heinzen}}]{Wineland1992}%
  \BibitemOpen
  \bibfield  {author} {\bibinfo {author} {\bibfnamefont {D.~J.}\ \bibnamefont
  {Wineland}}, \bibinfo {author} {\bibfnamefont {J.~J.}\ \bibnamefont
  {Bollinger}}, \bibinfo {author} {\bibfnamefont {W.~M.}\ \bibnamefont
  {Itano}}, \bibinfo {author} {\bibfnamefont {F.~L.}\ \bibnamefont {Moore}}, \
  and\ \bibinfo {author} {\bibfnamefont {D.~J.}\ \bibnamefont {Heinzen}},\
  }\bibfield  {title} {\enquote {\bibinfo {title} {Spin squeezing and reduced
  quantum noise in spectroscopy},}\ }\href {\doibase 10.1103/PhysRevA.46.R6797}
  {\bibfield  {journal} {\bibinfo  {journal} {Phys. Rev. A}\ }\textbf {\bibinfo
  {volume} {46}},\ \bibinfo {pages} {R6797} (\bibinfo {year}
  {1992})}\BibitemShut {NoStop}%
\bibitem [{\citenamefont {Giovannetti}\ \emph {et~al.}(2004)\citenamefont
  {Giovannetti}, \citenamefont {Lloyd},\ and\ \citenamefont
  {Maccone}}]{Giovannetti2004}%
  \BibitemOpen
  \bibfield  {author} {\bibinfo {author} {\bibfnamefont {V.}~\bibnamefont
  {Giovannetti}}, \bibinfo {author} {\bibfnamefont {S.}~\bibnamefont {Lloyd}},
  \ and\ \bibinfo {author} {\bibfnamefont {L.}~\bibnamefont {Maccone}},\
  }\bibfield  {title} {\enquote {\bibinfo {title} {Quantum-enhanced
  measurements: Beating the standard quantum limit},}\ }\href@noop {}
  {\bibfield  {journal} {\bibinfo  {journal} {Science}\ }\textbf {\bibinfo
  {volume} {306}},\ \bibinfo {pages} {1330} (\bibinfo {year}
  {2004})}\BibitemShut {NoStop}%
\bibitem [{\citenamefont {Saffman}\ and\ \citenamefont
  {Walker}(2005)}]{Saffman2005a}%
  \BibitemOpen
  \bibfield  {author} {\bibinfo {author} {\bibfnamefont {M.}~\bibnamefont
  {Saffman}}\ and\ \bibinfo {author} {\bibfnamefont {T.~G.}\ \bibnamefont
  {Walker}},\ }\bibfield  {title} {\enquote {\bibinfo {title} {Analysis of a
  quantum logic device based on dipole-dipole interactions of optically trapped
  {R}ydberg atoms},}\ }\href@noop {} {\bibfield  {journal} {\bibinfo  {journal}
  {Phys. Rev. A}\ }\textbf {\bibinfo {volume} {72}},\ \bibinfo {pages} {022347}
  (\bibinfo {year} {2005})}\BibitemShut {NoStop}%
\bibitem [{\citenamefont {Carr}\ and\ \citenamefont
  {Saffman}(2016)}]{Carr2016}%
  \BibitemOpen
  \bibfield  {author} {\bibinfo {author} {\bibfnamefont {A.~W.}\ \bibnamefont
  {Carr}}\ and\ \bibinfo {author} {\bibfnamefont {M.}~\bibnamefont {Saffman}},\
  }\bibfield  {title} {\enquote {\bibinfo {title} {Doubly magic optical
  trapping for {C}s atom hyperfine clock transitions},}\ }\href@noop {}
  {\bibfield  {journal} {\bibinfo  {journal} {Phys. Rev. Lett.}\ }\textbf
  {\bibinfo {volume} {117}},\ \bibinfo {pages} {150801} (\bibinfo {year}
  {2016})}\BibitemShut {NoStop}%
\bibitem [{\citenamefont {Monz}\ \emph {et~al.}(2011)\citenamefont {Monz},
  \citenamefont {Schindler}, \citenamefont {Barreiro}, \citenamefont {Chwalla},
  \citenamefont {Nigg}, \citenamefont {Coish}, \citenamefont {Harlander},
  \citenamefont {H\"ansel}, \citenamefont {Hennrich},\ and\ \citenamefont
  {Blatt}}]{Monz2011}%
  \BibitemOpen
  \bibfield  {author} {\bibinfo {author} {\bibfnamefont {Thomas}\ \bibnamefont
  {Monz}}, \bibinfo {author} {\bibfnamefont {Philipp}\ \bibnamefont
  {Schindler}}, \bibinfo {author} {\bibfnamefont {Julio~T.}\ \bibnamefont
  {Barreiro}}, \bibinfo {author} {\bibfnamefont {Michael}\ \bibnamefont
  {Chwalla}}, \bibinfo {author} {\bibfnamefont {Daniel}\ \bibnamefont {Nigg}},
  \bibinfo {author} {\bibfnamefont {William~A.}\ \bibnamefont {Coish}},
  \bibinfo {author} {\bibfnamefont {Maximilian}\ \bibnamefont {Harlander}},
  \bibinfo {author} {\bibfnamefont {Wolfgang}\ \bibnamefont {H\"ansel}},
  \bibinfo {author} {\bibfnamefont {Markus}\ \bibnamefont {Hennrich}}, \ and\
  \bibinfo {author} {\bibfnamefont {Rainer}\ \bibnamefont {Blatt}},\ }\bibfield
   {title} {\enquote {\bibinfo {title} {14-qubit entanglement: Creation and
  coherence},}\ }\href@noop {} {\bibfield  {journal} {\bibinfo  {journal}
  {Phys. Rev. Lett.}\ }\textbf {\bibinfo {volume} {106}},\ \bibinfo {pages}
  {130506} (\bibinfo {year} {2011})}\BibitemShut {NoStop}%
\bibitem [{\citenamefont {Gullion}\ \emph {et~al.}(1990)\citenamefont
  {Gullion}, \citenamefont {Baker},\ and\ \citenamefont
  {Conradi}}]{Gullion1990}%
  \BibitemOpen
  \bibfield  {author} {\bibinfo {author} {\bibfnamefont {T.}~\bibnamefont
  {Gullion}}, \bibinfo {author} {\bibfnamefont {D.~B.}\ \bibnamefont {Baker}},
  \ and\ \bibinfo {author} {\bibfnamefont {M.~S.}\ \bibnamefont {Conradi}},\
  }\bibfield  {title} {\enquote {\bibinfo {title} {New, compensated
  {C}arr-{P}urcell sequences},}\ }\href@noop {} {\bibfield  {journal} {\bibinfo
   {journal} {J. Mag. Res.}\ }\textbf {\bibinfo {volume} {89}},\ \bibinfo
  {pages} {479} (\bibinfo {year} {1990})}\BibitemShut {NoStop}%
\bibitem [{\citenamefont {Abrams}\ and\ \citenamefont
  {Lloyd}(1999)}]{Abrams1999}%
  \BibitemOpen
  \bibfield  {author} {\bibinfo {author} {\bibfnamefont {D.~S.}\ \bibnamefont
  {Abrams}}\ and\ \bibinfo {author} {\bibfnamefont {S.}~\bibnamefont {Lloyd}},\
  }\bibfield  {title} {\enquote {\bibinfo {title} {Quantum algorithm providing
  exponential speed increase for finding eigenvalues and eigenvectors},}\
  }\href@noop {} {\bibfield  {journal} {\bibinfo  {journal} {Phys. Rev. Lett.}\
  }\textbf {\bibinfo {volume} {83}},\ \bibinfo {pages} {5162} (\bibinfo {year}
  {1999})}\BibitemShut {NoStop}%
\bibitem [{\citenamefont {Bravyi}\ and\ \citenamefont
  {Kitaev}(2002)}]{Bravyi2002}%
  \BibitemOpen
  \bibfield  {author} {\bibinfo {author} {\bibfnamefont {S.~B.}\ \bibnamefont
  {Bravyi}}\ and\ \bibinfo {author} {\bibfnamefont {A.~Y.}\ \bibnamefont
  {Kitaev}},\ }\bibfield  {title} {\enquote {\bibinfo {title} {Fermionic
  quantum computation},}\ }\href@noop {} {\bibfield  {journal} {\bibinfo
  {journal} {Ann. Phys.}\ }\textbf {\bibinfo {volume} {298}},\ \bibinfo {pages}
  {210} (\bibinfo {year} {2002})}\BibitemShut {NoStop}%
\bibitem [{\citenamefont {Bravyi}\ \emph {et~al.}(2017)\citenamefont {Bravyi},
  \citenamefont {Gambetta}, \citenamefont {Mezzacapo},\ and\ \citenamefont
  {Temme}}]{Bravyi2017}%
  \BibitemOpen
  \bibfield  {author} {\bibinfo {author} {\bibfnamefont {S.}~\bibnamefont
  {Bravyi}}, \bibinfo {author} {\bibfnamefont {J.~M.}\ \bibnamefont
  {Gambetta}}, \bibinfo {author} {\bibfnamefont {A.}~\bibnamefont {Mezzacapo}},
  \ and\ \bibinfo {author} {\bibfnamefont {K.}~\bibnamefont {Temme}},\
  }\bibfield  {title} {\enquote {\bibinfo {title} {Tapering off qubits to
  simulate {F}ermionic {H}amiltonians},}\ }\href@noop {} {\bibfield  {journal}
  {\bibinfo  {journal} {arXiv:1701.08213}\ } (\bibinfo {year}
  {2017})}\BibitemShut {NoStop}%
\bibitem [{\citenamefont {Ko\l{}os}\ \emph {et~al.}(1986)\citenamefont
  {Ko\l{}os}, \citenamefont {Szalewicz},\ and\ \citenamefont
  {Monkhorst}}]{Kolos1986}%
  \BibitemOpen
  \bibfield  {author} {\bibinfo {author} {\bibfnamefont {W.}~\bibnamefont
  {Ko\l{}os}}, \bibinfo {author} {\bibfnamefont {K.}~\bibnamefont {Szalewicz}},
  \ and\ \bibinfo {author} {\bibfnamefont {H.~J.}\ \bibnamefont {Monkhorst}},\
  }\bibfield  {title} {\enquote {\bibinfo {title} {New {B}orn-{O}ppenhelmer
  potential energy curve and vibrational energies for the electronic ground
  state of the hydrogen molecule},}\ }\href@noop {} {\bibfield  {journal}
  {\bibinfo  {journal} {J. Chem. Phys.}\ }\textbf {\bibinfo {volume} {84}},\
  \bibinfo {pages} {3278} (\bibinfo {year} {1986})}\BibitemShut {NoStop}%
\bibitem [{\citenamefont {Preskill}(2018)}]{Preskill2018}%
  \BibitemOpen
  \bibfield  {author} {\bibinfo {author} {\bibfnamefont {J.}~\bibnamefont
  {Preskill}},\ }\bibfield  {title} {\enquote {\bibinfo {title} {Quantum
  {C}omputing in the {NISQ} era and beyond},}\ }\href {\doibase
  10.22331/q-2018-08-06-79} {\bibfield  {journal} {\bibinfo  {journal}
  {{Quantum}}\ }\textbf {\bibinfo {volume} {2}},\ \bibinfo {pages} {79}
  (\bibinfo {year} {2018})}\BibitemShut {NoStop}%
\bibitem [{\citenamefont {Peruzzo}\ \emph {et~al.}(2014)\citenamefont
  {Peruzzo}, \citenamefont {McClean}, \citenamefont {Shadbolt}, \citenamefont
  {Yung}, \citenamefont {Zhou}, \citenamefont {Love}, \citenamefont
  {Aspuru-Guzik},\ and\ \citenamefont {O'Brien}}]{Peruzzo2014}%
  \BibitemOpen
  \bibfield  {author} {\bibinfo {author} {\bibfnamefont {A.}~\bibnamefont
  {Peruzzo}}, \bibinfo {author} {\bibfnamefont {J.}~\bibnamefont {McClean}},
  \bibinfo {author} {\bibfnamefont {P.}~\bibnamefont {Shadbolt}}, \bibinfo
  {author} {\bibfnamefont {M.-H.}\ \bibnamefont {Yung}}, \bibinfo {author}
  {\bibfnamefont {X.-Q.}\ \bibnamefont {Zhou}}, \bibinfo {author}
  {\bibfnamefont {P.~J.}\ \bibnamefont {Love}}, \bibinfo {author}
  {\bibfnamefont {A.}~\bibnamefont {Aspuru-Guzik}}, \ and\ \bibinfo {author}
  {\bibfnamefont {J.~L.}\ \bibnamefont {O'Brien}},\ }\bibfield  {title}
  {\enquote {\bibinfo {title} {A variational eigenvalue solver on a photonic
  quantum processor},}\ }\href@noop {} {\bibfield  {journal} {\bibinfo
  {journal} {Nat. Commun.}\ }\textbf {\bibinfo {volume} {5}},\ \bibinfo {pages}
  {4213} (\bibinfo {year} {2014})}\BibitemShut {NoStop}%
\bibitem [{\citenamefont {Shor}(1994)}]{Shor1994}%
  \BibitemOpen
  \bibfield  {author} {\bibinfo {author} {\bibfnamefont {P.~W.}\ \bibnamefont
  {Shor}},\ }\bibfield  {title} {\enquote {\bibinfo {title} {Algorithms for
  quantum computation: Discrete logarithms and factoring},}\ }\href@noop {}
  {\bibfield  {journal} {\bibinfo  {journal} {in Proc. 35th Annual Symposium on
  Foundations of Computer Science, IEEE Computer Society Press}\ ,\ \bibinfo
  {pages} {124--134}} (\bibinfo {year} {1994})}\BibitemShut {NoStop}%
\bibitem [{\citenamefont {Harrow}\ \emph {et~al.}(2009)\citenamefont {Harrow},
  \citenamefont {Hassidim},\ and\ \citenamefont {Lloyd}}]{Harrow2009}%
  \BibitemOpen
  \bibfield  {author} {\bibinfo {author} {\bibfnamefont {A.~W.}\ \bibnamefont
  {Harrow}}, \bibinfo {author} {\bibfnamefont {A.}~\bibnamefont {Hassidim}}, \
  and\ \bibinfo {author} {\bibfnamefont {S.}~\bibnamefont {Lloyd}},\ }\bibfield
   {title} {\enquote {\bibinfo {title} {Quantum algorithm for linear systems of
  equations},}\ }\href@noop {} {\bibfield  {journal} {\bibinfo  {journal}
  {Phys. Rev. Lett.}\ }\textbf {\bibinfo {volume} {103}},\ \bibinfo {pages}
  {150502} (\bibinfo {year} {2009})}\BibitemShut {NoStop}%
\bibitem [{\citenamefont {O'Brien}\ \emph {et~al.}(2019)\citenamefont
  {O'Brien}, \citenamefont {Tarasinski},\ and\ \citenamefont
  {Terhal}}]{OBrien2019}%
  \BibitemOpen
  \bibfield  {author} {\bibinfo {author} {\bibfnamefont {T.~E.}\ \bibnamefont
  {O'Brien}}, \bibinfo {author} {\bibfnamefont {B.}~\bibnamefont {Tarasinski}},
  \ and\ \bibinfo {author} {\bibfnamefont {B.~M.}\ \bibnamefont {Terhal}},\
  }\bibfield  {title} {\enquote {\bibinfo {title} {Quantum phase estimation of
  multiple eigenvalues for small-scale (noisy) experiments},}\ }\href@noop {}
  {\bibfield  {journal} {\bibinfo  {journal} {New. J. Phys.}\ }\textbf
  {\bibinfo {volume} {21}},\ \bibinfo {pages} {023022} (\bibinfo {year}
  {2019})}\BibitemShut {NoStop}%
\bibitem [{\citenamefont {Endo}\ \emph {et~al.}(2021)\citenamefont {Endo},
  \citenamefont {Cai}, \citenamefont {Benjamin},\ and\ \citenamefont
  {Yuan}}]{Endo2021}%
  \BibitemOpen
  \bibfield  {author} {\bibinfo {author} {\bibfnamefont {S.}~\bibnamefont
  {Endo}}, \bibinfo {author} {\bibfnamefont {Z.}~\bibnamefont {Cai}}, \bibinfo
  {author} {\bibfnamefont {S.~C.}\ \bibnamefont {Benjamin}}, \ and\ \bibinfo
  {author} {\bibfnamefont {X.}~\bibnamefont {Yuan}},\ }\bibfield  {title}
  {\enquote {\bibinfo {title} {Hybrid quantum-classical algorithms and quantum
  error mitigation},}\ }\href {\doibase 10.7566/JPSJ.90.032001} {\bibfield
  {journal} {\bibinfo  {journal} {Jour. Phys. Soc. Japan}\ }\textbf {\bibinfo
  {volume} {90}},\ \bibinfo {pages} {032001} (\bibinfo {year}
  {2021})}\BibitemShut {NoStop}%
\bibitem [{\citenamefont {Robicheaux}\ \emph {et~al.}(2021)\citenamefont
  {Robicheaux}, \citenamefont {Graham},\ and\ \citenamefont
  {Saffman}}]{Robicheaux2021}%
  \BibitemOpen
  \bibfield  {author} {\bibinfo {author} {\bibfnamefont {F.}~\bibnamefont
  {Robicheaux}}, \bibinfo {author} {\bibfnamefont {T.}~\bibnamefont {Graham}},
  \ and\ \bibinfo {author} {\bibfnamefont {M.}~\bibnamefont {Saffman}},\
  }\bibfield  {title} {\enquote {\bibinfo {title} {Photon recoil and laser
  focusing limits to {R}ydberg gate fidelity},}\ }\href@noop {} {\bibfield
  {journal} {\bibinfo  {journal} {Phys. Rev. A}\ }\textbf {\bibinfo {volume}
  {103}},\ \bibinfo {pages} {022424} (\bibinfo {year} {2021})}\BibitemShut
  {NoStop}%
\bibitem [{\citenamefont {Bluvstein}\ \emph {et~al.}(2021)\citenamefont
  {Bluvstein}, \citenamefont {Levine}, \citenamefont {Semeghini}, \citenamefont
  {Wang}, \citenamefont {Ebadi}, \citenamefont {Kalinowski}, \citenamefont
  {Keesling}, \citenamefont {Maskara}, \citenamefont {Pichler}, \citenamefont
  {Greiner}, \citenamefont {Vuleti\'c},\ and\ \citenamefont
  {Lukin}}]{Bluvstein2021b}%
  \BibitemOpen
  \bibfield  {author} {\bibinfo {author} {\bibfnamefont {D.}~\bibnamefont
  {Bluvstein}}, \bibinfo {author} {\bibfnamefont {H.}~\bibnamefont {Levine}},
  \bibinfo {author} {\bibfnamefont {G.}~\bibnamefont {Semeghini}}, \bibinfo
  {author} {\bibfnamefont {T.~T.}\ \bibnamefont {Wang}}, \bibinfo {author}
  {\bibfnamefont {S.}~\bibnamefont {Ebadi}}, \bibinfo {author} {\bibfnamefont
  {M.}~\bibnamefont {Kalinowski}}, \bibinfo {author} {\bibfnamefont
  {A.}~\bibnamefont {Keesling}}, \bibinfo {author} {\bibfnamefont
  {N.}~\bibnamefont {Maskara}}, \bibinfo {author} {\bibfnamefont
  {H.}~\bibnamefont {Pichler}}, \bibinfo {author} {\bibfnamefont
  {M.}~\bibnamefont {Greiner}}, \bibinfo {author} {\bibfnamefont
  {V.}~\bibnamefont {Vuleti\'c}}, \ and\ \bibinfo {author} {\bibfnamefont
  {M.~D.}\ \bibnamefont {Lukin}},\ }\bibfield  {title} {\enquote {\bibinfo
  {title} {A quantum processor based on coherent transport of entangled atom
  arrays},}\ }\href@noop {} {\bibfield  {journal} {\bibinfo  {journal}
  {arXiv:2112.03923}\ } (\bibinfo {year} {2021})}\BibitemShut {NoStop}%
\bibitem [{\citenamefont {Hsiao}\ \emph {et~al.}(2018)\citenamefont {Hsiao},
  \citenamefont {Lin},\ and\ \citenamefont {Chen}}]{YFHsiao2018}%
  \BibitemOpen
  \bibfield  {author} {\bibinfo {author} {\bibfnamefont {Y.-F.}\ \bibnamefont
  {Hsiao}}, \bibinfo {author} {\bibfnamefont {Y.-J.}\ \bibnamefont {Lin}}, \
  and\ \bibinfo {author} {\bibfnamefont {Y.-C.}\ \bibnamefont {Chen}},\
  }\bibfield  {title} {\enquote {\bibinfo {title}
  {$\mathrm{\ensuremath{\Lambda}}$-enhanced gray-molasses cooling of cesium
  atoms on the ${D}_{2}$ line},}\ }\href@noop {} {\bibfield  {journal}
  {\bibinfo  {journal} {Phys. Rev. A}\ }\textbf {\bibinfo {volume} {98}},\
  \bibinfo {pages} {033419} (\bibinfo {year} {2018})}\BibitemShut {NoStop}%
\bibitem [{\citenamefont {Gillen-Christandl}\ \emph {et~al.}(2016)\citenamefont
  {Gillen-Christandl}, \citenamefont {Gillen}, \citenamefont {Piotrowicz},\
  and\ \citenamefont {Saffman}}]{Gillen-Christandl2016}%
  \BibitemOpen
  \bibfield  {author} {\bibinfo {author} {\bibfnamefont {K.}~\bibnamefont
  {Gillen-Christandl}}, \bibinfo {author} {\bibfnamefont {G.}~\bibnamefont
  {Gillen}}, \bibinfo {author} {\bibfnamefont {M.~J.}\ \bibnamefont
  {Piotrowicz}}, \ and\ \bibinfo {author} {\bibfnamefont {M.}~\bibnamefont
  {Saffman}},\ }\bibfield  {title} {\enquote {\bibinfo {title} {Comparison of
  {G}aussian and super {G}aussian laser beams for addressing atomic qubits},}\
  }\href@noop {} {\bibfield  {journal} {\bibinfo  {journal} {Appl. Phys. B}\
  }\textbf {\bibinfo {volume} {122}},\ \bibinfo {pages} {131} (\bibinfo {year}
  {2016})}\BibitemShut {NoStop}%
\bibitem [{\citenamefont {Kuhr}\ \emph {et~al.}(2005)\citenamefont {Kuhr},
  \citenamefont {Alt}, \citenamefont {Schrader}, \citenamefont {Dotsenko},
  \citenamefont {Miroshnychenko}, \citenamefont {Rauschenbeutel},\ and\
  \citenamefont {Meschede}}]{Kuhr2005}%
  \BibitemOpen
  \bibfield  {author} {\bibinfo {author} {\bibfnamefont {S.}~\bibnamefont
  {Kuhr}}, \bibinfo {author} {\bibfnamefont {W.}~\bibnamefont {Alt}}, \bibinfo
  {author} {\bibfnamefont {D.}~\bibnamefont {Schrader}}, \bibinfo {author}
  {\bibfnamefont {I.}~\bibnamefont {Dotsenko}}, \bibinfo {author}
  {\bibfnamefont {Y.}~\bibnamefont {Miroshnychenko}}, \bibinfo {author}
  {\bibfnamefont {A.}~\bibnamefont {Rauschenbeutel}}, \ and\ \bibinfo {author}
  {\bibfnamefont {D.}~\bibnamefont {Meschede}},\ }\bibfield  {title} {\enquote
  {\bibinfo {title} {Analysis of dephasing mechanisms in a standing-wave dipole
  trap},}\ }\href@noop {} {\bibfield  {journal} {\bibinfo  {journal} {Phys.
  Rev. A}\ }\textbf {\bibinfo {volume} {72}},\ \bibinfo {pages} {023406}
  (\bibinfo {year} {2005})}\BibitemShut {NoStop}%
\bibitem [{\citenamefont {Maller}\ \emph {et~al.}(2015)\citenamefont {Maller},
  \citenamefont {Lichtman}, \citenamefont {Xia}, \citenamefont {Sun},
  \citenamefont {Piotrowicz}, \citenamefont {Carr}, \citenamefont {Isenhower},\
  and\ \citenamefont {Saffman}}]{Maller2015}%
  \BibitemOpen
  \bibfield  {author} {\bibinfo {author} {\bibfnamefont {K.}~\bibnamefont
  {Maller}}, \bibinfo {author} {\bibfnamefont {M.~T.}\ \bibnamefont
  {Lichtman}}, \bibinfo {author} {\bibfnamefont {T.}~\bibnamefont {Xia}},
  \bibinfo {author} {\bibfnamefont {Y.}~\bibnamefont {Sun}}, \bibinfo {author}
  {\bibfnamefont {M.~J.}\ \bibnamefont {Piotrowicz}}, \bibinfo {author}
  {\bibfnamefont {A.~W.}\ \bibnamefont {Carr}}, \bibinfo {author}
  {\bibfnamefont {L.}~\bibnamefont {Isenhower}}, \ and\ \bibinfo {author}
  {\bibfnamefont {M.}~\bibnamefont {Saffman}},\ }\bibfield  {title} {\enquote
  {\bibinfo {title} {{R}ydberg-blockade controlled-{NOT} gate and entanglement
  in a two-dimensional array of neutral-atom qubits},}\ }\href@noop {}
  {\bibfield  {journal} {\bibinfo  {journal} {Phys. Rev. A}\ }\textbf {\bibinfo
  {volume} {92}},\ \bibinfo {pages} {022336} (\bibinfo {year}
  {2015})}\BibitemShut {NoStop}%
\bibitem [{\citenamefont {Levine}\ \emph {et~al.}(2019)\citenamefont {Levine},
  \citenamefont {Keesling}, \citenamefont {Semeghini}, \citenamefont {Omran},
  \citenamefont {Wang}, \citenamefont {Ebadi}, \citenamefont {Bernien},
  \citenamefont {Greiner}, \citenamefont {Vuleti\'c}, \citenamefont {Pichler},\
  and\ \citenamefont {Lukin}}]{Levine2019}%
  \BibitemOpen
  \bibfield  {author} {\bibinfo {author} {\bibfnamefont {H.}~\bibnamefont
  {Levine}}, \bibinfo {author} {\bibfnamefont {A.}~\bibnamefont {Keesling}},
  \bibinfo {author} {\bibfnamefont {G.}~\bibnamefont {Semeghini}}, \bibinfo
  {author} {\bibfnamefont {A.}~\bibnamefont {Omran}}, \bibinfo {author}
  {\bibfnamefont {T.~T.}\ \bibnamefont {Wang}}, \bibinfo {author}
  {\bibfnamefont {S.}~\bibnamefont {Ebadi}}, \bibinfo {author} {\bibfnamefont
  {H.}~\bibnamefont {Bernien}}, \bibinfo {author} {\bibfnamefont
  {M.}~\bibnamefont {Greiner}}, \bibinfo {author} {\bibfnamefont
  {V.}~\bibnamefont {Vuleti\'c}}, \bibinfo {author} {\bibfnamefont
  {H.}~\bibnamefont {Pichler}}, \ and\ \bibinfo {author} {\bibfnamefont
  {M.~D.}\ \bibnamefont {Lukin}},\ }\bibfield  {title} {\enquote {\bibinfo
  {title} {Parallel implementation of high-fidelity multiqubit gates with
  neutral atoms},}\ }\href@noop {} {\bibfield  {journal} {\bibinfo  {journal}
  {Phys. Rev. Lett.}\ }\textbf {\bibinfo {volume} {123}},\ \bibinfo {pages}
  {170503} (\bibinfo {year} {2019})}\BibitemShut {NoStop}%
\bibitem [{\citenamefont {Saffman}\ \emph {et~al.}(2020)\citenamefont
  {Saffman}, \citenamefont {Beterov}, \citenamefont {Dalal}, \citenamefont
  {Paez},\ and\ \citenamefont {Sanders}}]{Saffman2020}%
  \BibitemOpen
  \bibfield  {author} {\bibinfo {author} {\bibfnamefont {M.}~\bibnamefont
  {Saffman}}, \bibinfo {author} {\bibfnamefont {I.~I.}\ \bibnamefont
  {Beterov}}, \bibinfo {author} {\bibfnamefont {A.}~\bibnamefont {Dalal}},
  \bibinfo {author} {\bibfnamefont {E.~J.}\ \bibnamefont {Paez}}, \ and\
  \bibinfo {author} {\bibfnamefont {B.~C.}\ \bibnamefont {Sanders}},\
  }\bibfield  {title} {\enquote {\bibinfo {title} {Symmetric {R}ydberg
  controlled$-{Z}$ gates with adiabatic pulses},}\ }\href@noop {} {\bibfield
  {journal} {\bibinfo  {journal} {Phys. Rev. A}\ }\textbf {\bibinfo {volume}
  {101}},\ \bibinfo {pages} {062309} (\bibinfo {year} {2020})}\BibitemShut
  {NoStop}%
\bibitem [{\citenamefont {Zhang}\ \emph {et~al.}(2011)\citenamefont {Zhang},
  \citenamefont {Robicheaux},\ and\ \citenamefont {Saffman}}]{SZhang2011}%
  \BibitemOpen
  \bibfield  {author} {\bibinfo {author} {\bibfnamefont {S.}~\bibnamefont
  {Zhang}}, \bibinfo {author} {\bibfnamefont {F.}~\bibnamefont {Robicheaux}}, \
  and\ \bibinfo {author} {\bibfnamefont {M.}~\bibnamefont {Saffman}},\
  }\bibfield  {title} {\enquote {\bibinfo {title} {Magic-wavelength optical
  traps for {R}ydberg atoms},}\ }\href@noop {} {\bibfield  {journal} {\bibinfo
  {journal} {Phys. Rev. A}\ }\textbf {\bibinfo {volume} {84}},\ \bibinfo
  {pages} {043408} (\bibinfo {year} {2011})}\BibitemShut {NoStop}%
\end{thebibliography}%

\section*{Methods}
{\bf Experimental apparatus} 
The apparatus was similar to that used in\cite{Graham2019} with new capabilities as described in the following. 
Our platform is designed around a 2D blue-detuned optical trap array consisting of an array of crossed  lines shown in Fig. \ref{fig.Full_Assembly} at a wavelength of 825 nm. We shaped the  lines using a top-hat hologram fabricated by Holo/Or and a cylindrical lens telescope to adjust the line aspect ratio. Individual lines were split using an acousto-optic deflector (AOD) driven with multiple frequency tones. Using an acousto-optic deflector allowed the trap number and line spacing to be reconfigured dynamically.  Previous line array implementations suffered from atom detection noise arising from atomic traps being formed in Talbot planes of the array\cite{Graham2019}. In this implementation, each line has a different frequency shift from the AOD, thereby destroying the interference effects responsible for such out of plane trapping. This resulted in reduced noise in trap occupancy measurements. We combined eight horizontal lines with eight vertical lines using a polarizing beam splitter and imaged them onto the atom trapping region of a glass vacuum chamber forming a $7 \times 7$ grid of atom traps with a spacing of $3 ~\mu \rm m$.  Trap depths were set to about $0.3~\rm mK$ during circuit operation and increased to $0.7~\rm mK$ during readout. 

Optical access through the cell edges and the front face allowed 3D cooling of Cs atoms with 852 nm light with red-detuned polarization gradient cooling followed by  Raman lambda-grey molasses cooling with 895 nm light to reach temperatures below $5~\mu\rm K$\cite{YFHsiao2018}. We used a 1064 nm optical tweezer beam copropagating with the blue-detuned array for atomic rearrangement. We controlled the position of this tweezer beam using crossed, upstream acousto-optic deflectors to move atoms into a desired pattern using the Hungarian algorithm to determine the atomic move order. After rearrangement, we used atoms in sites separated by 3 array periods, or  $9~ \mu \rm m$, for circuit operations. Using a larger spacing reduced crosstalk from scattering in the optical system.  A $\pi$-polarized, 895 nm  optical pumping beam incident from the side of the cell pumped trapped atoms to the $\ket{1}=\ket{f=4,m=0}$ state. After optical pumping, the atomic temperature was typically $5~\mu\rm K$. A bias magnetic field of 1.6 mT was used during optical pumping and circuit operation. 

Atom occupancy was determined by imaging resonance fluorescence from  852 nm molasses beams detuned by -12$\gamma$ ($\gamma/2\pi=5.2~\rm MHz$ is the linewidth of the $6p_{3/2}$ state) onto the EMCCD camera. For quantum state measurements, atoms in $f=4$ were pushed out of the traps with a resonant beam, followed by an occupancy measurement. A dark(bright) signal indicated a  quantum state of $\ket{1} (\ket{0})$.

{\bf Qubit coherence} Trapped atom lifetimes limited by residual vacuum pressure were observed to be $\sim10~\rm s$ at the $1/e$ population decay point. The qubit $T_1$ time was $\sim 4 ~\rm s$ with approximately  equal lifetimes seen for the $\ket{0}\rightarrow \ket{1}$ and $\ket{1}\rightarrow \ket{0}$ transitions. The $T_2^*$ time as observed by Ramsey interference was typically 3.5 ms, although $T_2^* = 8 ~\rm ms$ was observed under conditions of optimized cooling. Using a sequence of ${\sf X}, -{\sf X}, {\sf X}, -{\sf X}, ...$ dynamical decoupling pulses, a homogeneous coherence time of $T_2\sim 50~\rm ms $ was observed. Using a XY8 pulse sequence\cite{Gullion1990} $T_2\sim 1~\rm s$ was observed. For the circuit data reported here, no dynamical decoupling was applied, so the limiting coherence time was $T_2^*$. 

{\bf Quantum gate set}
After qubit initialization, quantum circuits were run following compilation into the hardware native gate set. The physical pulse sequences corresponding to the GHZ, phase estimation, and QAOA circuits are given in \cite{SM2021circuits}.
 A universal set of quantum gates was provided by resonant microwaves and two narrow-band laser sources at 459 nm and 1040 nm. A pulsed 40 W microwave source resonant with the $\ket{0}=\ket{f=3,m=0} \rightarrow \ket{1}=\ket{f=4,m=0}$ transition provided a global ${\sf R}_{\phi}(\theta)$ rotations, where $\phi$ denotes a rotation axis in the $x-y$ plane of the Bloch sphere and $\theta$ denotes the rotation angle. The microwave driven Rabi frequency was 76.5 kHz. We applied single-site ${\sf R_Z}\left( \theta \right)$ rotations using a focused 459 nm laser  which was $+0.76$ GHz detuned from the $6s_{1/2},f=4 \rightarrow 7p_{1/2}$(center of mass)  transition. This laser provided a differential Stark shift of $+600$ kHz between $\ket{0}$ and $\ket{1}$. We applied the ${\sf R_Z}(\theta)$ gate  by pulsing the 459 nm laser for a time corresponding to the desired rotation angle.  With the combination of these two gates, we could apply  site-selective, single-qubit ${\sf R}_\phi(\theta)$ rotations.  
 
 To complete a universal gate set, we used simultaneous two-atom Rydberg excitation  to  implement a $\sf C_Z$ gate\cite{Levine2019}. For Rydberg excitation, we used two lasers at 459 nm and 1040 nm to induce a two-photon excitation to the 
 $75s_{1/2}$ Rydberg state. The 1040 nm (459 nm) Rydberg beams co-propagated (counter-propagated) with the 825 nm trap light.   We controlled the Rydberg beam pointing using crossed AODs. Beams were focused to $3~ \mu\rm m$ waists, which allowed single qubit addressing.  The 1040 and 459 nm beams were pointed at two sites simultaneously (horizontally or vertically displaced) by driving one of the scanner AODs for each color with two frequencies.  Since beams diffracted by an AOD receive frequency shifts equivalent to the AOD drive frequency, two-photon resonance was maintained at both sites by using diffraction orders of opposite sign for the 459 and 1040 nm scanners.  This required adjusting the magnification of the optical train after the  459 and 1040 nm  scanners such that the displacement in the image region versus AOD drive frequency was the same for both Rydberg beams. 
 
 The $\sf C_Z$ gate protocol used was the detuned two-pulse sequence introduced 
 in\cite{Levine2019}, with parameters modified for a relatively weak Rydberg interaction outside the blockade limit.  For the $\sf C_Z$ gate pulses, the one-atom Rydberg Rabi frequency was 1.7 MHz, and the Rydberg blockade shift was 3 MHz. Prior to the Rydberg pulses, we used two additional 459 pulses, one targeting each atom, to provide local ${\sf R_Z}(\theta)$ rotations needed to achieve  a canonical $\sf C_Z$ gate.  The gate fidelity was characterized by creating an entangled Bell state with $92.7(13) \%$ raw fidelity ($\sim 95.5 \%$ SPAM corrected).  See \cite{SM2021circuits} for a summary of all gate, state preparation, and measurement fidelities.
 {\sf CNOT} gates were implemented with the standard decomposition $\sf CNOT= (I\otimes H)C_Z(I\otimes H).$

{\bf Acknowledgements}  This work was supported by  
DARPA-ONISQ Contract No. HR001120C0068, 
NSF award PHY-1720220, 
NSF Award 2016136 for the QLCI center Hybrid Quantum
Architectures and Networks, 
U.S. Department of Energy under Award No. DE-SC0019465, and 
via Innovate UK's Sustainable Innovation Fund Small Business Research Initiative (SBRI).

{\bf Author contributions}  TMG, YS, JS, CP, KJ, PE, XJ, AM, BG, MK, ME, JC, MTL, MG, JG, DB, TB, TN, MS contributed to the design and construction of the apparatus, including the classical control system. TMG, YS, JS, KJ, LP, PE contributed to operating the apparatus, taking data, and analysing it. TMG, YS, CC, EDD, CP, LP, OC, NB, BR, MS contributed to circuit design, optimization, and simulation. The manuscript was written by TMG, YS, JS, OC, NB, BR, MS. The project was supervised by TN and MS. All authors discussed the results and contributed to the manuscript. 

{\bf Competing interests}
ME, JC, MTL, MG, JG, DB, TB, CC, EDD, TN, MS are  shareholders in ColdQuanta, Inc.  OC, NB, BR are employees of Riverlane, Ltd. 

\clearpage


\setcounter{page}{1}
\renewcommand\thepage{SM-\arabic{page}}

\setcounter{figure}{0}
\renewcommand\thefigure{SM-\arabic{figure}}

\setcounter{table}{0}
\renewcommand\thetable{SM-\Roman{table}}

\setcounter{section}{0}
\renewcommand\thesection{SM-\Roman{section}}

\setcounter{equation}{1}
\renewcommand\theequation{SM-\arabic{equation}}

\widetext

\section*{Supplementary material for: Demonstration of multi-qubit entanglement and  algorithms on a programmable neutral atom quantum computer}


\section{Experimental platform}

The experimental approach is based on a further development of that used in \cite{Graham2019}  where additional analysis can be found in the accompanying supplemental material. The overall layout is shown in Fig. 1 in the main text, and a basic description is provided in the Methods section. We provide additional  details below.

\subsection{Vacuum cell and imaging}

The vacuum system consists of a 
2D-MOT source region where a pre-cooled atomic sample is prepared. Cs atoms are then pushed through a differential pumping aperture into the science cell, which has a rectangular shape. The large facing windows through which trapping and control beams enter are separated by 1 cm so the atomic qubits are 5 mm from the nearest surface. 
The facing windows each have four electrodes which are controlled by low noise dc voltage supplies in order to cancel background fields. Cancellation was performed automatically by scanning electrode voltages to minimize the quadratic Stark shift of the $75s_{1/2}$
Rydberg state. Without compensation, background fields at the level of a few $\rm V/m$ were observed. 
Despite operating at a zero field condition, intermittent jumps of the Rydberg energy level were seen. The exact mechanism causing this is not known, but it is attributed to changes in adsorption of alkali atoms on the cell walls. To reduce the frequency of such jumps, we continuously illuminated the cell with 
410 nm light from a light emitting diode.

In order to increase photon collection efficiency for atom occupancy and quantum state measurements, dual sided imaging was used. A high numerical aperture NA$=0.7$ objective lens 
was mounted on each side of the cell. Using dichroic mirrors, dual images of the atom array were routed to adjacent regions of the EMCCD camera. With two NA$=0.7$ objectives, the theoretical light collection efficiency is $0.286$.
Accounting for various passive losses in optical components and camera quantum efficiency, the detection probability per photon scattered by an atom is about 0.08. Atom imaging and state measurements used 4 of the 6 MOT beams. Two of the beams share the same axis as the imaging camera and were turned off during measurement. Typical state readout parameters were 90 ms integration time at detuning $-12\gamma_{6p_{3/2}}$ with 4 beams in a plane each with $220~\mu\rm W$ power and  $2~\rm mm$ waist ($1/e^2$ intensity radius).

\subsection{Trap array}

 \begin{figure}[!t]
     \centering
     \includegraphics[width=.7\textwidth]{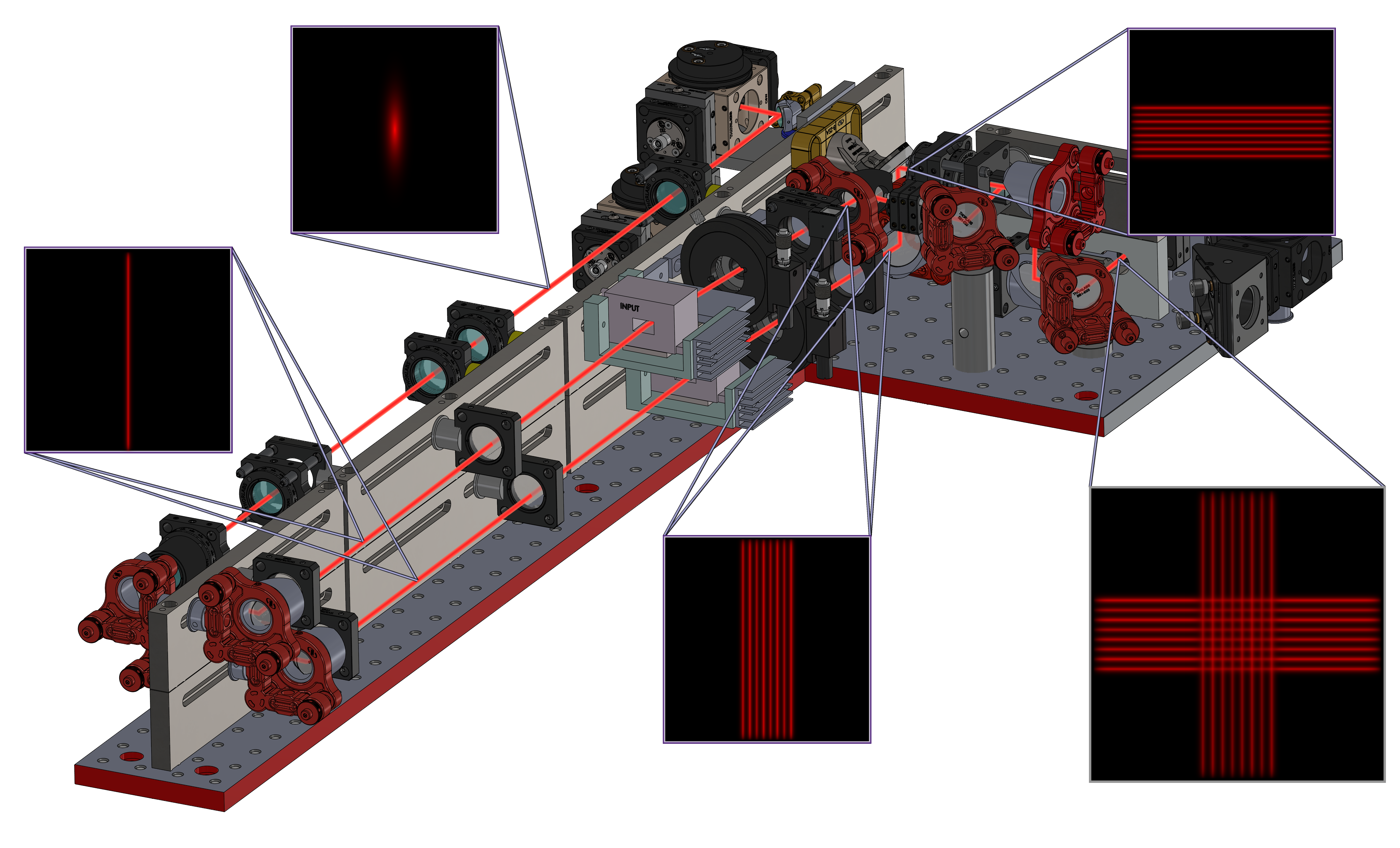}
     \caption{\label{fig.SM_TrapArrayModule} Layout of optical module for generation of trapping light. The light enters through a single mode fiber that is collimated and then converted into a uniform intensity line with a diffractive optical element. The line is then divided into $N$ lines by driving an AOD with $N$ radio-frequency tones. This procedure is performed in two paths to create arrays of horizontal and vertical lines. The arrays are then combined on a polarizing beamsplitter to create the grid of trap light which is imaged onto the atoms.      }
 \end{figure}

To trap the atoms in the array, we use a Far Off-Resonance Optical Trap (FORT) with a blue detuning. 
The optical module for generating the trap light is shown in Fig. \ref{fig.SM_TrapArrayModule}. 
The FORT is an array of vertical and horizontal, highly elliptical tophat beams (i.e. lines).  
To create the line array, a Diffractive Optical Element (DOE) first converts an elliptical beam into a single vertical tophat line. After some beam shaping, the tophat is focused through the AOD, which creates an array of tophat lines. Only the first order diffracted beam is used, and the array is created using a superposition of pure sinusoidal tones. The horizontal and vertical lines are made in separate beam paths, with the horizontal lines initially vertical lines, but rotated with a periscope, also giving an opposite polarization. These lines are combined on a polarizing beam splitter (PBS), further shaped, then travel through the objective lens to trap the atoms.
For the $z$-axis confinement, the atoms are confined by the divergence of the trap beams, since the line array only exists at the image plane of the objective lens. This approach does not suffer from Talbot planes because each line is a different frequency, meaning no interference effects arise. Thus, there are only  atoms trapped at the image plane. Furthermore, the horizontal and vertical lines are opposite polarization, making it so that these lines do not interfere either.
In addition, each set of horizontal and vertical lines was powered by a separate single frequency Ti:Sa laser. The lasers were tuned to have approximately the same 825 nm wavelength, but frequencies differing by a few hundreds of GHz to prevent any undesired interference effects with regards to atom cooling or trapping.

The tones of the RF signal sent to the AODs are all near the resonant frequency of the AOD (50 MHz) and are separated in frequency space by 1.2 MHz. These tones are generated using a software defined ratio (SDR), allowing for tuning of the relative phase and amplitudes of the individual tones. 
Using this approach we have created arrays with up to $24\times 24$ lines and trapped atoms in arrays with $15\times 15$ lines and 196 sites. 
For this work we used a $7\times 7$ site array with 8 lines per axis. The rf signal generating the lines goes through a series of non-linear devices, notably the amplifier and the AOD. Thus, tuning phase and amplitude is important, since there are non-linear compression effects if the instantaneous power approaches the compression point of the amplifier or linearity threshold of the AOD.
To tune the amplitude of the tones, a camera is set up at a pickoff of the trap array image plane (i.e., same image plane as the atoms). The goal is to have equal power in each line, and the tone power is adjusted until the array is balanced. To tune the phase, the power spectrum of the trap array is measured. The beat frequencies of the array lines with each other should only be seen at 1.2 MHz (and integer multiples thereof), but if the phases are badly tuned, there may be times where the instantaneous amplitude is too high such that the non-linear effects cause beating at other, lower frequencies (order 100 kHz). The phase of each tone is randomly assigned until an acceptable low-frequency beat spectrum is found.
At the atom plane (the image plane), the lines are separated by $3~\mu\rm m$, the beam waist (line width) is $1~\mu\rm m$, and the tophat length is $\sim40~\mu\rm m$. The measured trap vibration frequencies  were 19 kHz and 4 kHz in the radial and transverse dimensions, respectively, at a trap depth of 0.3 mK. 

\subsection{Trap array analysis }
\label{sec.trapanalysis}

The trap parameters can be described analytically in compact form assuming ideal lines with uniform intensity along the line, and Gaussian transverse profiles. Adding the contributions from the nearest lines in each unit cell, we find the following expressions for a square array with line spacing $d$, line waist ($1/e^2$ intensity radius) $w$, and aspect ratio $s=d/w$
\begin{eqnarray}
I_{\rm c}&=&I_{\rm d}\frac{4s}{\sqrt{2\pi}}  e^{-s^2/2},\nonumber\\
I_{\rm s}&=&I_{\rm d}\frac{s}{\sqrt{2\pi}} \left( 1+2  e^{-s^2/2}\right),\nonumber\\
I_{\rm t}&=&I_{\rm s}-I_{\rm c}=I_{\rm d}\frac{s}{\sqrt{2\pi}}\left( 1-2  e^{-s^2/2}\right),\nonumber
\end{eqnarray}
with $I_{\rm d}=I_{0} \sqrt{2\pi}/s=P/d^2$ 
where $I_0$ is the peak intensity of each line and $P$ is the optical power per unit cell. In these expressions, $I_{\rm c}$ is the intensity at the center of each cell, $I_{\rm s}$ is the intensity at the saddle point midway between corners of the cell on each line, and $I_{\rm t}$ is the effective trapping intensity. 

\begin{figure}[!t]
     \centering
          \includegraphics[width=.9\textwidth]{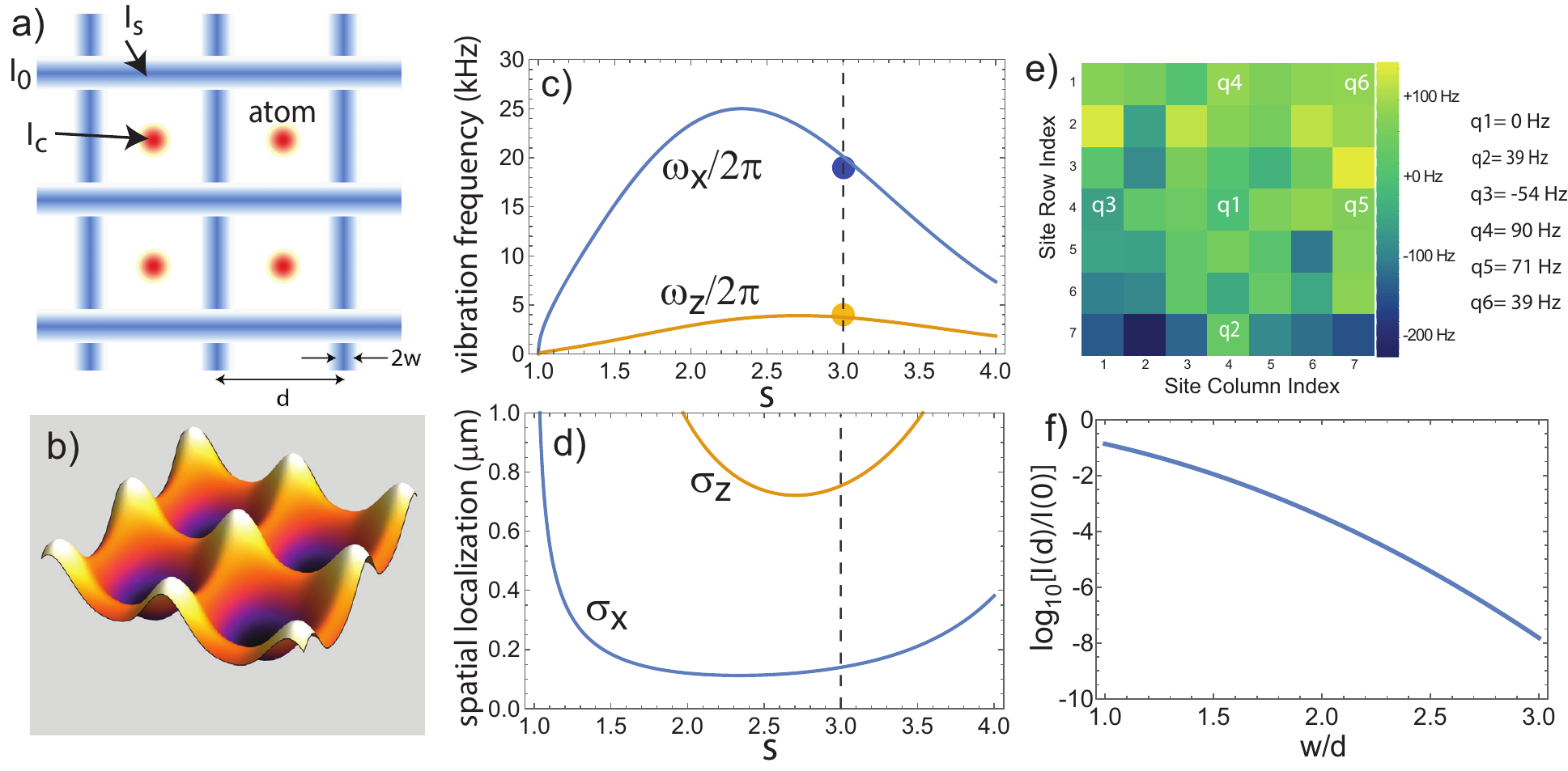}
     \caption{\label{fig.trapplots} Optical trap array.  a) Geometry of line array,  b) Intensity landscape of traps for $s=3$.  c) Vibrational frequencies.  The circles  show the measured trap vibration frequencies of 19 kHz radial and 4 kHz axial. Plots used parameters $d=3~\mu\rm m$, $s=3$, $U_{\rm d}=k_{\rm B}\times 300~\mu\rm K$, $T_{\rm a}=5~\mu\rm K$, $\lambda=825~\rm nm$.  d) Plots of spatial localization. e) Map of measured variation in qubit frequency across the trap array, together with the measured shifts for the sites that were used relative to site q1.  f) Ideal Gaussian beam intensity profile on a log scale.   }
\end{figure}

The above expressions describe the intensity in the plane of the array. The trapping intensity at distance $z$ perpendicular to the plane of the array is 
$ I_{{\rm t},z}(z)=I_{\rm c}(z)-I_{\rm c}(0)$. This takes on a maximum value at a distance $z_{\rm max}=L_R\sqrt{s^2-1}$ with $L_R=\pi w^2/\lambda$ where $\lambda$ is the wavelength of the trapping light.   The effective axial trapping intensity is 
$$ 
I_{{\rm t},z}(z_{\rm max})=I_{\rm d}\frac{4}{\sqrt{2\pi}}\left( \frac{1}{\sqrt e}-s e^{-s^2/2}\right).
$$
Note that as the aspect ratio $s$ increases, the transverse trap depth proportional to $I_{\rm t}$ increases without bound, but the axial trap depth $I_{{\rm t},z}(z_{\rm max})$ saturates at a maximum value of  $I_{{\rm t},z}(z_{\rm max})=I_{\rm d} 4/\sqrt{2\pi e}= 0.97 I_{\rm d}.$ At our operating point of $s=3$ we have, $I_{\rm t}/I_{\rm d}=1.17$ and  $I_{{\rm t},z}(z_{\rm max})/I_{\rm d}=0.91$, so the confinement barrier is about 20\% lower perpendicular to the plane of the array. 

Using these expressions, we can calculate the spring constants $\kappa$  and trap vibration frequencies $\omega=\sqrt{\kappa/m_{\rm a}}$ for an atom of mass $m_{\rm a}$. We find
\begin{eqnarray}
\kappa_x=\kappa_y&=&\frac{U_{\rm d}}{d^2}\frac{\sqrt{32}s^3}{\pi^{1/2}}\left(s^2-1\right)e^{-s^2/2},\\
\kappa_z&=&\frac{U_{\rm d}\lambda^2}{d^4}\frac{\sqrt{8}s^5}{\pi^{5/2}}\left(s^2-1\right)e^{-s^2/2}
\end{eqnarray}
where $U_{\rm d}=-\frac{\alpha I_{\rm d}}{2\epsilon_0 c}$ with $\alpha<0$ is the atomic polarizability at wavelength $\lambda$.  The factor $U_{\rm d}$ is the spatially averaged light shift across the array. For an atom at temperature $T_{\rm a}$, the corresponding spatial localization can be expressed in terms of variances given by 
\begin{eqnarray}
\sigma_x^2=\sigma_y^2&=&\frac{k_B T_{\rm a}d^2}{U_{\rm d}}  \frac{\pi^{1/2}  }{\sqrt{32}}\frac{e^{s^2/2}}{s^3\left(s^2-1\right)},\\
\sigma_z^2&=&\frac{k_B T_{\rm a}d^4}{U_{\rm d}\lambda^2}  \frac{\pi^{5/2} }{2^{3/2}}\frac{e^{s^2/2}}{s^5\left(s^2-1\right)}.
\end{eqnarray}

Accurate balancing of the line intensities, and thereby the residual intensity at the center of each trap,  is verified by measuring light induced shifts of the qubit frequencies with microwave spectroscopy. Typical data is shown in Fig. \ref{fig.trapplots} e). Overall, the resonance splitting across the array was on the order of $\pm 100~\rm Hz$. For the six sites used in the GHZ experiment, the total range of qubit frequencies was 
144 Hz. The circuit execution time for the $N=6$ GHZ state was $180~\mu\rm s$ corresponding to a maximum undesired phase differential of 9.3 deg.  The longest circuit we implemented was 4 qubit phase estimation for which  the execution time was  1.1 ms corresponding to 57 deg. of uncompensated  phase. For longer circuit execution times, these shifts can be canceled by periodic application of ${\sf R_Z}(\theta)$ gates. 

\subsection{Qubit addressing and crosstalk}

Operations on single sites are achieved with laser beams focused to waist $w$. In the ideal case of a perfect unaberrated Gaussian TEM$_{00}$ beam, this corresponds to an intensity profile in the focal plane of 
$I(r)=I_0 e^{-2d^2/w^2}$ with $w$ the beam waist and $d$ the distance from the beam center.  Figure \ref{fig.trapplots}f) shows the intensity spillover on a neighboring site as a function of the ratio $d/w$, where $d$ is the qubit spacing. In practice, a higher level of intensity spillover is seen. This is due to optical aberrations and unavoidable scattering from a large number of surfaces in the optical train.

For the experiments reported here, we have operated with $d=9~\mu\rm m$ and 
$w=3~\mu\rm m$. The ideal intensity crosstalk level is then $I(d)/I(0)= e^{-18}=1.5\times 10^{-8}.$
The crosstalk of the 459 nm beam was measured by aligning the beam to a site and measuring the qubit rotation rate (from a Ramsey interference experiment) at a site $9~\mu\rm m$ distant, compared to the rotation rate at the targeted site. The ratio of these rates gives the intensity crosstalk. At the $9~\mu\rm m$  spacing, the typical observed crosstalk value was $\ll 0.01$.   

The crosstalk of the 1040 nm beam was measured by adding 9.2 GHz sidebands to the beam with an electro-optic modulator so it could directly drive $\sf R_X(\theta)$ rotations. We then measured the intensity crosstalk in the same way as for the 459 nm beam. A typical observed crosstalk value was  also $\ll 0.01$.

There are tradeoffs between optical crosstalk, qubit spacing, and beam waist. Although crosstalk can be reduced by operating with a smaller beam waist, doing so increases sensitivity to optical alignment and atom motion. The sensitivity to beam profile can be mitigated using shaped beams with a flat top\cite{Gillen-Christandl2016}, as has been effectively demonstrated in experiments that used global optical addressing beams\cite{Ebadi2021}.

\subsection{Two-qubit simultaneous addressing}
\label{SM.collective_addressing}
Consider scanning an optical beam with an acousto-optic deflector (AOD) to address an atomic transition in atoms at different spatial locations. Since the optical frequency varies with the scan angle, resonance cannot simultaneously be achieved at multiple locations with a single laser frequency. This limitation can be overcome by using a two-photon transition with the frequency shifts of the photons arranged to cancel each other. 

To be explicit, assume we are driving a resonance using beams of wavelength $\lambda_1, \lambda_2$.
The beams are deflected to positions $x_j$ using a configuration of AOD - distance $f$ - lens focal length $f$ - distance $f$, followed by an imaging magnification $M_j$ for each beam. The beam position in the output plane using diffraction order $m_j$ and diffraction angle $\theta_d$ is 
$$
x_j\simeq f \theta_d\simeq m_j\frac{\lambda_j f f_{aj}}{n_j v_a}M_j, ~~~j=1,2.
$$
The acoustic velocity is $v_a$, the applied frequency is $f_{aj}$, $n_j$ is the index of refraction of the modulator, and
$\lambda_j$ is the vacuum wavelength of beam $j$. 
We can invert this to write 
$$
f_{aj}\simeq \frac{n_jv_a }{m_j \lambda_j M_j f}x_j.
$$
Putting $x_1=x_2=x$ and imposing the resonance condition  $f_{a1}+f_{a2}=0$, we get 
$$
\frac{n_1 }{m_1 M_1 \lambda_1 }+\frac{n_2 }{m_2 M_2\lambda_2 }=0.$$
Choosing $m_1=1, m_2=-1$, we can satisfy this relation using 
\begin{equation} 
\frac{M_2}{M_1}
=\frac{\lambda_1 n_2}{\lambda_2 n_1}.
\label{eq.aomfmatch}
\end{equation}
We may also want the sizes of the scanned beams to be identical. If each beam has waist $w_j$ at the AOD, the waist at the output plane is 
$$
w_j^{\rm out}=M_j \frac{\lambda_j f}{\pi w_j}.
$$
Setting $w_1^{\rm out}=
w_2^{\rm out}$ gives the condition 
\begin{equation} 
\frac{w_2}{w_1}=\frac{\lambda_2 M_2}{\lambda_1 M_1}.
\label{eq.aomwmatch}
\end{equation}

Combining Eqs. (\ref{eq.aomfmatch},\ref{eq.aomwmatch}) gives
$$
\frac{w_2}{w_1}=\frac{n_2}{n_1}.
$$

As an example using $\lambda_1=1.038, \lambda_2=0.459$ and $n_1=n_2$, we find
$$
\frac{M_2}{M_1}=0.442
$$
and 
$$
w_2=w_1.$$

We have implemented this approach to enable simultaneous addressing of pairs of sites that are in the same row or same column of the qubit array. Fine adjustment of the ratio $M_2/M_1$ was achieved with a zoom lens mounted in the 459 nm optical train. 
Sites that are in a different row and a different column (diagonally opposite corners of a rectangle) can be addressed, but undesired beams will also appear at the other corners of the rectangle. With this beam steering system, $\sf C_Z$ gates are therefore constrained to qubits in the same row or column. This constraint can be relaxed by implementing more advanced beam steering devices such as spatial light modulators. 

\begin{figure}[!t]
     \centering
  \includegraphics[width=.95\textwidth]{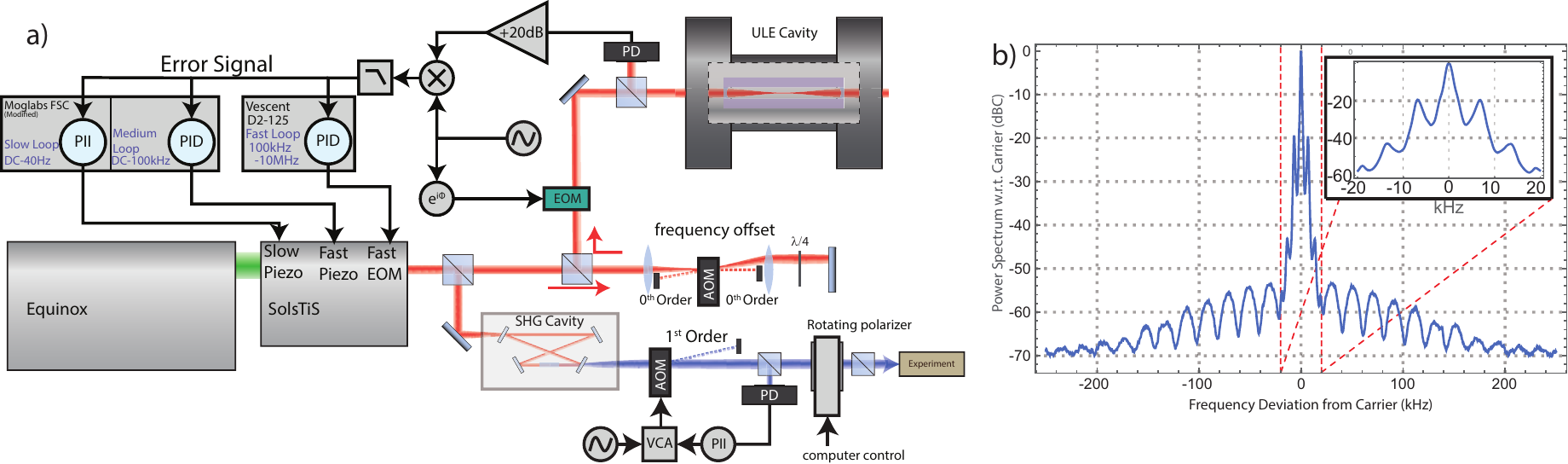}
     \caption{\label{fig.Rydberglasers}Rydberg laser stabilization. a) Schematic of locking setup. b) Measured self-heterodyne spectrum of 918 nm laser with a 10 km fiber delay line.    }
\end{figure}

An unexpected issue was encountered when implementing this dual-site addressing scheme. While the sum of the two laser frequencies at each addressed site is constant the individual frequencies of the 459 and 1040 nm beams differ from site to site. Thus a small amount of intensity spillover from the edge of the Gaussian beam or diffuse scattering from optical surfaces leads to time dependent modulation of the intensity of each color, since the two frequency components of each color are coherent with each other. This time dependent modulation can lead to large qubit control errors even for intensity crosstalk at  the 1\% level. For this reason we operated at a qubit spacing of $9~\mu\rm m$ which was sufficient to reduce the crosstalk to $\ll 0.01$. A modified beam scanning system that is being implemented will provide simultaneous addressing with exactly the same frequencies at both sites and remove this issue, thereby enabling operation at smaller spacings.

\subsection{Rydberg lasers}

Rydberg states are excited with a two-photon transition. A first photon at 459 nm couples the $6s_{1/2}$ ground state to $7p_{1/2}$. A second photon near 1040 nm couples $7p_{1/2}$ to the Rydberg $75s_{1/2}$ state. 
The 459 nm light is prepared by frequency doubling an M-Squared SolsTis Ti:Sa system pumped by an M-Squared Equinox pump laser\footnote{We mention commercial vendor names for technical reference and are not endorsing any commercial products.}.  The 918 nm light is frequency stabilized and locked to a high finesse ultra low expansion (ULE) glass reference cavity in a temperature stabilized vacuum can using a Pound Drever Hall (PDH) locking scheme. The reference cavity has a free spectral range of 1.5 GHz and a linewidth of about 10 kHz.   A set of AOMs are used to fine-tune the frequency of the laser light relative to the fixed frequency of the ULE cavity mode. Frequency doubling occurs in a home built resonant ring doubler with an LBO crystal. The singly resonant doubling cavity is stabilized to the 918 nm light with a H\"ansch-Couillaud lock. The intensity of the light is then stabilized using an AOM based noise eater that operates in the dc-100 kHz range and a slow stabilization loop based on a rotatable waveplate and a polarizer. The light is then coupled into a single mode fiber for transport to the science cell. 
The 1040 nm light is generated and stabilized in a similar fashion with an M-squared pump laser and Ti:Sa laser operating at 1040 nm. 

Both locking schemes involve three feedback loops. A fast loop sends feedback to an electro-optic modulator (EOM) inside the SolsTiS cavity, which is responsible for feedback in the frequency range of 100 kHz - 10 MHz. This loop involves a PID using the Vescent D2-125 Laser Servo. The medium loop feeds back to the fast piezo in the SolsTiS, which has a bandwidth of DC-100kHz. This loop is a PID loop controlled by the fast loop in a modified Moglabs FSC (for larger output range). Lastly, there is a slow loop ($\sim 40~\rm Hz$ bandwidth), which controls the slow piezo in the SolsTiS, and is a PII loop, controlled with the same modified Moglabs FSC. The FSC units were modified to increase the voltage range of the integrator for long term locking (servos were modified for larger integrator rails). A diagram of the scheme is shown in  Fig. \ref{fig.Rydberglasers}. 

This locking scheme allows us to achieve a very narrow linewidth for the Rydberg lasers, with servo resonance peaks less than -50 dBC for frequencies greater than 20 kHz from the carrier. To measure the noise spectrum of the lasers we use a fiber based self-heterodyne system. This system beats light shifted by a 100 MHz AOM with a  time-delayed beam, split from the original laser and sent through a 10 km fiber. The beat signal is measured with a photodiode  to determine the laser spectrum. A measurement of the 918 nm spectrum is seen in Fig. \ref{fig.Rydberglasers}. Although we did not directly measure the carrier linewidth of the stabilized lasers, previous tests involving beating two systems constructed in a similar fashion indicate linewidths $\sim 200~\rm Hz$.

\section{Qubit coherence}

Trapped atom lifetimes limited by residual vacuum pressure were observed to be $\sim10~\rm s$ at the $1/e$ population decay point. The qubit $T_1$ time was $\sim 4 ~\rm s$ with approximately  equal lifetimes seen for the $\ket{0}\rightarrow \ket{1}$ and $\ket{1}\rightarrow \ket{0}$ transitions.

\begin{figure}[!t]
     \centering
     \includegraphics[width=.4\textwidth]{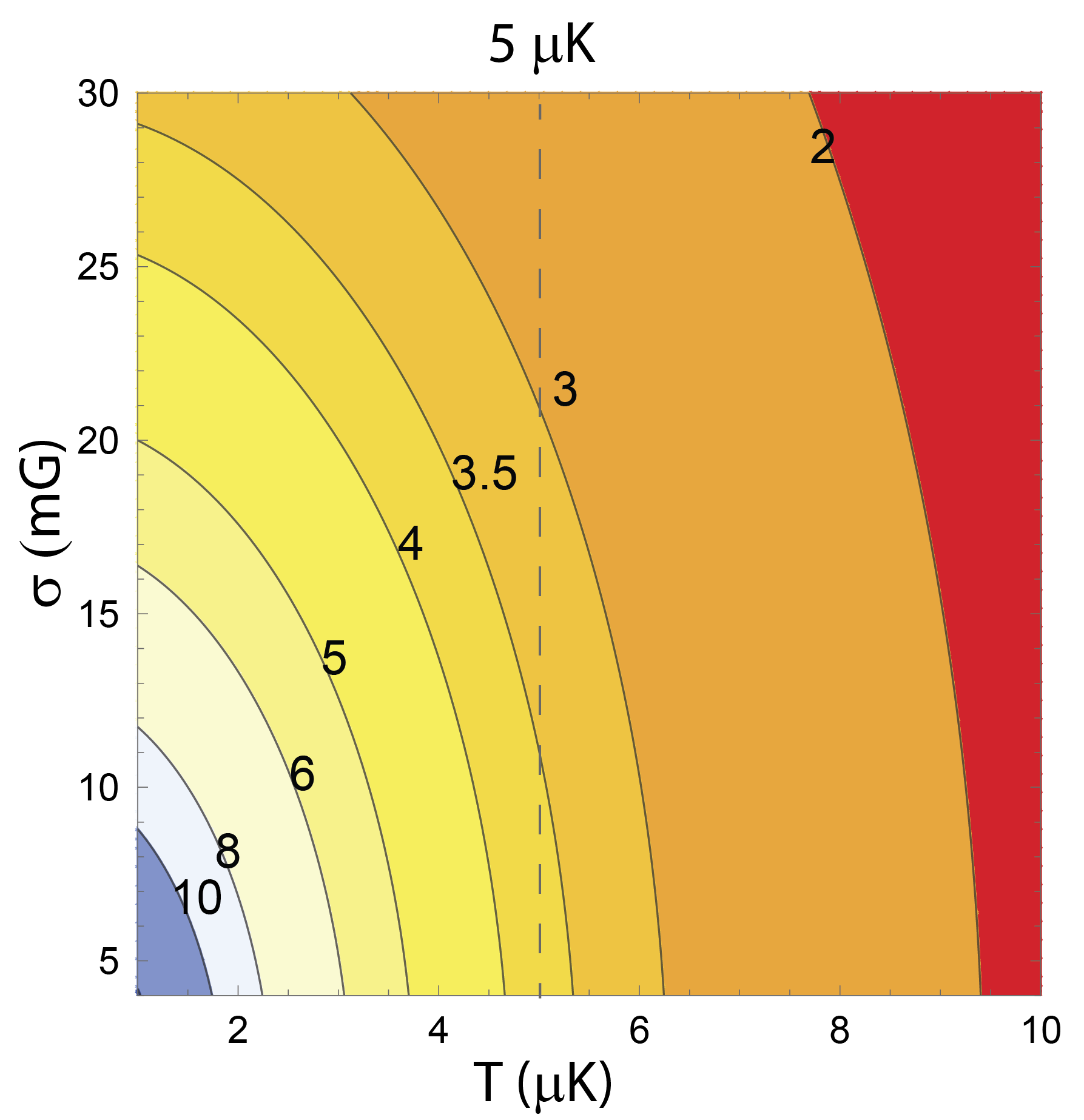}
     \caption{\label{fig.T2star} Calculated $T_2^*$ as a function of atom temperature and magnetic field noise for 825 nm trap light and a bias magnetic field of 1.6 mT. The contours are labeled with the value of $T_2^*$ in ms. The estimated magnetic noise based on magnetometer measurements is $\sigma < 20 ~\rm mG$, and the average measured coherence time is $\langle T_2^*\rangle = 3.5~\rm ms$, which indicates an atomic temperature near $5~\mu\rm K$. }
\end{figure}

The primary contributions to transverse qubit coherence are magnetic noise, intensity noise of the trap light, and atom motion causing time dependent qubit dephasing\cite{Saffman2005a,Kuhr2005}. We introduce a magnetic dephasing time $T_{2,B}^*$ and a motional dephasing time $T_{2,\rm motion}^*$. 
In a Gaussian approximation these can be combined to give 
\begin{equation}
T_{2}^*=\frac{T_{2,B}^*T_{2,\rm motion}^*}{\left[ \left(T_{2,B}^* \right)^2+\left(T_{2,\rm motion}^* \right)^2\right]^{1/2}}.
\label{eq.T2star}
\end{equation}
 
Assuming Gaussian magnetic noise with variance $\sigma^2$, the coherence time is 
$$
T_{2,B}^*=\frac{2^{1/2} \pi^2 \hbar^2 \nu_{\rm clock} }{\mu_B^2 B_0\sigma}
$$
with $\hbar$ Planck's constant, $\mu_B$ the Bohr magneton, $B_0$ the bias magnetic field, and $\nu_{\rm clock}=9192631770~\rm Hz$ the Cs clock frequency. A semi-classical approximation to the atomic motion gives for the motional coherence time
$$
T_{2,\rm motion}^*=1.947 \frac{\hbar}{k_B T_a |\eta|}
$$
with $k_B$ the Boltzmann constant, $T_a$ the atomic temperature, and $\eta$ a parameter that characterizes the differential Stark shift exerted on the qubit states by the trapping light. For 825 nm trapping light,  $\eta\simeq -0.00079$.
Although the motional dephasing rate is not strictly Gaussian, it can be well approximated as such, which leads to Eq. (\ref{eq.T2star}) for the combined magnetic and motional dephasing.

Qubit coherence was measured using Ramsey interference with microwave pulses. The average over the six sites used for the GHZ preparation experiment was $\langle T_2^*\rangle = 3.5~\rm ms$. The calculated $T_2^*$ from Eq. (\ref{eq.T2star}) is shown in Fig. \ref {fig.T2star}. On the basis of magnetic noise measurements and measured coherence time, we estimate the atomic temperature to be $T_a\simeq 5~\mu\rm K$. Independent temperature measurements based on trap drop and recapture are similar but tend to give a value $1-2~\mu\rm K$ higher. 

The calculated $T_2^*$ neglects any contribution from trap laser intensity noise. The blue detuned line array localizes atoms at local minima of the intensity, which reduces the sensitivity to trap laser noise.
For the parameters chosen in the experiment, the analysis in Sec. \ref{sec.trapanalysis} shows that the intensity seen by a cold trapped atom is about a factor of 22 smaller than the intensity corresponding to the trapping potential. This implies a factor of 22 reduction in sensitivity to intensity noise, compared to a red detuned trapping modality. The acceptable agreement between calculated and observed $T_2^*$ and $T_a$ values suggests that the contribution to the qubit coherence from intensity noise was not significant. This was the case even though the trap lasers were free running (two Ti:Sa lasers) without any additional stabilization or noise eating.

\section{Quantum gate set}

The native gate set used for circuits was a global ${\sf R}_\phi^{(G)}(\theta)$ gate implemented with microwaves, a local 
${\sf R_Z}(\theta)$ gate, and a $\sf C_Z$ gate acting on pairs of qubits. A local   ${\sf R}_\phi(\theta)$ gate is synthesized from 
${\sf R}_\phi^{(G)}(\theta)$ and local 
${\sf R_Z}(\theta)$ as explained below. 

\subsection{Global ${\sf R}_\phi^{(G)}(\theta)$ gates}

Global gates are driven by microwaves resonant with the Cs clock transition $6s\ket{f=4,m=0}-6s\ket{f=3,m=0}$.
 The microwave signal is derived from mixing a stable 9 GHz oscillator with the output of an arbitrary waveform generator (AWG).
Both the 9 GHz generator and the AWG clock are referenced to a 10 MHz timing signal that is derived from a GPS stabilized crystal oscillator. The negative sideband of the mixer is filtered out leaving a control signal centered at the 9.1926 GHz clock frequency. Changing the duration and phase of the  AWG output allows for arbitrary ${\sf R}_\phi(\theta)$ rotations with $\phi$ denoting the axis in the $x-y$ plane of the Bloch sphere about which the qubit is rotated and $\theta$ the pulse area.
Using a 40 W amplifier and a standard microwave horn located a few cm from the vacuum cell we achieve a Rabi frequency of 76.5 kHz.

\subsection{Local ${\sf R_Z}(\theta)$ gates}
\label{SM.LocalGates}

Single qubit ${\sf R_Z}(\theta)$ gates are implemented by addressing a qubit with 459 nm light that is detuned by $\Delta$ from the $6s_{1/2},f=4 - 7p_{1/2}$ transition. We can approximately describe the differential Stark shift on the qubit by ignoring the hyperfine structure of $7p_{1/2}$. In this approximation, we find 
a phase accumulation 
\begin{equation} 
\theta=\left[\frac{|\Omega_{459}|^2}{4\Delta}-
\frac{|\Omega_{459}|^2}{4(\Delta-\omega_q)}\right] t\equiv \Omega t
\end{equation} 
where $\Omega_{459}$ is the $6s_{1/2}-7p_{1/2}$ Rabi frequency, $\Omega$ is the effective qubit Rabi frequency,  $\omega_q$ is the qubit frequency (Cs clock frequency), and $t$ is the pulse duration.

\subsection{Local ${\sf R}_\phi(\theta)$ gates}

Global ${\sf R}_\phi^{(G)}(\theta)$ rotations and local ${\sf R_Z}(\theta)$ rotations are combined to give 
local ${\sf R}_\phi(\theta)$ gates
on individual qubits using the construction 
\begin{equation}
{\sf R}_\phi(\theta)={\sf R}_{\phi+\pi/2}^{(G)}(\pi/2){\sf R_Z}(\theta){\sf R}_{\phi+\pi/2}^{(G)}(-\pi/2).
\label{eq.localR}
\end{equation}

When compiling circuits, sequential appearance of ${\sf R}_\phi^{(G)}(\theta)$ and 
${\sf R}_\phi^{(G)}(-\theta)$
operations can be eliminated to reduce the gate count and circuit duration. 

\subsection{$\sf C_Z$ gates}

To complete a  universal gate set we implement $\sf C_Z$ gates using Rydberg interactions\cite{Jaksch2000,Saffman2010}. We have used the protocol introduced in \cite{Levine2019} combined with local Hadamard rotations to provide a $\sf CNOT$ gate. Tuning and characterization of the gate are described in detail in Sec. \ref{sec.symmetric}.

\section{One-qubit gate fidelity}

\begin{figure}[!t]
     \centering
 \includegraphics[width=.9\textwidth]{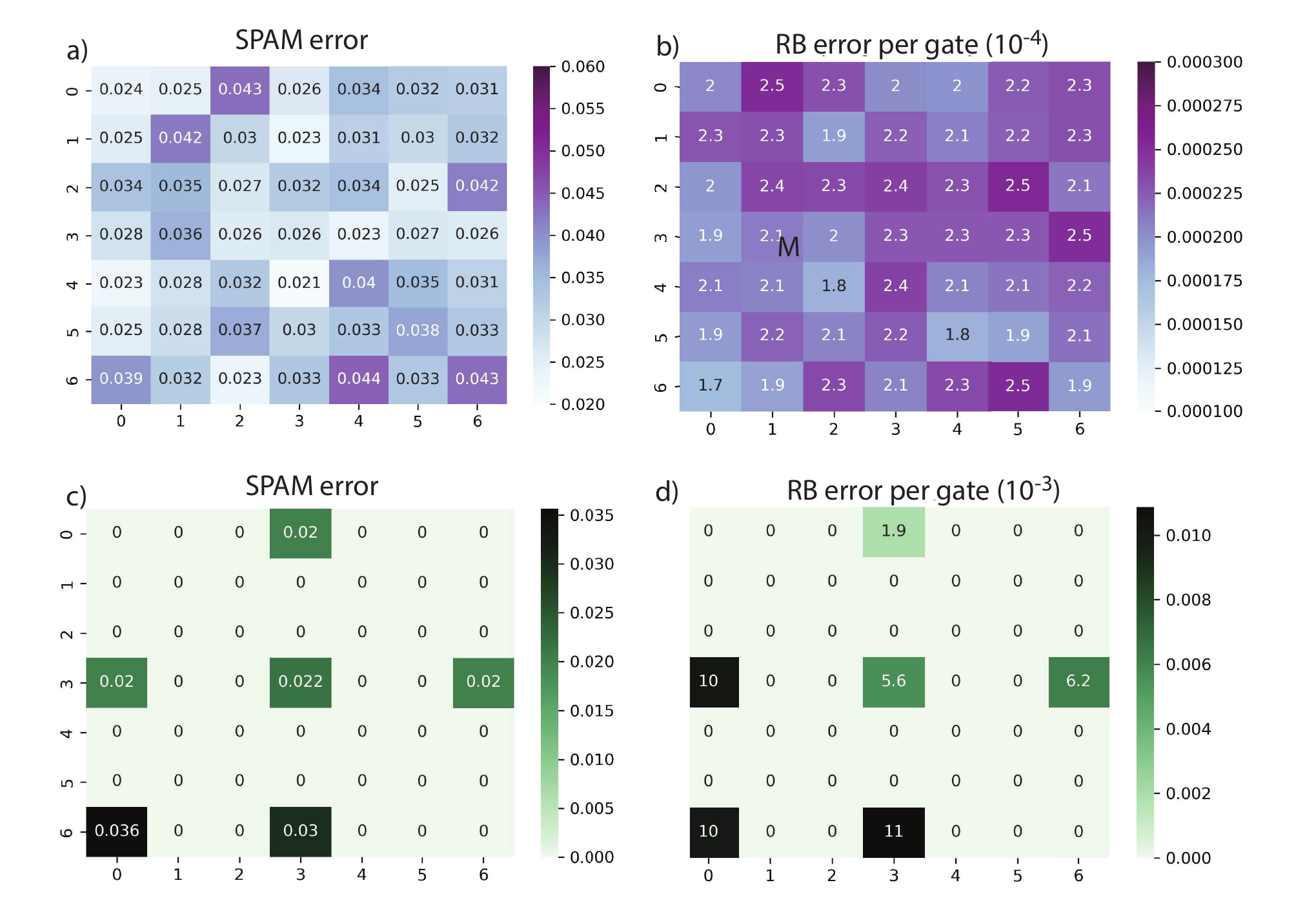}     \caption{\label{fig.RB} a) Characterization of SPAM errors in a 49 qubit array. The array averaged SPAM error is 3.1\%. b) ${\sf R}_\phi^{(G)}(\theta)$ gate fidelity for Clifford gates.  The array averaged error per gate is $2.2\times 10^{-3}$. c) Characterization of SPAM errors  for the 6 sites used in the main text.  The  average SPAM error is 2.5\%. d) Fidelity of ${\sf R_Z}(\theta)$ and  ${\sf R}_\phi(\theta)$ gates. The average error per gate    is $7.5\times 10^{-3}$. }
\end{figure}

Global microwave rotation gates have been shown in our earlier
work to have a fidelity of $0.998$ from randomized benchmarking experiments\cite{Xia2015}. The primary difference between the earlier work and the present experiment is the introduction of a higher power microwave amplifier in order to increase the Rabi frequency to 76.5 kHz. The fidelity of  the ${\sf R}_\phi^{(G)}(\theta)$ gates was characterized at each site of a  $7\times 7= 49$ qubit array using randomized benchmarking over the Clifford group. The results for the  SPAM error per qubit and gate fidelity are shown in Fig. \ref{fig.RB}. 

Local ${\sf R_Z}(\theta)$ gates use a detuned laser pulse to impart a differential Stark shift on the qubit states. The gate rotation angle is proportional to the integrated intensity at the atom during the pulse.  These gates are sensitive to several primary error mechanisms. The first is fluctuations in the pulse intensity on time scales slow compared to the duration of a single pulse. This mechanism is analyzed in Sec. \ref{sec.SM_1q_intensity noise}. The second is variations in the intensity seen by the atom due to position variations under the Gaussian envelope of the addressing beam. The third is photon scattering from the detuned laser pulse. 

The fidelity of  local single qubit  ${\sf R_Z}(\theta)$ and  ${\sf R}_\phi(\theta)$ gates was characterized at the 6 sites used for GHZ state preparation and algorithm demonstrations using randomized benchmarking over the Clifford group. The results for the  SPAM error per qubit and gate fidelity are shown in Fig. \ref{fig.RB}.

\subsection{Dephasing from low frequency intensity noise}
\label{sec.SM_1q_intensity noise}

Shot to shot variations in the intensity lead to dephasing of the qubit rotations. Let the optical intensity be normally distributed according to $p(I)=\frac{I_0}{\sqrt{2\pi\sigma_I^2}}e^{-(I-I_0)^2/2\sigma_I^2}$ with standard deviation $\sigma_I$. Assuming $\Delta$ is large compared to the linewidth of the 459 nm light, we can write $\Omega =a I$ with $a$ a constant. It follows that the gate phase (pulse area $\theta$ of the ${\sf R_Z}(\theta)$ gate) is 
$$
\theta =  a I t = a I \frac{\theta_0}{\Omega_0} 
$$
where $\theta_0= \Omega_0 t$. 
Assuming Gaussian intensity noise, the phase is distributed as
$$
p(\theta)=\frac{1}{\sqrt{2\pi\sigma_\theta^2}}e^{-(\theta-\theta_0)^2/2\sigma_\theta^2}
$$
where
$$
\sigma_\theta=  \frac{a\theta_0}{\Omega_0}\sigma_I= \theta_0 \frac{\sigma_I}{I_0}.
$$
We see that the phase uncertainty increases with $\theta_0$,  which implies a decreasing oscillation amplitude  proportional to the pulse area.

\begin{figure}[!t]
     \centering
 \includegraphics[width=1.\textwidth]{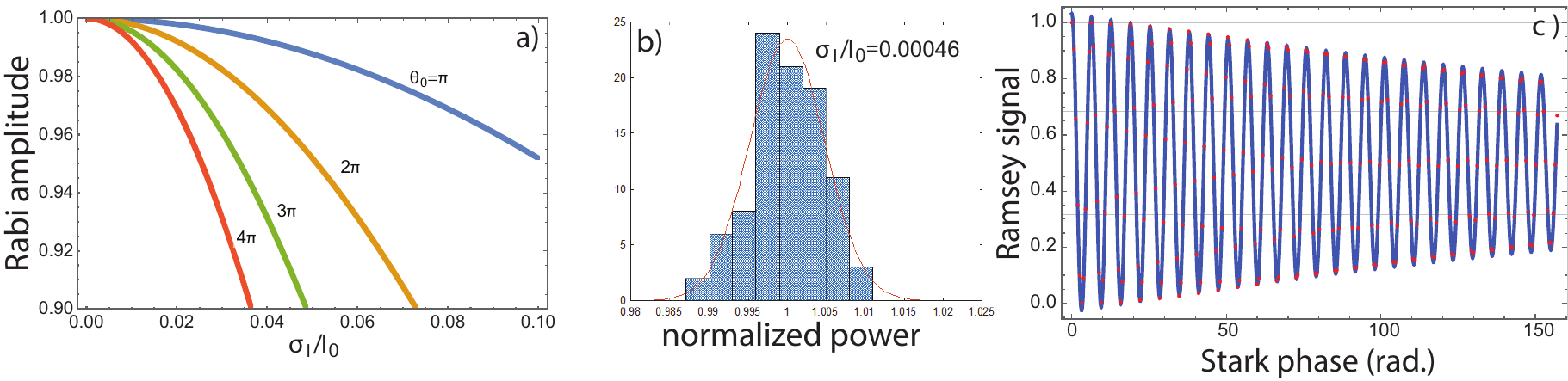}
     \caption{\label{fig.Rabiphaseerror}One-qubit gate analysis {\bf a)} Amplitude of Rabi oscillation due to shot to shot intensity noise  for several values of the pulse area. {\bf b)} Measured shot to shot variation of the 459 pulse power at a location after the AOD scanners in front of the vacuum cell. {\bf c)} Simulated Ramsey signal with ${\sf R_Z}(\phi)$ gate (red dots) and fit to
     $a_1+ a_2\cos(a_3 \phi)e^{-\phi/a_4} $ (blue lines), which gives $f\tau=46$. The numerical simulation integrated over the 3D atom and light distributions using $\lambda_{\rm trap}=825~\rm nm$, $w_{\rm line}=1~\mu\rm m$, $d=3~\mu\rm m$, $U_{\rm d}=300~\mu\rm K$,  $w_{459}=3~\mu\rm m$, $T_{\rm a}=5~\mu\rm K$.    }
\end{figure}

The Rabi oscillation amplitude is proportional to $\langle e^{\imath\theta}\rangle$, and assuming a Gaussian intensity distribution, the oscillation amplitude will decay as 
\begin{equation} 
\langle e^{\imath\theta}\rangle
=\int_{-\infty}^{\infty} d\theta\, e^{\imath\theta} \frac{1}{\sqrt{2\pi\sigma_\theta^2}}e^{-(\theta-\theta_0)^2/2\sigma_\theta^2}
=e^{\imath\theta_0}e^{-\sigma_\theta^2/2}
=e^{\imath\theta_0}e^{-\theta_0^2\sigma_I^2/2 I_0^2}.
\label{eq.Rabidephase}
\end{equation}

The Rabi amplitude as a function of intensity noise is shown in Fig. \ref{fig.Rabiphaseerror} together with measured shot to shot power fluctuations. To maintain good pulse stability for extended operation times, the optical power inside the science cell is periodically sampled to generate an error signal that is fed back to a rotatable waveplate and polarizer combination on the laser table. The observed  fluctuations imply an expected error of  $10^{-4}$ for a $\pi$ pulse, which is negligible compared to the  observed gate fidelity of $\sim0.01$ for a $\pi$ pulse. 

\subsection{Dephasing from atom position variations}

The observed gate infidelity is dominated by the second and third error  mechanisms. Atomic motion causes the atom to see a slightly different intensity for each shot. The time scale of the motion ($\sim 1/(20 ~\rm kHz)$) is long compared to the gate time so we may assume the intensity is constant during the gate. This effect can be described analytically or numerically\cite{Graham2019,Robicheaux2021} and leads to an exponential decay of the Rabi amplitude with the length of the pulse. We define a figure of merit as $f\tau$ the product of the Rabi frequency $f$ in the time $\tau$ for the amplitude to decay to $1/e$. This implies an error per $\pi$ pulse of $\epsilon=(1/e)/(2f\tau)=0.0046.$ A numerical simulation using experimental parameters, see Fig. \ref{fig.Rabiphaseerror}, gives 
$f\tau=46$ and an error per $\pi$ pulse of $\epsilon=0.0040.$

\subsection{Light scattering}

The final error contribution is due to spontaneous scattering from the $7p_{1/2}$ level. Since the detuning from $7p_{1/2}$ is small compared to the qubit frequency the scattering error is negligible for atoms in $\ket{0}=\ket{f=3,m=0}$ and for an atom in $\ket{1}=\ket{f=4,m=0}$ the scattering probability in a $\pi$ pulse is approximately 
$$
\epsilon_{\rm scatter}=\left(\frac{1}{2}\right)\frac{\pi/\Omega}{\tau_{7p}} \frac{|\Omega_{459}|^2}{2\Delta^2} =\left(\frac{1}{2}\right)\frac{2\pi}{\Delta\tau_{7p_{1/2}}}.
$$
This expression uses the standard result of $\Omega^2/2\Delta^2$ for the time averaged excited state population of a two-level system with detuned drive multiplied by a prefactor of $1/2$, which accounts for the coherence decay in the limit of a long pulse time compared to the excited state lifetime. 
Using $\Delta=2\pi\times 760 ~\rm MHz$ and $\tau_{7p_{1/2}}=155~\rm ns$ we find $\epsilon_{\rm scatter}=0.0042$.  This error can be reduced by operating at larger detuning.  

To summarize this section, we estimate the errors for the three mechanisms as $\epsilon_{\rm intensity}=0.0001,$
$\epsilon_{\rm position}=0.0040$, and 
$\epsilon_{\rm scatter}=0.0042$.
Adding these errors in quadrature gives an estimate of $\epsilon=0.0044$, which would correspond to $f\tau=42$.
Experimental tests of the Ramsey amplitude as a function of the length of an embedded 
Stark pulse show up to $f\tau =40$, which is consistent with these estimates. 
The qubits used in the main text had a somewhat higher  average gate error from ramdomized benchmarking of $0.0075$ as is shown in Fig. \ref{fig.RB}.
In order to improve the local ${\sf R_Z}(\theta)$ gate fidelity further, tighter confinement from lower temperature or deeper traps, as well as larger detuning to reduce light scattering 
will be needed.

\section{$\sf C_Z$ gate tuning and characterization}

\label{sec.symmetric}

\begin{figure}[!t]
\centering
\includegraphics[width=.95\textwidth]{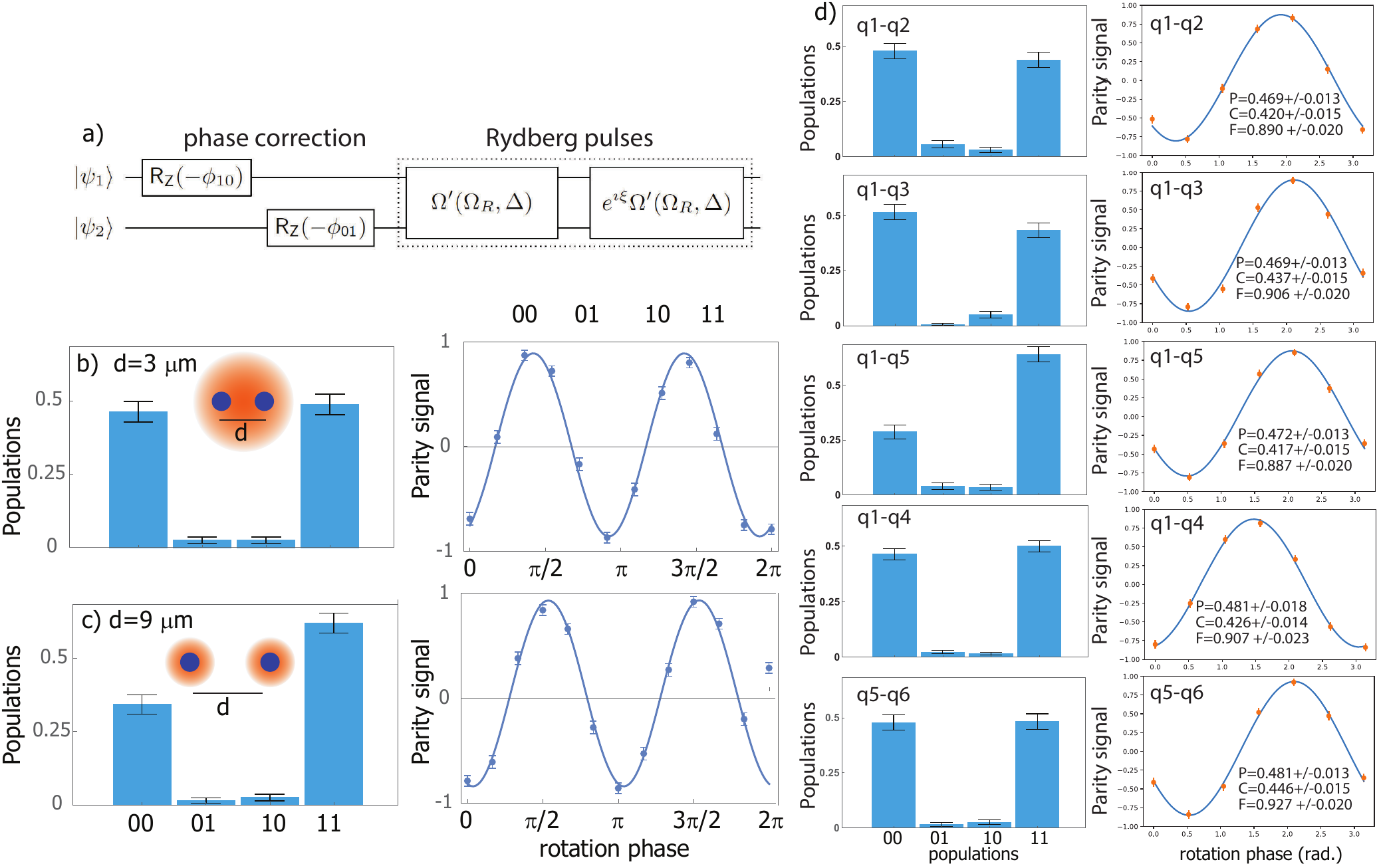}
\caption{$\sf C_Z$ gate implementation and characterization by preparation of Bell states. a) Symmetric $\sf C_Z$ gate with phase correction pulses. The gate in the dashed box is the uncorrected $\sf C_Z'$ that is composed of two Rydberg pulses that collectively drive both atoms.  Pulses are detuned from resonance by $\Delta$ and have a single-atom resonant Rabi frequency of $\Omega_R$. The detuned, single atom Rabi frequency is $\Omega'=\sqrt{\Omega_R^2 + \Delta^2}$. The second pulse is driven with a phase $\xi$ relative to the first pulse.  The two phase compensation pulses correct residual single-atom phases, $\phi_{01}$ and $\phi_{10}$, and are used to transform the gate into a canonical $\sf C_Z$. In principle the  compensation phases should be the same, but in practice we find better gate performance is achieved by allowing the phases to differ, which compensates for lack of perfect balance between the Rydberg pulses at the two sites.  The phase compensation pulses can be applied before or after the Rydberg pulses.  b) Two qubits spaced by $d=3~\mu\rm m$ addressed with large beams of waist $w=7.5~\mu\rm m$ focused halfway in between the qubits. At $d=3~\mu\rm m$ there is a strong blockade of ${\sf B}/2\pi=1.03~\rm GHz$.  Measured values were $P_{\rm Bell}=0.475(0.01)$, parity amplitude $C=0.440(0.01)$ and fidelity $F_{\rm Bell}=0.914(0.014)$.   c) Two qubits spaced by $d=9~\mu\rm m$ addressed with separate beams with waist $w=3~\mu\rm m$. At this spacing the blockade is weak, ${\sf B}/2\pi=3.0~\rm MHz$.  Measured values were 
     $P_{\rm Bell}=0.483(0.009)$, parity amplitude $C=0.444(0.010)$ and fidelity $F_{\rm Bell}=0.927(0.013)$. d) Characterization of $\sf C_Z$ gate fidelity for five qubit pairs. Reported values are without SPAM correction and the average fidelity is 0.90.} 
\label{fig.CZgate}
\end{figure}

We use the symmetric $\sf C_Z$ gate described by Levine et al.\cite{Levine2019}.  This gate is composed of two detuned Rydberg excitation pulses collectively driving two selected sites. Each pulse is designed to give the $\ket{11}$ state a $2\pi$ rotation. The $\ket{10}$ and $\ket{01}$ states receive only a partial rotation from each pulse. The relative phases of the  two pulses is adjusted such that these states return to the ground state at the end of the second pulse, see Fig. \ref{fig.CZgate}a). The phase that each state acquires during these gate pulses depends on the area enclosed on the Bloch sphere during the state evolution.  By adjusting the detuning and phase between the two pulses, the phase acquired by each of these terms can be tuned such that $\phi_{11} - \phi_{10} - \phi_{01} = (2n+1)\pi$, where $\phi_{ij}$ is the phase state $\ket{ij}$ acquires during the gate pulses and $n$ is an integer. Provided this condition is satisfied, the gate is maximally entangling and can be converted to a canonical $\sf C_Z$ gate with local phase rotations. 

Our optical control architecture is different from that in Ref. \cite{Levine2019}, so our gate calibration and characterization protocols are also somewhat different. In Ref. \cite{Levine2019} Rydberg beams with large waists $\sim 20~\mu\rm m$ propagated along  a line of atoms such that each atom saw essentially the same intensity.  In our implementation, Rydberg excitation beams are tightly focused to $w=3~\mu\rm m$ and propagate perpendicular to the plane of the qubit array. This allows for individual control of each atom, but also requires additional calibration to ensure uniform coupling to each atom when implementing a $\sf C_Z$ gate. Using multiple tones driving the scanner AODs, we simultaneously  drive Rydberg transitions on both atoms with the same two-photon excitation frequency, see Sec. \ref{SM.collective_addressing} for more details.  To symmetrically illuminate both atoms, the power of both Rydberg beams was balanced by tuning the power in each AOD tone such that the diffracted beam powers were balanced when viewed on a monitor camera.  Fine tuning for the intensity balance of the 459 nm beam was performed using $\sf R_Z$ rotations on both sites. The rotation angle on each site was measured in a ground state Ramsey experiment, confirming  the 459 nm beam intensities were matched to within $2\%$.  Fine tuning for the power balance of the 1040 nm beam was accomplished by adding  $9.2$ GHz sidebands to the beam to drive Raman transitions. By balancing the Raman Rabi frequency on both sites, we confirmed that the 1040 nm intensity on both sites was balanced to within $2\%$.  Once beam powers were tuned using the method described above, the 2-photon Rydberg Rabi frequency was matched to within $5\%$.  

For the circuits demonstrated in the main text, we used a qubit separation of $d=9~\mu\rm m$.
We  have also demonstrated a  $\sf C_Z$ gate with two sites which were separated by only $3 ~\mu \rm m$. In this configuration, the Rydberg beams were reconfigured to have a $7.5~ \mu \rm m$ waist that was focused midway between the two selected sites. To symmetrically illuminate both sites in this configuration, the beam alignment was scanned by adjusting the AOD frequency until  the intensity on both sites was equal to within $2\%$. As described above, $\sf R_Z$ rotations and Raman Rabi oscillations were used to balance 459 nm and 1040 nm intensities on the two sites. Note that this configuration is not compatible with single-site addressing and was not used in circuit experiments but demonstrates the ability to tune gate parameters to operate with very different qubit spacings and very different Rydberg interaction strengths.

After the intensities addressing the two atoms were balanced, we calibrated the $\sf C_Z$ gate pulses. The first step in this process was calculating the optimal detuning, $\Delta$ for each pulse such that the two-atom states acquire the correct phases as described above. The Rydberg blockade shift between selected sites was $3$ MHz for 
$d=9~ \mu \rm m$  ($1.03~\rm  GHz$ for $d=3~ \mu \rm m$) using the $75s_{1/2}$ Rydberg state. We set the single-atom resonant Rydberg Rabi frequency to be  $\Omega_R/2\pi = 1.7~\rm MHz$.  Given these parameters, we calculated that optimal gate pulses should be detuned by $\Delta = -0.250 \Omega_R$ for  $d=9~ \mu \rm m$ ($-0.377$ for $d=3 ~\mu \rm m$ ). Calculations were performed by numerically solving the time-dependent Hamiltonian as described in \cite{Saffman2020} and selecting optimal gate parameters by inspection. The pulse length, $\tau$, and the relative phase between the two pulses, $\xi$, were then fine-tuned using Rydberg excitation experiments. We tuned $\tau$ using a single $\Delta$ detuned Rydberg pulse to drive $\ket{11}$.  The pulse length time, $\tau$, was scanned about the calculated $2 \pi$ to optimize the population returning to $\ket{11}$. Once $\tau$ was optimized, we drove the state, $\ket{10}$, with two gate pulses while scanning the phase between them, $\xi$, to maximize the single atom return to ground.

After optimal $\Delta$, $\tau$, and $\xi$ were determined, the phases on the $\ket{10}$ and $\ket{01}$ states, $\phi_{01}$ and $\phi_{10}$ respectively, were compensated to obtain a canonical $\sf C_Z$ gate. We performed this compensation  using local $\sf R_Z$ gates with focused  459 nm pulses as described in Sec. \ref{SM.LocalGates}. The compensation pulse lengths were calibrated with ground Ramsey experiments that had a $\sf C_Z$ gate (with compensation pulses) sandwiched between two global $\pi/2$ pulses (see Sec. \ref{sec.circuits}). In these Ramsey experiments, only one atom of the selected pair was loaded into the array. The phase compensation pulse time was scanned to maximize the atom retention. This condition corresponds to compensating the $\phi_{01}$ and $\phi_{10}$ phases.

After the $\sf C_Z$ gates were calibrated, we measured their performance by preparing  Bell states $\ket{\psi} = \frac{\ket{00}+\ket{11}}{\sqrt2}$ using the circuit listed in Sec. \ref{SM.GHZCircuits} and measuring the Bell state fidelity. We performed this characterization by measuring the parity and Bell state populations as described in the main text. The parity and populations were used to calculate the fidelity of a two-qubit Bell state (see Fig. \ref{fig.CZgate}), which gave a maximum observed fidelity of $F_{\rm Bell}=0.927(0.013)$ without SPAM correction. A similar calibration procedure was performed for each gate pair; the average fidelity measured without SPAM correction was $0.90$ (see Fig. \ref{fig.CZgate}). 

SPAM errors significantly contribute to the observed raw fidelity. We calibrate this error based on several experiments. The measurement error is dominated by atom loss during the readout process; this loss was measured to be $\sim 1.5\%$ per atom. Imperfect optical pumping to the $\ket{1}$ state was found to be the main source of state preparation errors, contributing between $0.0\%$ and $0.5\%$ per atom depending on the atom site. The SPAM errors shown in Fig. \ref{fig.RB} which were extracted from randomized benchmarking were 2.5\% per qubit on average. Simply subtracting the SPAM errors from the raw infidelity overestimates the corrected gate fidelity.

To get a more accurate estimate of the SPAM corrected fidelity we use the measured SPAM values with a quantum process 
analysis\cite{Graham2019}. The analysis models  how state preparation and measurement errors affect the measured output state through a two-qubit quantum process formalism. Imperfect retention is modeled as loss that is split between the two atom readout periods. Similarly, atoms which are not pumped to the $m=0$ Zeeman state of the $f=4$ hyperfine manifold are modeled as atom loss out of the qubit basis during the state preparation. We  then propagate the initial state through these error channels and an ideal $\sf C_Z$ gate, and  observe how much the SPAM affects the population and parity oscillations.  We estimate that SPAM errors contribute between $2.2-3.1\%$ error to the measured Bell state fidelities. Thus, the maximum observed fidelity with SPAM correction was between  0.949 and 0.958, and the average SPAM corrected Bell state fidelity was between $0.921$ and $0.931$. Note that these methods do not include some of the subtleties of how blowaway based state measurement biases the measurements in the $\ket{1}$ state \cite{Levine2019}. Correcting this effect requires measurements which were not performed during the gate characterization. 
Note that the traps were turned off during Rydberg excitation pulses for each $\sf C_Z$ gate. This prevented position- and trap power- dependent dephasing. Future experiments using magic trapping of Rydberg states should remove the need for turning off the trap light during Rydberg experiments \cite{SZhang2011}.

\clearpage 

\section{Circuit diagrams}
\label{sec.circuits}

In this section we list the circuit diagrams for Figs. 2,3,4 in the main text compiled down to the native set of  
${\sf R}_\phi^{(G)}(\theta)$, ${\sf R_Z}(\theta)$, and $\sf C_Z$ gates. 
The notation in the diagrams is 
\begin{eqnarray}
{\rm PhX}(x)^y &=& {\sf R}_{x \pi}(y \pi),\nonumber\\
{\rm S} &=& {\sf R_Z}(\pi/2),~~~{\rm S}^{-1} = {\sf R_Z}(-\pi/2),\\
{\rm T }&=&{\sf R_Z}(\pi/4),~~~{\rm T}^{-1} = {\sf R_Z}(-\pi/4).\nonumber
\end{eqnarray}
In these definitions irrelevant global phases are ignored. Note that negative pulse areas can be interpreted with the identity ${\sf R}_\phi(-\theta)={\sf R}_{\phi+\pi}(\theta)$. 
Some of the  circuit diagrams are presented with different single qubit ${\sf R}_\phi(\theta)$ or ${\sf R_Z}(\theta)$ rotations in the same time slice. In the actual implementation these are unrolled into multiple time slices. Also for compactness some of the local ${\sf R}_\phi(\theta)$ gates have not been unrolled in the diagrams into local $\sf R_Z$ and global gates using Eq. (\ref{eq.localR}).

\subsection{GHZ circuits}
\label{SM.GHZCircuits}

The qubit lines are labelled with the indices given in the image of the qubits. The circuit execution times for $N=2-6$ were $40, 90, 120,150,180~\mu\rm s.$ \\

 \includegraphics[width=1.\textwidth]{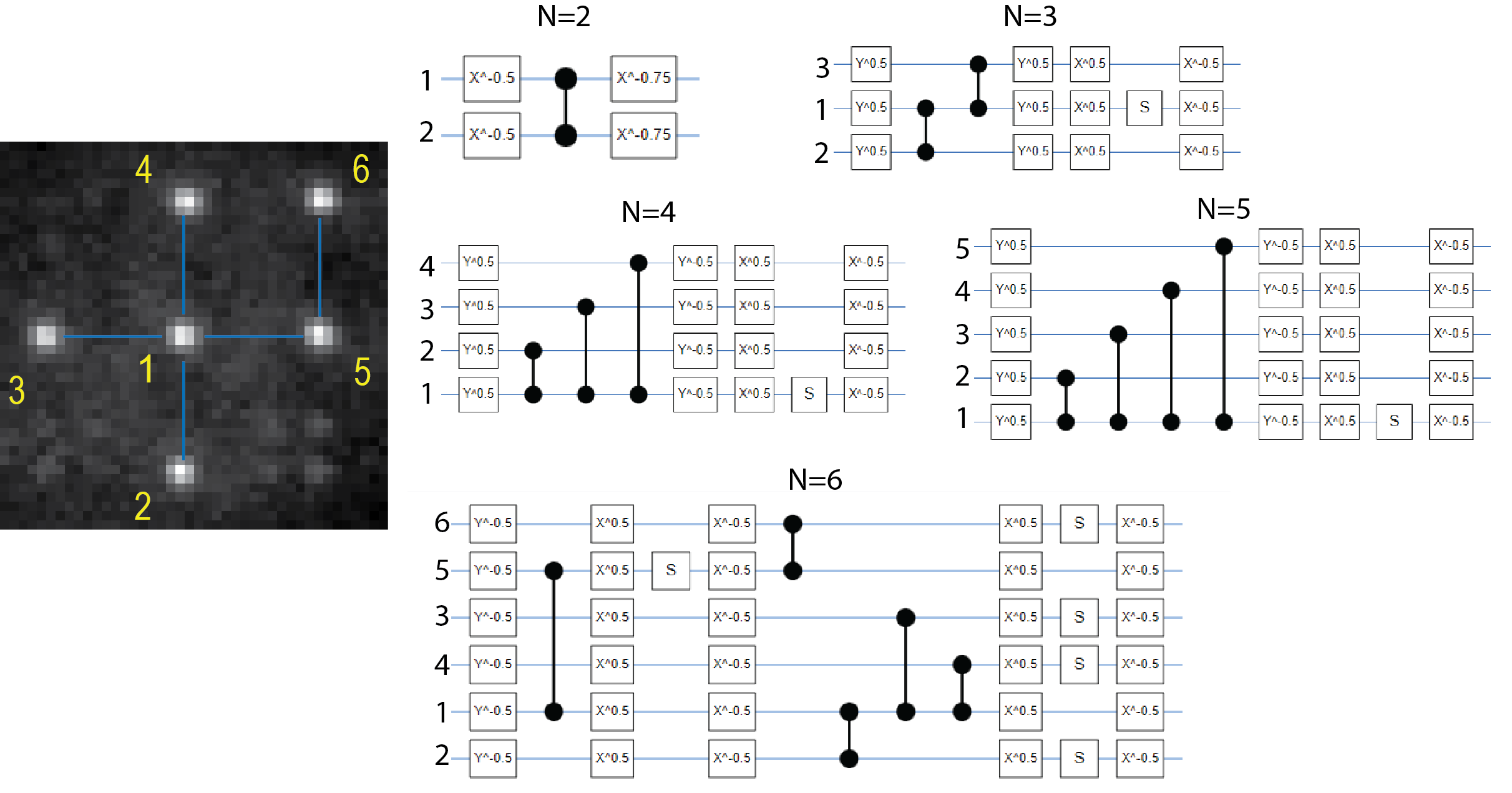}\\

\subsection{Phase estimation circuits}

The execution times for the three qubit circuits
with ${\sf U}= {\sf I}, {\sf Z}^{1/2}, {\sf Z}, {\sf Z}^{3/2}$ were $390, 460, 420, 430~\mu\rm s.$\\ 

${\sf U}=\sf I$: 

 \includegraphics[width=.8\textwidth]{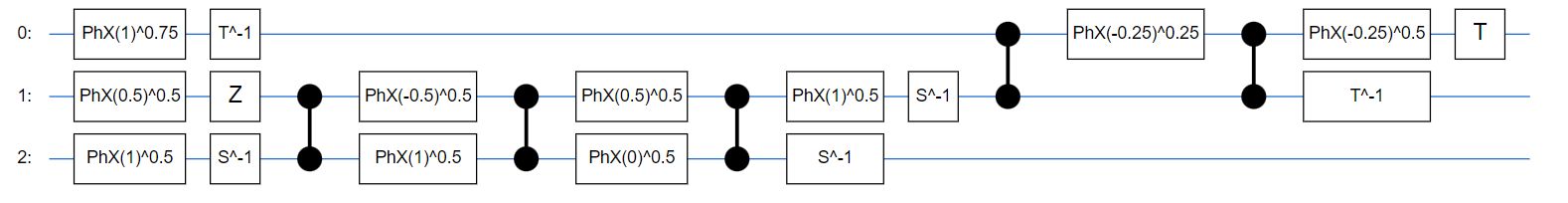}\\
 
 ${\sf U}={\sf Z}^{1/2}$:
 
  \includegraphics[width=.8\textwidth]{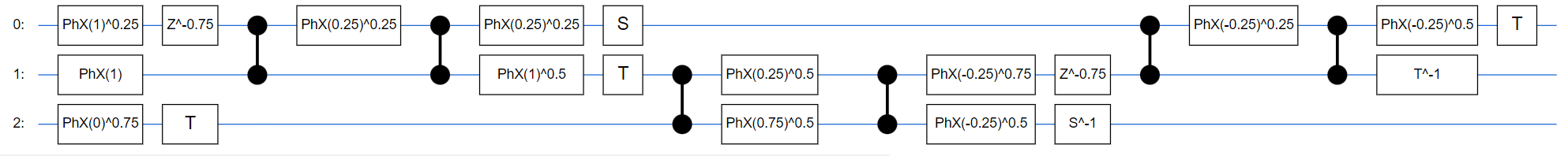}\\

${\sf U}=\sf Z$:

 \includegraphics[width=.8\textwidth]{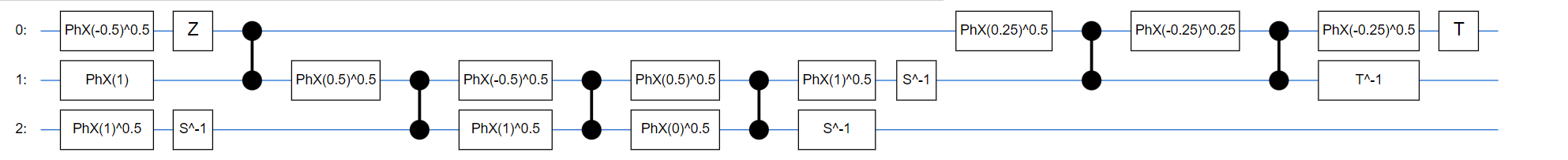}\\

${\sf U}={\sf Z}^{3/2}$:

 \includegraphics[width=.8\textwidth]{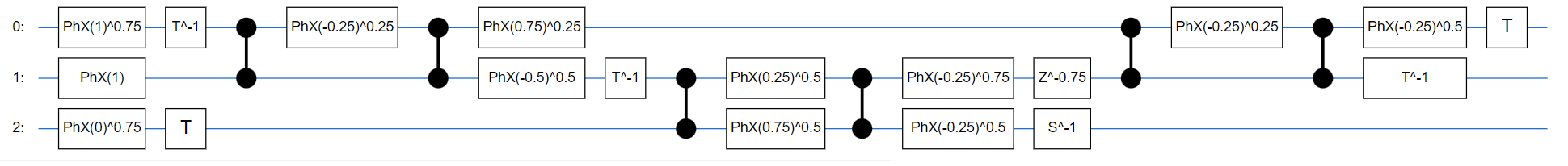}\\

Four qubit phase estimation circuit with
${\sf U}=e^{\imath 0.59{\sf X}}e^{\imath2.55\sf{Z}}$ (execution time $1100~\mu\rm s$):\\

 \includegraphics[width=.9\textwidth]{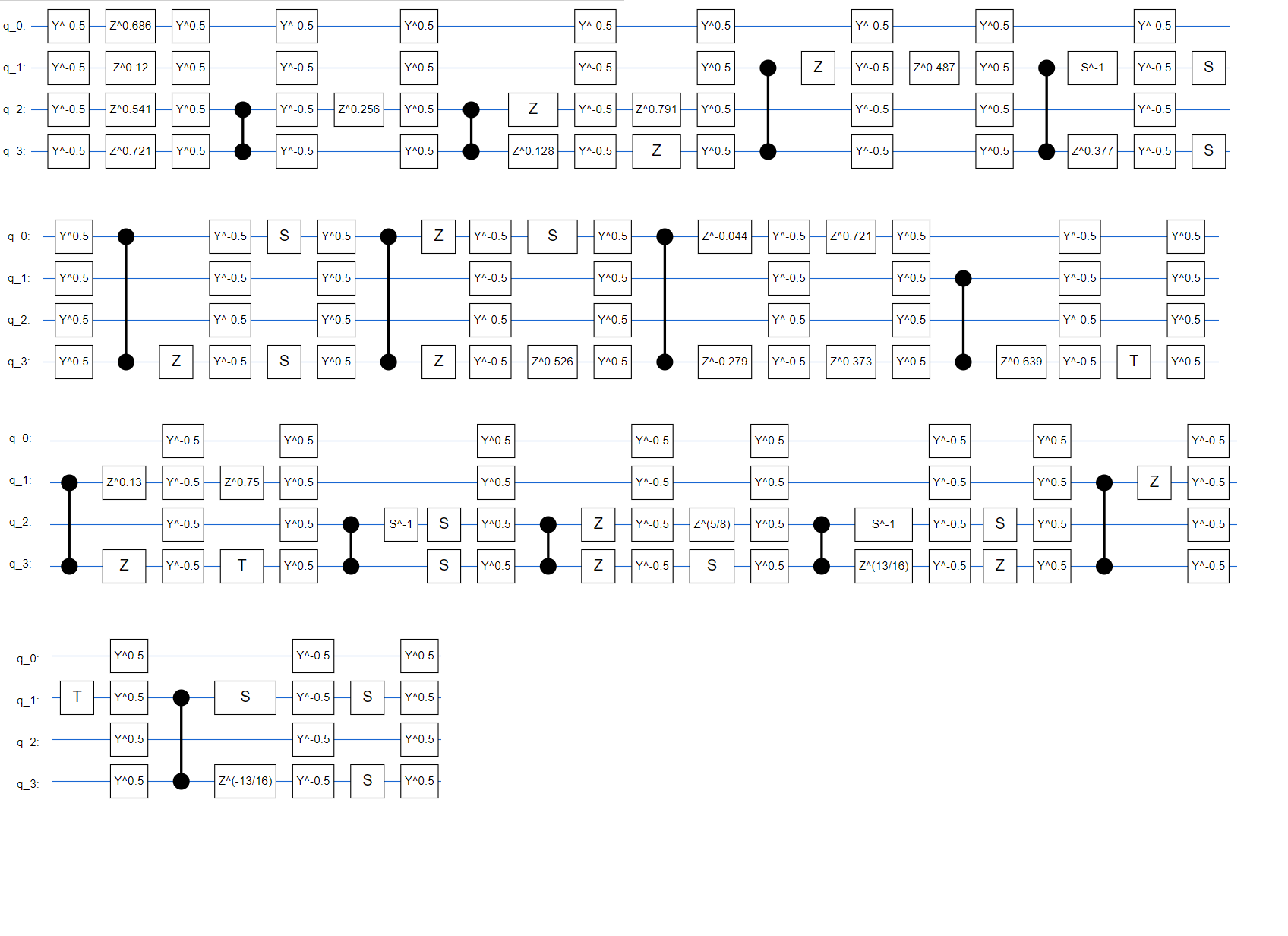}\\

\vspace{-2.cm}

\subsection{QAOA circuits}

Line graph with 3 qubits, $p=1$ circuit, execution time $240~\mu\rm s$,   $\beta=1.25, \gamma=1.67$:\\

 \includegraphics[width=.7\textwidth]{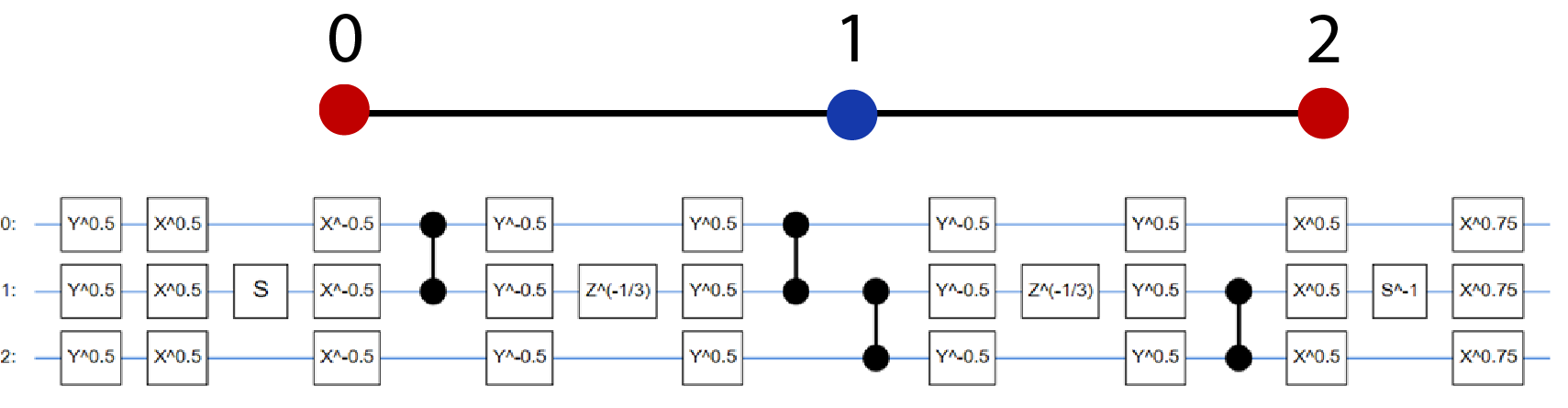}\\
 
Line graph with 3 qubits, $p=2$ circuit, execution time $440~\mu\rm s$, 
${\boldsymbol \beta}=(0.331,0.229)$, ${\boldsymbol \gamma}=(1.66,1.44)$:\\

 \includegraphics[width=.9\textwidth]{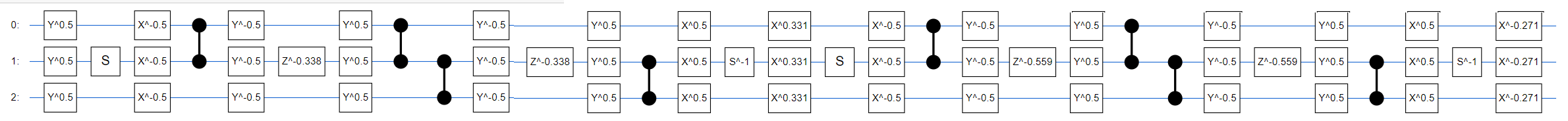}\\

Graph in ``T" geometry with 4 qubits, $p=1$ circuit, execution time $320~\mu\rm s$.  $\beta=0.750, \gamma=0.696$:\\

\includegraphics[width=.65\textwidth]{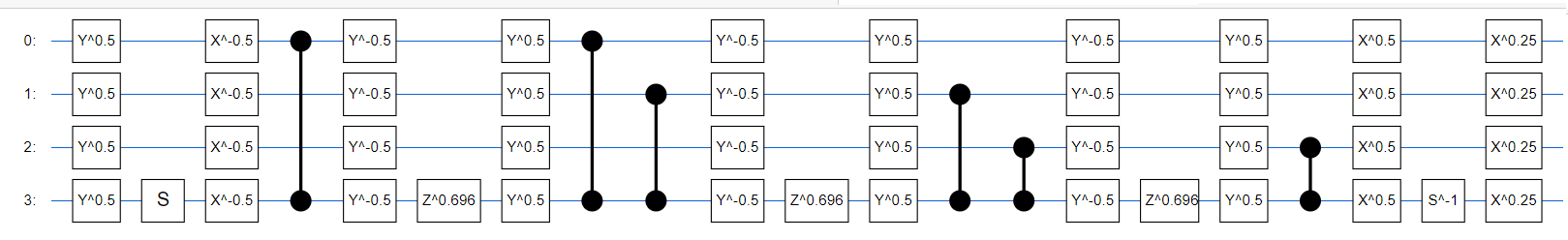}\\
 
Graph in ``T" geometry with 4 qubits, $p=2$ circuit, execution time $610~\mu\rm s$, 
${\boldsymbol \beta}=(1.71,1.19)$, ${\boldsymbol \gamma}=(0.700,0.624)$:\\

\includegraphics[width=1.\textwidth]{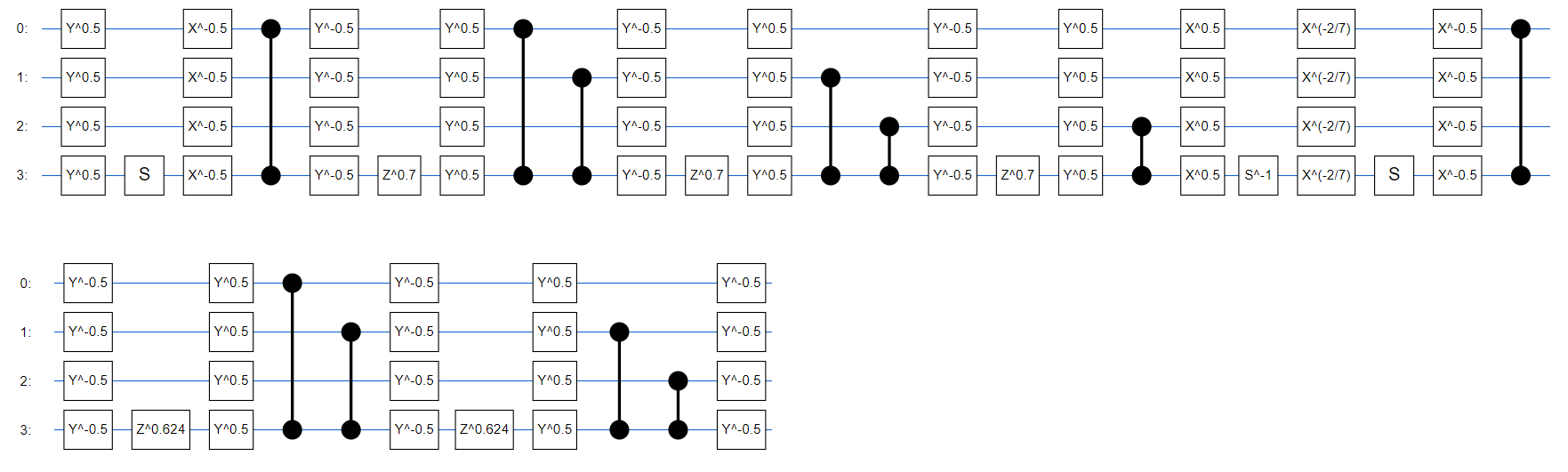}\\

Graph in ``T" geometry with 4 qubits, $p=3$ circuit, execution time $900~\mu\rm s$,
${\boldsymbol \beta}=(1.63,1.77,0.172)$, ${\boldsymbol \gamma}=(0.194,0.424,1.39)$:\\

\includegraphics[width=.9\textwidth]{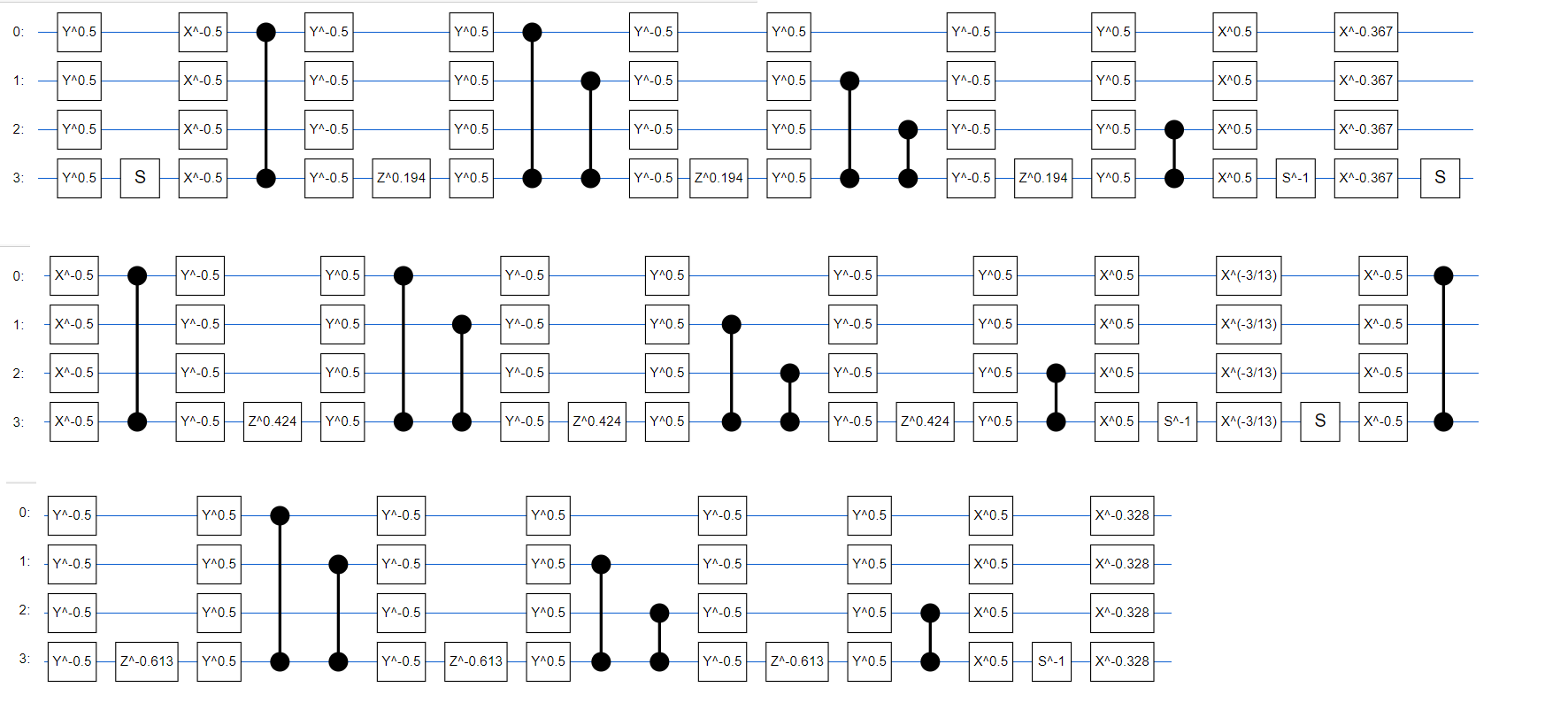}\\

\clearpage

\end{document}